\documentclass[11pt]{article}
\pdfoutput=1
\usepackage{amsmath,epsf,amssymb,latexsym,array,color}
\usepackage{graphicx,wasysym}
\usepackage{cancel}
\usepackage{tikz}
\newcommand*\circled[1]{\tikz[baseline=(char.base)]{
            \node[shape=circle,draw,inner sep=2pt] (char) {#1};}}

\def\be{\begin{equation}}
\def\ee{\end{equation}}
\def\ba{\begin{array}}
\def\ea{\end{array}}
\newcommand{\beq}{\begin{equation}}
\newcommand{\eeq}[1]{\label{#1}\end{equation}}
\newcommand{\bea}{\begin{eqnarray}}
\newcommand{\eea}[1]{\label{#1}\end{eqnarray}}

\usepackage{ifpdf}
\ifx\pdfoutput\undefined
   \pdffalse
\else
   \pdfoutput=1
   \pdftrue
  \usepackage[pdftex]{hyperref}
\numberwithin{equation}{subsection}

\begin{document}


\begin{titlepage}

\hskip 1.5cm

\begin{center}
{\huge \bf{Component Actions of Liberated $\mathcal{N}=1$ Supergravity and New Fayet-Iliopoulos Terms in Superconformal Tensor Calculus}}
\vskip 0.8cm  
{\bf \large Hun Jang\footnote{hun.jang@nyu.edu} and Massimo Porrati\footnote{massimo.porrati@nyu.edu}}  
\vskip 0.75cm

{\em Center for Cosmology and Particle Physics\\
	Department of Physics, New York University \\
	726 Broadway, New York, NY 10003, USA}
	
	\vspace{12pt}

\end{center}

\begin{abstract}
We explicitly compute the component action of certain recently discovered new $\mathcal{N}=1$ supergravity actions which enlarge the space of scalar potentials allowed by supersymmetry and also contain fermionic
interaction terms that become singular when supersymmetry is unbroken. They are the ``Liberated Supergravity'' introduced by
Farakos, Kehagias and Riotto, and supergravities with a new K\"{a}hler-invariant Fayet-Iliopoulos term proposed by Antoniadis, Chatrabhuti, Isono, and Knoops. This paper is complementary to our previous papers [Phys. Rev. D 103 (2021), 025008 and 105006], in which new constraints on the coupling constants of those new theories were found.
 In this paper we spell out many details that were left out of our previous papers.  
\end{abstract}

\vskip 1 cm
\vspace{24pt}
\end{titlepage}

\tableofcontents

\section{Introduction}
The results of LHC run 2 show no significant deviations from Standard Model predictions thereby pushing the energy scale 
of new physics --and in particular low-scale supersymmetry~\cite{lhc}-- to higher and higher values. 
This empirical observation combines with 
theoretical difficulties such as the eta problem that seems to plagues many supergravity models of inflation (see e.g.~\cite{fvp} for a review) to make alternative scenarios worth considering. One easy way out of all problems posed by supersymmetry is to assume 
that the world is not supersymmetric at all. On the other hand the high-energy completion of any field theory containing gravity 
provided by perturbative superstring theory implies that at some very high energy scale the world is supersymmetric (and 
10-dimensional). In superstring theory the largest and best understood set of models of low-energy physics --such as Calabi-Yau or 
flux compactifications-- are such that extra dimensions are compactified at energy scales which are much above the supersymmetry 
breaking scale. It is therefore reasonable to study four-dimensional models in which supersymmetry is broken at a high scale, e.g.
at a scale $M_S \gg 1\,\mathrm{TeV}$.
While a large $M_S$ means that supersymmetry does not help solving the Standard Model naturalness problem it does open
up many new possibilities for writing consistent actions.

Among these possibilities, some recent proposals use explicitly supersymmetric terms that look at first sight singular,
because when written in superspace they contain inverse powers of superfields. What makes possible to write such terms is that
they in fact they depend at most on inverse powers {\em of the auxiliary fields}. So they are well defined when supersymmetry is
broken. We will study in particular the theory called ``liberated $\mathcal{N}=1$ supergravity,'' proposed in Ref. \cite{fkr}
and the new Fayet-Iliopoulos (FI) term introduced in~\cite{acik,ar,oldACIK,cftp,akk,Kuzenko}. These modifications of the 
standard supergravity action relax the highly constrained form of scalar potentials in supergravity. 
(recall that the scalar potential has the universal form $V=V_F+V_D$, $V_F = e^G(G_AG^{A\Bar{B}}G_{\Bar{B}}-3)$ and $V_D$, for some function $G$ of the chiral multipet scalars.). The cosmological applications to inflation of such models have been recently investigated in~\cite{ar,jp2}.

Supergravity actions contain many nonrenormalizable operators that depend on the fermionic fields. Their existence 
defines an ultraviolet (UV) cutoff scale of the theory. While this property is generic in locally supersymmetic theories, it is not dangerous when the UV cutoff is the Planck scale $M_{pl}$, which is in any case the ultimate cutoff for supergravity, and of course also for Einstein gravity.  In the models studied in Refs.~\cite{jp1,jp2} instead, the UV cutoff is connected to the supersymmetry breaking scale, so it becomes an important 
parameter that constraints the domain of validity of those models.

The main purpose of this work is to show a systematic analysis of how a cutoff scale of an effective supergravity theory is determined by its fermionic nonrenormalizable terms and what constraints on the cutoff arise from the nonrenormalizable terms. These fermionic nonrenormalizable Lagrangians are conventionally represented by $\mathcal{L}_F \sim \mathcal{C}^{(4-\delta)}\mathcal{O}_F^{(\delta)}$ where $\delta > 4$ is the mass dimension of the fermionic nonrenormalizable operators $\mathcal{O}_F^{(\delta)}$. Then, we find constraints on the expansion coefficients $\mathcal{C}^{(4-\delta)}$ because these coefficients must be suppressed by a cutoff scale $\Lambda_{cut}$ of a theory for this to be valid as an effective field theory in range of some energies of interest below the cutoff; that is,
\begin{eqnarray}
\mathcal{C}^{(4-\delta)} \lesssim \frac{1}{\Lambda_{cut}^{\delta-4}}.
\end{eqnarray}
By taking this inequality into account, we are able to determine in what range of energies a low energy supergravity theory can be considered as an effective field theory. To do so, it is essential to explore all of the fermionic nonrenormalizable terms as we have just seen above. Checking these terms may be difficult if we use the superspace formulation as we have to fully calculate them and the
superspace fields contain many more auxiliary components than needed for the off-shell closure of the supersymmetry algebra. However, we point out that using the superconformal tensor calculus enables us to compute both bosonic and fermionic terms in a systematical and economical way. Hence, we take advantage of the superconformal formalism in this work.

Since this paper is rather technical, the reader chiefly interested in finding the constraints on  new supergravity terms can
skip ahead to Section 4.
The paper is organized as follows. In Sec. 2, we review how to embed liberated $\mathcal{N}=1$ supergravity in the superconformal tensor calculus, which was firstly introduced in Ref. \cite{jp1}, and explicitly compute the component action of the liberated supergravity using the tensor calculus in detail\footnote{Section 2 provides the detail calculations of the liberated supergravity, which was omitted in Ref. \cite{jp1}.}. In Sec. 3, we review the component action of a new FI term, which is K\"{a}hler-invariant, proposed by Antoniadis, Chatrabhuti, Isono, and Knoops (ACIK)\footnote{We shall call the new FI term proposed by the authors as ``ACIK-FI term'' to distinguish it from other types FI terms. Sec. 3 also provides the detailed calculations of the ACIK-FI term, which were omitted in Ref. \cite{jp2}.}. In Sec. 4, we recall the constraints on the liberated supergravity terms derived in~\cite{jp1} and derive the constraints on the new FI terms 
used in~\cite{jp2}. Both sets of constraints are based on an analysis of fermionic nonrenormalizable interactions. In Sec. 5, we briefly summarize our findings.

\section{Component action of liberated \texorpdfstring{$\mathcal{N}=1$}{} supergravity in superconformal tensor calculus}

In this section, we compute the component action of liberated $\mathcal{N}=1$ supergravity in superconformal tensor calculus \cite{cfgvnv,Linear}.  
We find that the superconformal Lagrangian of liberated $\mathcal{N}=1$ supergravity \cite{fkr} can be written by a D-term
\begin{eqnarray}
\mathcal{L}_{NEW} \equiv \bigg[ \mathcal{Y}^2 \frac{w^2\bar{w}^2}{T(\bar{w}^2)\bar{T}(w^2)} \mathcal{U}(Z,\Bar{Z})\bigg]_D,
\end{eqnarray}
in which we define $\mathcal{Y} \equiv S_0\bar{S}_0 e^{-K/3}$ where $S_0$ is a conformal compensator with Weyl/chiral weights $(1,1)$ and $K$ is a K\"{a}hler potential with the weights $(2,0)$. The notation for the other fields is as follows: $w^2 \equiv \frac{\mathcal{W}^2(K)}{\mathcal{Y}^2}, \quad \bar{w}^2 \equiv \frac{\bar{\mathcal{W}}^2(K)}{\mathcal{Y}^2}$, $\mathcal{W}^2(K)\equiv \mathcal{W}_{\alpha}(K)\mathcal{W}^{\alpha}(K)$ where $\mathcal{W}_{\alpha}(K)$ is a field strength multiplet with respect to the K\"{a}hler potential; $\mathcal{U}(Z,\Bar{Z})$ is a general function of matter multiplets $Z$, and $T(\Bar{w}^2),\Bar{T}(w^2)$ are chiral projection of $\Bar{w}^2$ and its conjugate respectively.

\subsection{Embedding Super-Weyl-K\"{a}hler transformations as an abelian gauge symmetry into the superconformal formalism}

In liberated $\mathcal{N}=1$ supergravity \cite{fkr}, a key idea is that the Super-Weyl-K\"{a}hler transformation can be promoted to an Abelian gauge symmetry. Liberated supergravity was constructed in~\cite{fkr} using the superspace formalism, where a K\"{a}hler transformation is introduced to compensate the variation of the action under a super-Weyl rescaling. In this work instead, we want to construct the equivalent liberated supergravity using the superconformal tensor calculus to analyze the fermionic interactions in a systematical and economical way. To do so, we introduce a conformal compensator multiplet, called $S_0$, removing the variation while maintaining the K\"{a}hler potential invariant under superconformal symmetry. Therefore, unlike the superspace formalism, it is essential to define such a gauge transformation independently of superconformal symmetry.

To find the Super-Weyl-K\"{a}hler transformations that are compatible with the superconformal formalism, we recall first 
the Super-Weyl-K\"{a}hler transformations that are used in the superspace formalism. A K\"{a}hler function $K(z,\bar{z})$, whose arguments have the vanishing Weyl/chiral weights, is defined up to a chiral gauge parameter $\Sigma$. The 
redefinition by $\Sigma$ acts on the components of the K\"{a}haler multiplet as
\begin{eqnarray}
&& K \rightarrow K + 6\Sigma+6\bar{\Sigma},\\
&& W \rightarrow We^{-6\Sigma}, \qquad \bar{W} \rightarrow \bar{W} e^{-6\bar{\Sigma}}\\
&& T \rightarrow e^{-4\Sigma+2\bar{\Sigma}}T,\qquad \bar{T} \rightarrow e^{2\Sigma-4\bar{\Sigma}}\bar{T},\\
&& \mathcal{D}_{\alpha}K \rightarrow e^{\Sigma-2\bar{\Sigma}}\mathcal{D}_{\alpha}K, \qquad \mathcal{W}_{\alpha} \rightarrow e^{-3\Sigma}\mathcal{W}_{\alpha}, \qquad \mathcal{W}^2 \rightarrow e^{-6\Sigma}\mathcal{W}^2,\\
&& T(\bar{\mathcal{W}}^2) \rightarrow T(\bar{\mathcal{W}}^2) e^{-4\Sigma -4\bar{\Sigma}},\\
&& \mathcal{D}^{\alpha}\mathcal{W}_{\alpha} \rightarrow e^{-2\Sigma-2\bar{\Sigma}}\mathcal{D}^{\alpha}\mathcal{W}_{\alpha},\\
&& E \rightarrow Ee^{2\Sigma + 2\bar{\Sigma}}, \qquad \mathcal{E} \rightarrow \mathcal{E} e^{6\Sigma}.
\end{eqnarray}
where $E$ and $\mathcal{E}$ are the D- and F-term densities, respectively.

Next, it may be useful to recall the relation between the superspace and superconformal formalisms. The invariant actions from the superconformal formalism are identified with those from the superspace calculus as follows \cite{kyy}:
\begin{eqnarray}
&& [\mathcal{V}]_D = 2 \int d^4\theta E \mathcal{V}, \\
&& [\mathcal{S}]_F = \int d^2\theta \mathcal{E}\mathcal{S} + \int d^2 \bar{\theta} \bar{\mathcal{E}} \bar{\mathcal{S}},
\end{eqnarray}
where $\mathcal{V}$ is a superconformal real multiplet with the Weyl/chirial weights (2,0) and $\mathcal{S}$ is a superconformal chiral multiplet with Weyl/chiral weights (3,3). To make the action invariant under the super-Weyl-K\"{a}hler transformations we should impose that the corresponding superconformal multiplets transform as 
\begin{eqnarray}
\mathcal{V} \rightarrow \mathcal{V} e^{-2\Sigma-2\bar{\Sigma}},\qquad  \mathcal{S} \rightarrow \mathcal{S} e^{-6\Sigma}.
\end{eqnarray}

Instead of considering the K\"{a}hler transformation of the K\"{a}hler potential a superconformal compensator is introduced 
in the superconformal formalism to eliminate the variation of the action transformed by a super-Weyl rescaling (also called Howe-Tucker transformation \cite{SUGRAprimer}). Thus, the compensator must transform as   
\begin{eqnarray}
S_0 \rightarrow S_0 e^{-2\Sigma}, \qquad \bar{S}_0 \rightarrow \bar{S}_0 e^{-2\bar{\Sigma}},
\end{eqnarray}
resulting in
\begin{eqnarray}
S_0\bar{S}_0 e^{-K/3} \rightarrow S_0\bar{S}_0 e^{-K/3} e^{-2\Sigma-2\bar{\Sigma}},
\end{eqnarray}
where $K$ is invariant under the superconformal symmetry (i.e. super-Weyl rescaling), so that the action can be invariant as desired. 

At this point, differently from the usual story of the superconformal symmetry, we require a ``K\"{a}hler transformation'' of the K\"{a}hler potential in order to construct a ``liberated'' supergravity that is invariant under the same Super-Weyl-K\"{a}hler transformations as an abelian gauge symmetry used in \cite{fkr}. Therefore, we assume that the superconformal compensators are inert under the super-Weyl-K\"{a}hler transformations 
\begin{eqnarray}
S_0 \rightarrow S_0, \qquad \bar{S}_0 \rightarrow \bar{S}_0,
\end{eqnarray}
while the K\"{a}hler potential {\it does} transform under the same transformations as above, namely as $K \rightarrow K + 6\Sigma + 6\Bar{\Sigma}$, so that
\begin{eqnarray}
&& \mathcal{Y} \rightarrow \mathcal{Y} e^{-2(\Sigma + \bar{\Sigma})},\\
&& w^2 \equiv \frac{\mathcal{W}^2(K)}{(e^{-K/3} S_0\bar{S}_0 )^2} \rightarrow w^2 e^{-2\Sigma+4\bar{\Sigma}},\\
&& \bar{T} \left(w^2\right) \rightarrow \bar{T}\left(w^2e^{-2\Sigma+4\bar{\Sigma}}\right)  e^{2\Sigma-4\bar{\Sigma}}=\bar{T}\left(w^2\right).
\end{eqnarray}

\subsection{List of superconformal multiplets}

In this section, we present all the superconformal multiplets of the liberated $\mathcal{N}=1$ supergravity following the notations and multiplication laws used in \cite{Linear,fvp}.

\subsubsection{K\"{a}hler potential multiplet}

Let us consider $n$ physical chiral multiplets of matter $z^I \equiv \{ z^I , P_L\chi^I, F^I\}$ where $I = 1,2,3,\cdots,n$ and  their anti-chiral multiplets $\bar{z}^{\bar{I}} \equiv \{\bar{z}^{\bar{I}}, P_R\chi^{\bar{I}},\bar{F}^{\bar{I}}\}$.\footnote{The complex conjugates are $\bar{z}^{\bar{I}} \equiv (z^I)^{*}$, $\chi^{\bar{I}} \equiv (\chi^I)^C $, $\bar{\chi}^{\bar{I}} \equiv (\bar{\chi}^I)^C$, and $\bar{F}^{\bar{I}} \equiv (F^I)^{*}$ (The barred index is the complex conjugate index, so that the handedness of fermion becomes opposite, i.e. $(P_{L/R}\chi)^C=(P_{L/R})^C(\chi)^C=P_{R/L}(\chi)^C$.). The chiralities of fermion are specified as $\chi^I \equiv P_L \chi^I$ and  $\chi^{\bar{I}} \equiv P_R \chi^{\bar{I}}$. The Majorana conjugates are $\overline{(P_{L/R}\chi)} =\bar{\chi} P_{L/R}$ (The handedness is preserved.).} Then, according to the superconformal tensor calculus, the K\"{a}hler potential multiplet can be written as follows:

\begin{eqnarray}
K(z,\bar{z}) = \{C_K, \mathcal{Z}_K , \mathcal{H}_K, \mathcal{K}_K, \mathcal{B}_{\mu}^K, \Lambda_K, \mathcal{D}_K \}
\end{eqnarray}
where
\begin{eqnarray}
C_K &=& K(z,\bar{z}),\\
\mathcal{Z}_K  &=& i\sqrt{2}(-K_I \chi^I + K_{\bar{I}}\chi^{\bar{I}}),\\
\mathcal{H}_K &=& -2K_{I}F^I + K_{IJ}\bar{\chi}^I\chi^J,\\
\mathcal{K}_K &=& -2K_{\bar{I}}\bar{F}^{\bar{I}} + K_{\bar{I}\bar{J}}\bar{\chi}^{\bar{I}}\chi^{\bar{J}} = \mathcal{H}_K^{*},\\
\mathcal{B}_{\mu}^K &=& 
iK_I\mathcal{D}_{\mu} z^I -iK_{\bar{I}}\mathcal{D}_{\mu} \bar{z}^{\bar{I}} + i K_{I\bar{J}}\bar{\chi}^I \gamma_{\mu} \chi^{\bar{J}},\\
\Lambda_K &=& P_L\Lambda_K + P_R\Lambda_K 
\\
P_L \Lambda_K &=& -\sqrt{2}iK_{\bar{I}J}[(\cancel{\mathcal{D}}z^{J})\chi^{\bar{I}}-\bar{F}^{\bar{I}}\chi^J] -\frac{i}{\sqrt{2}}K_{\bar{I}\bar{J}K}\chi^{K}\bar{\chi}^{\bar{I}}\chi^J,\\
P_R \Lambda_K &=& \sqrt{2}iK_{I\bar{J}}[(\cancel{\mathcal{D}}\bar{z}^{\bar{J}})\chi^{I}-F^{I}\chi^{\bar{J}}] +\frac{i}{\sqrt{2}}K_{IJ\bar{K}}\chi^{\bar{K}}\bar{\chi}^{I}\chi^{\bar{J}},\\
\mathcal{D}_K &=& 2K_{I\bar{J}} 
\bigg( -\mathcal{D}_{\mu} z^I \mathcal{D}^{\mu} \bar{z}^{\bar{J}} -\frac{1}{2} \bar{\chi}^I P_L \cancel{\mathcal{D}}\chi^{\bar{J}} -\frac{1}{2} \bar{\chi}^{\bar{J}} P_R \cancel{\mathcal{D}} \chi^I + F^I\bar{F}^{\bar{J}}\bigg) \nonumber\\
&&+ K_{IJ\bar{K}} \bigg( -\bar{\chi}^I \chi^J \bar{F}^{\bar{K}} + \bar{\chi}^I (\cancel{\mathcal{D}}z^J)\chi^{\bar{K}}         \bigg) + K_{\bar{I}\bar{J}K} \bigg( -\bar{\chi}^{\bar{I}} \chi^{\bar{J}} F^{K} + \bar{\chi}^{\bar{I}} (\cancel{\mathcal{D}}\bar{z}^{\bar{J}})\chi^{K}\bigg) \nonumber\\
&& + \frac{1}{2} K_{IJ\bar{K}\bar{L}} (\bar{\chi}^IP_L\chi^J )(\bar{\chi}^{\bar{K}} P_R \chi^{\bar{L}}) .
\end{eqnarray}
Here the covariant derivatives\footnote{From Eq. (16.34) in Ref. \cite{fvp} we find that for a general superconformal chiral multiplet $(z^I,P_L\chi^I,F^I)$ with Weyl/chiral weights $(w,c=w)$ and gauge symmetries with Killing vector fields $k_A^I$, the full superconformal covariant derivatives $\mathcal{D}_{a}$ are given by
\begin{eqnarray}
 \mathcal{D}_{a} z^I &=& e^{\mu}_a\Big[(\partial_{\mu} -wb_{\mu} -wiA_{\mu})z^I -\frac{1}{\sqrt{2}}\bar{\psi}_{\mu} \chi^I -A^A_{\mu} k_A^I\Big],\nonumber\\
 \mathcal{D}_{a} P_L\chi^I &=& e^{\mu}_aP_L\bigg[ \left(\partial_{\mu} +\frac{1}{4}\omega_{\mu}^{ab}\gamma_{ab}-(w+1/2)b_{\mu} + (w-3/2)iA_{\mu}\right)\chi^I -\frac{1}{\sqrt{2}}(\cancel{\mathcal{D}}z^I + F^I)\psi_{\mu} \nonumber\\
&&-\sqrt{2}wz^I  \phi_{\mu} -A^A_{\mu} \chi^J\partial_J k_A^I \bigg]. \nonumber
\end{eqnarray}} 
of chiral multiplets of matter with the weights $(0,0)$ are given by
\begin{eqnarray}
 \mathcal{D}_{a} z^I &=& e^{\mu}_a\Big[\partial_{\mu}z^I -\frac{1}{\sqrt{2}}\bar{\psi}_{\mu} \chi^I\Big],\\
 \mathcal{D}_{a} P_L\chi^I &=& e^{\mu}_aP_L\bigg[ \left(\partial_{\mu} +\frac{1}{4}\omega_{\mu}^{ab}\gamma_{ab}-\frac{1}{2}b_{\mu} -\frac{3}{2}iA_{\mu}\right)\chi^I -\frac{1}{\sqrt{2}}(\cancel{\mathcal{D}}z^I + F^I)\psi_{\mu} \bigg].\nonumber\\{}
\end{eqnarray}

Note that the only bosonic contribution to  $\mathcal{D}_K$ is given by
\begin{eqnarray}
\mathcal{D}_K|_{\textrm{boson}} = 2K_{I\bar{J}} \left( -\partial_{\mu} z^I \partial^{\mu} \bar{z}^{\bar{J}} +F^I\bar{F}^{\bar{J}}\right) \equiv \tilde{\mathcal{F}}
\end{eqnarray}
and we especially denote this by $\tilde{\mathcal{F}}$. We also note that the $\tilde{\mathcal{F}}$ is positive definite up to terms containing spatial gradients; so,  for small spatial gradients,
\begin{eqnarray}
\mathcal{D}_K|_{\textrm{boson}}\equiv \tilde{\mathcal{F}} \sim 2K_{I\bar{J}} \left( \dot{z}^I \dot{\bar{z}}^{\bar{J}} +F^I\bar{F}^{\bar{J}}\right) >0.
\end{eqnarray}

\subsubsection{Compensator multiplet}

Chiral compensators $S_0,\bar{S}_0$ with the Weyl/chiral weights (1,1) are defined as follows. The chiral supermultiplets $S_0 = \{s_0, P_L\chi^0, F_0\}$ and $\bar{S}_0 = \{s_0^{*}, P_R\chi^0, F_0^{*}\}$ can be embedded into the superconformal formalism as
    \begin{eqnarray}
    && S_0 \equiv \{ s_0, -i\sqrt{2}P_L\chi^0, -2F_0 , 0 , i\mathcal{D}_{\mu} s_0, 0,0 \},\\
    && \bar{S}_0 \equiv \{ s_0^{*}, i\sqrt{2}P_R\chi^0 , 0 ,-2F_0^{*}, -i\mathcal{D}_{\mu} s_0^{*}, 0,0 \}.
    \end{eqnarray}
   Then, the composite real compensator $S_0\bar{S}_0$ is
    \begin{eqnarray}
    S_0\bar{S}_0 = \{ \mathcal{C}_0, \mathcal{Z}_0, \mathcal{H}_0, \mathcal{K}_0, \mathcal{B}_{\mu}^0, \Lambda_0, \mathcal{D}_0 \},
    \end{eqnarray}
    where \footnote{ $P_{L,R}\gamma_{\textrm{odd indices}} =\gamma_{\textrm{odd indices}}P_{R,L}$}
    \begin{eqnarray}
    \mathcal{C}_0 &=& s_0s_0^{*}|_{0f},\\
    \mathcal{Z}_0 &=& i\sqrt{2}(-s_0^{*} P_L\chi^0 + s_0 P_R \chi^0) |_{ 1f},\\
    \mathcal{H}_0 &=& -2 s_0^{*} F^0 |_{0f},\\
    \mathcal{K}_0 &=& -2s_0 F^{*}_0 |_{0f},\\
    \mathcal{B}_{\mu}^0 &=& is_0^{*} \mathcal{D}_{\mu} s_0 - is_0 \mathcal{D}_{\mu} s_0^{*} + i \bar{\chi}_0\gamma_{\mu} P_R \chi^0 =  is_0^{*}\partial_{\mu} s_0 - is_0 \partial_{\mu} s_0^{*}|_{0f}+\cdots,\\
    P_L \Lambda_0 &=& -\sqrt{2}i [(\cancel{\mathcal{D}}s_0^{*})P_R\chi^0 - F_0^{*} P_L\chi^0]=-\sqrt{2}i [(\cancel{\partial}s_0^{*})P_R\chi^0 - F_0^{*} P_L\chi^0]|_{1f}+\cdots,\\
    P_R \Lambda_0 &=& \sqrt{2}i[(\cancel{\mathcal{D}}s_0)P_L\chi^0 - F_0 P_R\chi^0]=\sqrt{2}i[(\cancel{\partial}s_0)P_L\chi^0 - F_0 P_R\chi^0]|_{1f}+\cdots,\\
    \mathcal{D}_0 &=& 2\left( -\mathcal{D}_{\mu}s_0 \mathcal{D}^{\mu}s_0^{*} - \frac{1}{2}\bar{\chi}_0 P_L \cancel{\mathcal{D}}\chi^0-\frac{1}{2}\bar{\chi}_0 P_R \cancel{\mathcal{D}}\chi^0  + F_0F_0^{*} \right) = 2( -\partial_{\mu}s_0 \partial^{\mu}s_0^{*} + F_0F_0^{*} )|_{0f}+\cdots,\nonumber\\{}
    \end{eqnarray}
where the covariant derivatives of the conformal compensator $(s_0,P_L\chi^0,F^0)$ with Weyl/chiral weights $(1,1)$ are given by
\begin{eqnarray}
  \mathcal{D}_{a} s_0 &=&  e^{\mu}_a (\partial_{\mu}s_0 -\frac{1}{\sqrt{2}}\bar{\psi}_{\mu} \chi^0),\\
 \mathcal{D}_{a} P_L\chi^0 &=& e^{\mu}_aP_L\bigg[ \left(\partial_{\mu} +\frac{1}{4}\omega_{\mu}^{ab}\gamma_{ab}-\frac{3}{2}b_{\mu} -\frac{1}{2}iA_{\mu}\right)\chi^0 -\frac{1}{\sqrt{2}}(\cancel{\mathcal{D}}s_0 + F^0)\psi_{\mu} -\sqrt{2}s_0  \phi_{\mu}  \bigg].~\qquad
\end{eqnarray}

The composite real compensator $\Upsilon \equiv S_0\bar{S}_0 e^{-K/3}$
\begin{eqnarray}
\Upsilon = \{ \mathcal{C}_{\Upsilon}, \mathcal{Z}_{\Upsilon}, \mathcal{H}_{\Upsilon}, \mathcal{K}_{\Upsilon}, \mathcal{B}_{\mu}^{\Upsilon}, \Lambda_{\Upsilon}, \mathcal{D}_{\Upsilon} \}
\end{eqnarray}
has components 
\begin{eqnarray}
 \mathcal{C}_{\Upsilon} &=& \Upsilon = s_0\bar{s}_0 e^{-K/3}|_{0f},\\
 \mathcal{Z}_{\Upsilon} &=& e^{-K/3} \mathcal{Z}_0  +\frac{ i\sqrt{2}}{3}\Upsilon(K_I \chi^I - K_{\bar{I}}\chi^{\bar{I}})\nonumber\\
 &=&  i\sqrt{2}\Upsilon(-\frac{1}{s_0} P_L\chi^0 + \frac{1}{s_0^{*}} P_R \chi^0 + \frac{1}{3}K_I \chi^I - \frac{1}{3}K_{\bar{I}}\chi^{\bar{I}})|_{1f},\\
 \mathcal{H}_{\Upsilon} &=& e^{-K/3} \mathcal{H}_0+ \Upsilon \Big(\frac{2}{3}K_{I}F^I + (\frac{1}{9}K_{I}K_{J}-\frac{1}{3}K_{IJ})\bar{\chi}^I\chi^J\Big) -\frac{1}{2}[\bar{\mathcal{Z}_0}P_L\mathcal{Z}_{\Upsilon}+\bar{\mathcal{Z}}_{\Upsilon}P_L\mathcal{Z}_0]\nonumber\\
 &=&-2\Upsilon ( \frac{F^0}{s_0} -\frac{1}{3} K_IF^I) |_{0f}+\cdots,\quad \\
 \mathcal{K}_{\Upsilon} &=& e^{-K/3} \mathcal{H}_0^{*}+ \Upsilon \Big(\frac{2}{3}K_{\bar{I}}\bar{F}^{\bar{I}} + (\frac{1}{9}K_{\bar{I}}K_{\bar{J}}-\frac{1}{3}K_{\bar{I}\bar{J}})\bar{\chi}^{\bar{I}}\chi^{\bar{J}}\Big) -\frac{1}{2}[\bar{\mathcal{Z}_0}P_R\mathcal{Z}_{\Upsilon}+\bar{\mathcal{Z}}_{\Upsilon}P_R\mathcal{Z}_0],\nonumber\\
 &=&-2\Upsilon ( \frac{F^{0*}}{s_0^{*}} -\frac{1}{3} K_{\bar{I}}\bar{F}^{\bar{I}}) |_{0f}+\cdots, \\
 \mathcal{B}_{\mu}^{\Upsilon} &=& e^{-K/3} \mathcal{B}_{\mu}^0 -i\Upsilon \Big(\frac{1}{3}K_I\mathcal{D}_{\mu} z^I -\frac{1}{3}K_{\bar{I}}\mathcal{D}_{\mu} \bar{z}^{\bar{I}} -  (\frac{1}{9}K_{I}K_{\bar{J}}-\frac{1}{3}K_{I\bar{J}})\bar{\chi}^I \gamma_{\mu} \chi^{\bar{J}}\Big)\nonumber\\&&
 + \frac{1}{4}i [\bar{\mathcal{Z}_0}\gamma_*\gamma_{\mu}\mathcal{Z}_{\Upsilon}+\bar{\mathcal{Z}}_{\Upsilon}\gamma_*\gamma_{\mu}\mathcal{Z}_0],\nonumber\\
 &=& i\Upsilon(\frac{1}{s_0}\partial_{\mu} s_0 - \frac{1}{s_0^{*}} \partial_{\mu} s_0^{*} - \frac{1}{3}K_{I}\partial_{\mu}z^I+\frac{1}{3}K_{\bar{I}}\partial_{\mu}\bar{z}^{\bar{I}})|_{0f} + \cdots.
 \end{eqnarray}
 
 \begin{eqnarray}
 \Lambda_{\Upsilon} &=& e^{-K/3}\Lambda_0-i\sqrt{2}\Upsilon\bigg( (\frac{1}{9}K_{\bar{I}}K_{J}-\frac{1}{3}K_{\bar{I}J})[(\cancel{\mathcal{D}}z^{J})\chi^{\bar{I}}-\bar{F}^{\bar{I}}\chi^J] \nonumber\\
&&+\frac{1}{2} (-\frac{1}{27}K_{\bar{I}}K_{\bar{J}}K_K+\frac{1}{9}(K_{\bar{I}\bar{J}}K_K+K_{\bar{I}K}K_{\bar{J}}+K_{\bar{I}}K_{\bar{J}K})-\frac{1}{3}K_{\bar{I}\bar{J}K})\chi^{K}\bar{\chi}^{\bar{I}}\chi^J
\nonumber\\
&&- (\frac{1}{9}K_{I}K_{\bar{J}}-\frac{1}{3}K_{I\bar{J}})[(\cancel{\mathcal{D}}\bar{z}^{\bar{J}})\chi^{I}-F^{I}\chi^{\bar{J}}] \nonumber\\
&&-\frac{1}{2} (-\frac{1}{27}K_{I}K_{J}K_{\bar{K}}+\frac{1}{9}(K_{IJ}K_{\bar{K}}+K_{I\bar{K}}K_{J}+K_{I}K_{J\bar{K}})-\frac{1}{3}K_{IJ\bar{K}})\chi^{\bar{K}}\bar{\chi}^{I}\chi^{\bar{J}} \bigg) \nonumber\\
&&+ \frac{1}{2} \Big( [i\gamma_* \cancel{\mathcal{B}_0}+\textrm{Re}\mathcal{H}_0-i\gamma_* \textrm{Im}\mathcal{H}_0-\cancel{\mathcal{D}}(s_0s_0^{*})]\mathcal{Z}_{\Upsilon}\nonumber\\
 &&+[i\gamma_* \cancel{\mathcal{B}_{\Upsilon}}+\textrm{Re}\mathcal{H}_{\Upsilon}-i\gamma_* \textrm{Im}\mathcal{H}_{\Upsilon}-\cancel{\mathcal{D}}(\Upsilon)]\mathcal{Z}_{0}\Big),\nonumber\\
 &=& -\sqrt{2}ie^{-K/3} [(\cancel{\partial}s_0^{*})P_R\chi^0 - F_0^{*} P_L\chi^0]+\sqrt{2}ie^{-K/3}[(\cancel{\partial}s_0)P_L\chi^0 - F_0 P_R\chi^0]\nonumber\\
 &&-i\sqrt{2}\Upsilon\bigg( (\frac{1}{9}K_{\bar{I}}K_{J}-\frac{1}{3}K_{\bar{I}J})[(\cancel{\partial}z^{J})\chi^{\bar{I}}-\bar{F}^{\bar{I}}\chi^J] - (\frac{1}{9}K_{I}K_{\bar{J}}-\frac{1}{3}K_{I\bar{J}})[(\cancel{\partial}\bar{z}^{\bar{J}})\chi^{I}-F^{I}\chi^{\bar{J}}] \bigg) \nonumber\\
&& + \frac{1}{2}\bigg[ \Big\{ i\gamma_*(is_0^{*}\cancel{\partial}s_0-is_0\cancel{\partial}s_0^{*}) + \textrm{Re}[-2s_0^{*}F^0] -i\gamma_* \textrm{Im}[-2s_0^{*}F^0] -\cancel{\partial}(s_0s_0^{*}) \Big\}\nonumber\\
&&\qquad \times\Big\{ i\sqrt{2}\Upsilon(-s_0^{-1}P_L\chi^0 + s_0^{-1*}P_R\chi^0 + \frac{1}{3}K_{I}\chi^{I}-\frac{1}{3}K_{\bar{I}}\chi^{\bar{I}}) \Big\}  \nonumber\\
&&\qquad  + \Big\{ i\gamma_*\big(i\Upsilon(s_0^{-1}\cancel{\partial}s_0 - s_0^{-1*} \cancel{\partial}s_0^{*} -\frac{1}{3}K_{I}\cancel{\partial}z^{I} + \frac{1}{3}K_{\bar{I}}\cancel{\partial} \bar{z}^{\bar{I}} ) \big)+\textrm{Re}[-2\Upsilon(F^0s_0^{-1}-\frac{1}{3}K_{I}F^{I})] \nonumber\\
&&\qquad\qquad  -i\gamma_* \big(-2\Upsilon (F^0 s_0^{-1}) -\frac{1}{3} K_{I}F^{I}\big) - \cancel{\partial}\Upsilon \Big\} \Big\{ i\sqrt{2}(-s_0^{*} P_L\chi^0 + s_0 P_R \chi^0) \Big\} \bigg] \bigg|_{1f} +\cdots.\nonumber\\{}
 \end{eqnarray}
 
 \begin{eqnarray}
 \mathcal{D}_{\Upsilon} &=& e^{-K/3} \mathcal{D}_0+\Upsilon \bigg[2(\frac{1}{9}K_{I}K_{\bar{J}}-\frac{1}{3}K_{I\bar{J}}) 
\bigg( -\mathcal{D}_{\mu} z^I \mathcal{D}^{\mu} \bar{z}^{\bar{J}} -\frac{1}{2} \bar{\chi}^I P_L \cancel{\mathcal{D}}\chi^{\bar{J}} \nonumber\\&&
-\frac{1}{2} \bar{\chi}^{\bar{J}} P_R \cancel{\mathcal{D}} \chi^I + F^I\bar{F}^{\bar{J}}\bigg) +  (-\frac{1}{27}K_{I}K_{J}K_{\bar{K}}+\frac{1}{9}(K_{IJ}K_{\bar{K}}+K_{I\bar{K}}K_{J}+K_{I}K_{J\bar{K}})\nonumber\\&&-\frac{1}{3}K_{IJ\bar{K}}) \bigg( -\bar{\chi}^I \chi^J \bar{F}^{\bar{K}} + \bar{\chi}^I (\cancel{\mathcal{D}}z^J)\chi^{\bar{K}}         \bigg)
\nonumber\\
&&+  (-\frac{1}{27}K_{\bar{I}}K_{\bar{J}}K_K+\frac{1}{9}(K_{\bar{I}\bar{J}}K_K+K_{\bar{I}K}K_{\bar{J}}+K_{\bar{I}}K_{\bar{J}K})-\frac{1}{3}K_{\bar{I}\bar{J}K}) \nonumber\\&&\times \bigg( -\bar{\chi}^{\bar{I}} \chi^{\bar{J}} F^{K} + \bar{\chi}^{\bar{I}} (\cancel{\mathcal{D}}\bar{z}^{\bar{J}})\chi^{K}\bigg)  + \frac{1}{2}  \bigg\{-\frac{1}{3}K_{IJ\bar{K}\bar{L}}+\frac{1}{81}K_{I}K_{J}K_{\bar{K}}K_{\bar{L}}
\nonumber\\
&&-\frac{1}{27}[K_{IJ}K_{\bar{K}}K_{\bar{L}}+K_{I\bar{K}}K_{J}K_{\bar{L}}+K_{I}K_{J\bar{K}}K_{\bar{L}}+K_{I\bar{L}}K_{J}K_{\bar{K}}+K_{I}K_{J\bar{L}}K_{\bar{K}}+K_{I}K_{J}K_{\bar{K}\bar{L}}] 
\nonumber\\
&&+\frac{1}{9}[K_{IJ\bar{K}}+K_{IJ\bar{L}}K_{\bar{K}}+K_{IJ}K_{\bar{K}\bar{L}}+K_{I\bar{K}\bar{L}}K_{J}+K_{I\bar{K}}K_{J\bar{L}}+K_{I\bar{L}}K_{J\bar{K}}+K_{I}K_{J\bar{K}\bar{L}}] \bigg\} \nonumber\\
&&\times (\bar{\chi}^IP_L\chi^J )(\bar{\chi}^{\bar{K}} P_R \chi^{\bar{L}})\bigg] \nonumber\\
&&+ \frac{1}{2}(\mathcal{H}_0\mathcal{H}^{*}_{\Upsilon}+\mathcal{H}_0^{*}\mathcal{H}_{\Upsilon}-2\mathcal{B}_{\mu}^0\mathcal{B}^{\mu}_{\Upsilon} - 2\mathcal{D}(s_0s_0^{*}) \cdot \mathcal{D} e^{-K/3} - 2\bar{\Lambda}_0\mathcal{Z}_{\Upsilon} -2\bar{\Lambda}_{\Upsilon}\mathcal{Z}_{0}- \bar{\mathcal{Z}}_0 \cancel{\mathcal{D}} \mathcal{Z}_{\Upsilon}-\bar{\mathcal{Z}}_{\Upsilon} \cancel{\mathcal{D}} \mathcal{Z}_{0}).\nonumber\\
&=& \bigg[ e^{-K/3}2( -\partial_{\mu}s_0 \partial^{\mu}s_0^{*} + F_0F_0^{*} ) + 2\Upsilon (\frac{1}{9}K_{I}K_{\bar{J}}-\frac{1}{3}K_{I\bar{J}})(-\partial_{\mu}z^I\partial^{\mu}\bar{z}^{\bar{J}}+F^I\bar{F}^{\bar{J}}) \nonumber\\
&& + \frac{1}{2}(-2s_0^{*}F^0)\Big(-2\Upsilon ( \frac{F^{0*}}{s_0^{*}} -\frac{1}{3} K_{\bar{I}}\bar{F}^{\bar{I}})\Big)+ \frac{1}{2}(-2s_0F^{0*})\Big(-2\Upsilon ( \frac{F^0}{s_0} -\frac{1}{3} K_{I}F^{I})\Big)\nonumber\\
&&-(is_0^{*}\partial_{\mu} s_0 - is_0 \partial_{\mu} s_0^{*})\Big(i\Upsilon(\frac{1}{s_0}\partial^{\mu} s_0 - \frac{1}{s_0^{*}} \partial^{\mu} s_0^{*} - \frac{1}{3}K_{I}\partial^{\mu}z^I+\frac{1}{3}K_{\bar{I}}\partial^{\mu}\bar{z}^{\bar{I}})\Big) \nonumber\\
&&- (s_0^{*}\partial s_0+s_0\partial s_0^{*})\Big( -\frac{1}{3}e^{-K/3}(K_I\partial z^I + K_{\bar{I}}\partial\bar{z}^{\bar{I}})\Big)  \bigg]_{0f}+\cdots.
\end{eqnarray}

The conventional superconformal gauge is defined by the choice
\begin{eqnarray}
&&\mathcal{C}_{\Upsilon} = \Upsilon = 1,\\ &&\mathcal{Z}_{\Upsilon} = 0 \implies P_L\chi^0 - \frac{1}{3}e^{K/6} K_I P_L\chi^I =0,\\
&& b_{\mu} = 0.
\end{eqnarray}

\subsubsection{\texorpdfstring{$\mathcal{W}^2(K) \equiv \mathcal{W}_{\alpha}(K)\mathcal{W}^{\alpha}(K)$}{} composite chiral multiplet: (Weyl/Chiral) weights \texorpdfstring{$=(3,3)$}{}}

Let us define 
\begin{eqnarray}
&& \mathcal{W}^2(K) \equiv \{\mathcal{C}_{W},\mathcal{Z}_{W},\mathcal{H}_{W},\mathcal{K}_{W},\mathcal{B}^{W}_{\mu},\Lambda_{W},\mathcal{D}_{W} \},\\
&& \bar{\mathcal{W}}^{2}(K) \equiv \{\mathcal{C}_{W}^{*},\mathcal{Z}_{W}^C,\mathcal{K}_{W}^{*},\mathcal{H}_{W}^{*},(\mathcal{B}^{W}_{\mu})^{*},\Lambda_{W}^C,\mathcal{D}_{W}^{*} \}
\end{eqnarray}
where
\begin{eqnarray}
\mathcal{C}_{W} &=& \Bar{\Lambda}_K P_L \Lambda_K \nonumber\\
&=&
-2\Big[ K_{\bar{I}'J'}\Bar{\chi}^{\bar{I}'}\cancel{\partial}z^{J'} K_{\bar{I}J}\cancel{\partial}z^J\chi^{\bar{I}} - K_{\bar{I}'J'}\Bar{\chi}^{\bar{I}'}\cancel{\partial}z^{J'} K_{\bar{I}J}\bar{F}^{\bar{I}}\chi^J\nonumber\\
&&-K_{\bar{I}'J'}\bar{F}^{\bar{I}'}\Bar{\chi}^{J'}K_{\bar{I}J}\cancel{\partial}z^J\chi^{\bar{I}}+ K_{\bar{I}'J'}\bar{F}^{\bar{I}'}\Bar{\chi}^{J'}K_{\bar{I}J}\bar{F}^{\bar{I}}\chi^J \Big]\Big|_{2f}
 + \cdots,\\
\mathcal{Z}_{W} &=&  -i2 P_L (-\frac{1}{2}\gamma\cdot \hat{F}_K + i\mathcal{D}_K)\Lambda_K = -2\sqrt{2}i\tilde{\mathcal{F}}K_{\bar{I}J}P_L[\cancel{\partial}z^J\chi^{\bar{I}}-\bar{F}^{\bar{I}}\chi^J]\Big|_{1f}+\cdots,\\
\mathcal{H}_{W} &=& -2(2\Bar{\Lambda}_KP_L \cancel{\mathcal{D}}\Lambda_K + \hat{F}_K^-\cdot \hat{F}_K^- - \mathcal{D}^2_K) = 2\tilde{\mathcal{F}}^2|_{0f} +\cdots \equiv -2F^W,\\
\mathcal{K}_{W} &=& 0,\\
\mathcal{B}^{W}_{\mu} &=& i\mathcal{D}_{\mu}(\Bar{\Lambda}_K P_L \Lambda_K) = i\partial_{\mu}(\Bar{\Lambda}_K P_L \Lambda_K)|_{2f}+\cdots,\\
\Lambda_{W} &=& 0,\\
\mathcal{D}_{W} &=& 0. \label{Multiplet_W1}
\end{eqnarray}
Note that $\mathcal{C}_{W} = \mathcal{C}_{W}|_{2f}+\cdots, \mathcal{Z}_{W} = \mathcal{Z}_{W}|_{1f} +\cdots, \mathcal{H}_{W} = \mathcal{H}_{W}|_{0f} + \cdots, \mathcal{K}_{W} = 0, \mathcal{B}^{W}_{\mu} = \mathcal{B}^{W}_{\mu}|_{2f} + \cdots, \Lambda_{W}=\mathcal{D}_{W} =0$.

and
\begin{eqnarray}
\mathcal{C}_{\bar{W}}&=& \mathcal{C}_{W}^{*} = \Bar{\Lambda}_K P_R \Lambda_K,\\
\mathcal{Z}_{\bar{W}}&=&\mathcal{Z}_{W}^C = i2 P_R (-\frac{1}{2}\gamma\cdot \hat{F}_K -i\mathcal{D}_K)\Lambda_K,\\
\mathcal{H}_{\bar{W}}&=&\mathcal{K}_{W}^{*} = 0,\\
\mathcal{K}_{\bar{W}}&=&\mathcal{H}_{W}^{*} = -2(2\Bar{\Lambda}_KP_R \cancel{\mathcal{D}}\Lambda_K + \hat{F}_K^+\cdot \hat{F}_K^+ - \mathcal{D}^2_K) \equiv -2\bar{F}^{\bar{W}},\\
\mathcal{B}^{\bar{Q}}_{\mu}&=&(\mathcal{B}^{W}_{\mu})^{*} = -i\mathcal{D}_{\mu}(\Bar{\Lambda}_K P_R \Lambda_K),\\
\Lambda_{\bar{W}}&=&\Lambda_{W}^C = 0,\\
\mathcal{D}_{\bar{W}}&=&\mathcal{D}_{W}^{*} = 0.\label{Multiplet_W2}
\end{eqnarray}
We also defined:
\begin{eqnarray}
&& \hat{F}_{ab}^K \equiv e_a^{\mu}e_b^{\nu} (2\partial_{[\mu }\mathcal{B}^K_{\nu]} + \bar{\psi}_{[\mu}\gamma_{\nu]}\Lambda_K),\\
&& \Tilde{\hat{F}}_{ab} \equiv -i\frac{1}{2} \varepsilon_{abcd}\hat{F}^{cd}, \qquad \hat{F}_{ab}^{\pm} \equiv \frac{1}{2} (\hat{F}_{ab} \pm \tilde{\hat{F}}_{ab}),\qquad   (\hat{F}_{ab}^{\pm})^{*}= \hat{F}_{ab}^{\mp}\\
&& \cancel{\mathcal{D}}\Lambda_K \equiv \gamma \cdot \mathcal{D} \Lambda_K,\\
&& \mathcal{D}_{\mu} \Lambda_K \equiv \Big(\partial_{\mu} -\frac{3}{2}b_{\mu} + \frac{1}{4}\omega_{\mu}^{ab} \gamma_{ab} -\frac{3}{2}i\gamma_* \mathcal{A}_{\mu} \Big) \Lambda_K - \Big(\frac{1}{4}\gamma^{ab} \hat{F}_{ab}^K + \frac{1}{2}i\gamma_* \mathcal{D}_K \Big)\psi_{\mu}.
\end{eqnarray}

Next, we shall consider a gauge fixing that is equivalent to the gauge condition given by $\eta =0$ in Ref. \cite{fkr}. In fact, this can be obtained by imposing $\Lambda_K =0$. We will call this gauge the ``Liberated SUGRA gauge''. Then, the only non-vanishing superconformal components of the multiplets $\mathcal{W}^2(K),\bar{\mathcal{W}}^2(K)$ in this gauge are given by  $\mathcal{H}_{W} = -2(\hat{F}_K^-\cdot\hat{F}_K^- -\mathcal{D}_K^2)$ and $\mathcal{K}_{\bar{W}} = -2(\hat{F}_K^+\cdot\hat{F}_K^+ -\mathcal{D}_K^2) = \mathcal{H}_{W}^{*}$.

Another representation of the chiral multiplet is
\begin{eqnarray}
\mathcal{W}^2(K) &=& \left(\mathcal{C}_{W}, \frac{i}{\sqrt{2}}\mathcal{Z}_{W},-\frac{1}{2}\mathcal{H}_{W}\right) \nonumber\\
&=& \left( \Bar{\Lambda}_K P_L \Lambda_K,~ \sqrt{2} P_L (-\gamma\cdot \hat{F}_K + 2i\mathcal{D}_K)\Lambda_K,~ 2\Bar{\Lambda}_KP_L \cancel{\mathcal{D}}\Lambda_K + \hat{F}_K^-\cdot \hat{F}_K^- - \mathcal{D}^2_K\right) \nonumber\\
&\equiv& \left( X^W, P_L\chi^W, F^W \right),\\
\bar{\mathcal{W}}^2(K) &=& \left(\mathcal{C}_{W}^{*}, -\frac{i}{\sqrt{2}}\mathcal{Z}_{W}^C,-\frac{1}{2}\mathcal{H}_{W}^{*}\right) \nonumber\\
&=& \left( \Bar{\Lambda}_K P_R \Lambda_K,~ \sqrt{2} P_R (-\gamma\cdot \hat{F}_K - 2i\mathcal{D}_K^{*})\Lambda_K^C,~ 2\Bar{\Lambda}_KP_R \cancel{\mathcal{D}}\Lambda_K + \hat{F}_K^+ \cdot \hat{F}_K^+ - (\mathcal{D}^{*}_K)^2\right) \nonumber\\
&\equiv& \left( \bar{X}^{\bar{W}}, P_R\chi^{\bar{W}}, \bar{F}^{\bar{W}} \right)
\end{eqnarray}

We will also need the following definitions:
\begin{eqnarray}
X^W = \bar{\Lambda}_K P_L \Lambda_K \equiv W &=&  -2K_{\bar{I}J}[\bar{\chi}^{J}(\overline{\cancel{\mathcal{D}}z^{\bar{I}}})-\bar{F}^{\bar{I}}\bar{\chi}^{J}]K_{{\bar{I}}'J'}[(\cancel{\mathcal{D}}z^{{\bar{I}}'})\chi^{J'}-F^{J'}\chi^{{\bar{I}}'}] \nonumber\\
&& - K_{{\bar{I}}J}[\bar{\chi}^{J}(\overline{\cancel{\mathcal{D}}z^{\bar{I}}})-\bar{F}^{\bar{I}}\bar{\chi}^{J}] K_{\bar{I}'\bar{J}'K'}[\chi^{K'}\bar{\chi}^{\bar{I}'}\chi^{\bar{J}'}] \nonumber\\
&& - K_{\bar{I}\bar{J}K}[\bar{\chi}^{\bar{J}}\chi^{\bar{I}}\bar{\chi}^{K}]K_{{\bar{I}}'J'}[(\cancel{\mathcal{D}}z^{{\bar{I}}'})\chi^{J'}-F^{J'}\chi^{{\bar{I}}'}] \nonumber\\
&& - \frac{1}{2}K_{\bar{I}\bar{J}K}[\bar{\chi}^{\bar{J}}\chi^{\bar{I}}\bar{\chi}^{K}]K_{\bar{I}'\bar{J}'K'}[\chi^{K'}\bar{\chi}^{\bar{I}'}\chi^{\bar{J}'}],\\
\bar{X}^{\bar{W}} =  \bar{\Lambda}_K P_R \Lambda_K \equiv \bar{W} &=& (\bar{\Lambda}_K P_L \Lambda_K)^C,
\end{eqnarray}
and
\begin{eqnarray}
\mathcal{D}_KP_L\Lambda_K &=& \bigg[2K_{I\bar{J}} 
\bigg( -\mathcal{D}_{\mu} z^I \mathcal{D}^{\mu} \bar{z}^{\bar{J}} -\frac{1}{2} \bar{\chi}^I P_L \cancel{\mathcal{D}}\chi^{\bar{J}} -\frac{1}{2} \bar{\chi}^{\bar{J}} P_R \cancel{\mathcal{D}} \chi^I + F^I\bar{F}^{\bar{J}}\bigg) \nonumber\\
&&+ K_{IJ\bar{K}} \bigg( -\bar{\chi}^I \chi^J \bar{F}^{\bar{K}} + \bar{\chi}^I (\cancel{\mathcal{D}}z^J)\chi^{\bar{K}}         \bigg) + K_{\bar{I}\bar{J}K} \bigg( -\bar{\chi}^{\bar{I}} \chi^{\bar{J}} F^{K} + \bar{\chi}^{\bar{I}} (\cancel{\mathcal{D}}\bar{z}^{\bar{J}})\chi^{K}\bigg) \nonumber\\
&& + \frac{1}{2} K_{IJ\bar{K}\bar{L}} (\bar{\chi}^IP_L\chi^J )(\bar{\chi}^{\bar{K}} P_R \chi^{\bar{L}})\bigg]
\nonumber\\&&\times
\bigg[-\sqrt{2}iK_{\bar{I}'J'}[(\cancel{\mathcal{D}}z^{J'})\chi^{\bar{I}'}-\bar{F}^{\bar{I}'}\chi^{J'}] -\frac{i}{\sqrt{2}}K_{\bar{I}'\bar{J}'K'}\chi^{K'}\bar{\chi}^{\bar{I}'}\chi^{J'}\bigg].
\end{eqnarray}
\begin{eqnarray}
\mathcal{D}_K^2 &=& \bigg[2K_{I\bar{J}} 
\bigg( -\mathcal{D}_{\mu} z^I \mathcal{D}^{\mu} \bar{z}^{\bar{J}} -\frac{1}{2} \bar{\chi}^I P_L \cancel{\mathcal{D}}\chi^{\bar{J}} -\frac{1}{2} \bar{\chi}^{\bar{J}} P_R \cancel{\mathcal{D}} \chi^I + F^I\bar{F}^{\bar{J}}\bigg) \nonumber\\
&&+ K_{IJ\bar{K}} \bigg( -\bar{\chi}^I \chi^J \bar{F}^{\bar{K}} + \bar{\chi}^I (\cancel{\mathcal{D}}z^J)\chi^{\bar{K}}         \bigg) + K_{\bar{I}\bar{J}K} \bigg( -\bar{\chi}^{\bar{I}} \chi^{\bar{J}} F^{K} + \bar{\chi}^{\bar{I}} (\cancel{\mathcal{D}}\bar{z}^{\bar{J}})\chi^{K}\bigg) \nonumber\\
&& + \frac{1}{2} K_{IJ\bar{K}\bar{L}} (\bar{\chi}^IP_L\chi^J )(\bar{\chi}^{\bar{K}} P_R \chi^{\bar{L}})\bigg]^2.
\end{eqnarray}

\subsubsection{\texorpdfstring{$w^2,\bar{w}^2$}{} Composite Complex Multiplets: (Weyl/Chiral) weights \texorpdfstring{$=(-1,\pm 3)$}{}}

\begin{eqnarray}
&&  w^2 \equiv \dfrac{\mathcal{W}^2(K)}{\Upsilon^2} 
= \{\mathcal{C}_w,\mathcal{Z}_w,\mathcal{H}_w,\mathcal{K}_w,\mathcal{B}^w_{\mu},\Lambda_w,\mathcal{D}_w\},\\
&&   \bar{w}^2 \equiv \dfrac{\bar{\mathcal{W}}^2(K)}{\Upsilon^2} 
= \{\mathcal{C}_{\bar{w}},\mathcal{Z}_{\bar{w}},\mathcal{H}_{\bar{w}},\mathcal{K}_{\bar{w}},\mathcal{B}^{\bar{w}}_{\mu},\Lambda_{\bar{w}},\mathcal{D}_{\bar{w}}\}
\end{eqnarray}
where
\begin{eqnarray}
\mathcal{C}_{w} &=& \frac{\mathcal{C}_{W}}{\Upsilon^2},\\
\mathcal{Z}_{w} &=& \frac{1}{\Upsilon^2}\mathcal{Z}_{W} -2\frac{\mathcal{C}_{W}}{\Upsilon^3}\mathcal{Z}_{\Upsilon},\\
\mathcal{H}_{w} &=&  \frac{1}{\Upsilon^2}\mathcal{H}_{W} -2\frac{\mathcal{C}_{W}}{\Upsilon^3}\mathcal{H}_{\Upsilon} -\frac{1}{2}\bigg[ -4\frac{1}{\Upsilon^3} (\bar{\mathcal{Z}}_{W}P_L\mathcal{Z}_{\Upsilon}+\bar{\mathcal{Z}}_{\Upsilon}P_L\mathcal{Z}_{W})+6\frac{\mathcal{C}_{W}}{\Upsilon^4}\bar{\mathcal{Z}}_{\Upsilon}P_L\mathcal{Z}_{\Upsilon}\bigg],\\
\mathcal{K}_{w} &=&  \frac{1}{\Upsilon^2}\mathcal{K}_{W} -2\frac{\mathcal{C}_{W}}{\Upsilon^3}\mathcal{K}_{\Upsilon} -\frac{1}{2}\bigg[ -4\frac{1}{\Upsilon^3} (\bar{\mathcal{Z}}_{W}P_R\mathcal{Z}_{\Upsilon}+\bar{\mathcal{Z}}_{\Upsilon}P_R\mathcal{Z}_{W})+6\frac{\mathcal{C}_{W}}{\Upsilon^4}\bar{\mathcal{Z}}_{\Upsilon}P_R\mathcal{Z}_{\Upsilon}\bigg],\\
\mathcal{B}^{w}_{\mu} &=& \frac{1}{\Upsilon^2}\mathcal{B}^{W}_{\mu} -2\frac{\mathcal{C}_{W}}{\Upsilon^3}\mathcal{B}^{\Upsilon}_{\mu} + \frac{1}{2}i  \bigg[ -4\frac{1}{\Upsilon^3} (\bar{\mathcal{Z}}_{W}P_L\gamma_{\mu}\mathcal{Z}_{\Upsilon}+\bar{\mathcal{Z}}_{\Upsilon}P_L\gamma_{\mu}\mathcal{Z}_{W})+6\frac{\mathcal{C}_{W}}{\Upsilon^4}\bar{\mathcal{Z}}_{\Upsilon}P_L\gamma_{\mu}\mathcal{Z}_{\Upsilon}\bigg],\qquad\quad\\
\Lambda_{w} &=& \frac{1}{\Upsilon^2}\Lambda_{W} -2\frac{\mathcal{C}_{W}}{\Upsilon^3}\Lambda_{\Upsilon} + \frac{1}{2} 
\bigg[ -\frac{2}{\Upsilon^3} (i\gamma_*\cancel{\mathcal{B}}_{W}+P_L\mathcal{K}_{W}+P_R\mathcal{H}_{W}-\cancel{\mathcal{D}}\mathcal{C}_{W})\mathcal{Z}_{\Upsilon} \nonumber\\
&&-\frac{2}{\Upsilon^3}(i\gamma_*\cancel{\mathcal{B}}_{\Upsilon}+P_L\mathcal{K}_{\Upsilon}+P_R\mathcal{H}_{\Upsilon}-\cancel{\mathcal{D}}\mathcal{C}_{\Upsilon})\mathcal{Z}_{W} 
+ 6\frac{\mathcal{C}_{W}}{\Upsilon^4}(i\gamma_*\cancel{\mathcal{B}}_{\Upsilon}+P_L\mathcal{K}_{\Upsilon}+P_R\mathcal{H}_{\Upsilon}-\cancel{\mathcal{D}}\mathcal{C}_{\Upsilon})\mathcal{Z}_{\Upsilon}   \bigg],\nonumber\\
&& -\frac{1}{4}\bigg[ \frac{6}{\Upsilon^4}\frac{3!}{2!} \mathcal{Z}_{(W}\bar{\mathcal{Z}}_{\Upsilon}\mathcal{Z}_{\Upsilon)} -24 \frac{\mathcal{C}_{W}}{\Upsilon^5}\mathcal{Z}_{\Upsilon}\bar{\mathcal{Z}}_{\Upsilon}\mathcal{Z}_{\Upsilon}  \bigg],\\ 
\mathcal{D}_{w} &=& \frac{1}{\Upsilon^2}\mathcal{D}_{W} -2\frac{\mathcal{C}_{W}}{\Upsilon^3}\mathcal{D}_{\Upsilon} + \frac{1}{2}  \bigg[ -\frac{2}{\Upsilon^3} 2!(\mathcal{K}_{(W}\mathcal{H}_{\Upsilon)}-\mathcal{B}_{(W}\cdot \mathcal{B}_{\Upsilon)} -\mathcal{D}\mathcal{C}_{(W}\cdot \mathcal{D}\mathcal{C}_{\Upsilon)} - 2\bar{\Lambda}_{(W} \mathcal{Z}_{\Upsilon)} - \bar{\mathcal{Z}}_{(W} \cancel{\mathcal{D}}\mathcal{Z}_{\Upsilon)})\nonumber\\
&&+6\frac{\mathcal{C}_{W}}{\Upsilon^4}(\mathcal{K}_{\Upsilon}\mathcal{H}_{\Upsilon}-\mathcal{B}_{\Upsilon}\cdot \mathcal{B}_{\Upsilon} -\mathcal{D}\mathcal{C}_{\Upsilon}\cdot \mathcal{D}\mathcal{C}_{\Upsilon} - 2\bar{\Lambda}_{\Upsilon} \mathcal{Z}_{\Upsilon} - \bar{\mathcal{Z}}_{\Upsilon} \cancel{\mathcal{D}}\mathcal{Z}_{\Upsilon}) \bigg]\nonumber\\
&& -\frac{1}{4}\bigg[ \frac{6}{\Upsilon^4}\frac{3!}{2!} \bar{\mathcal{Z}}_{(W}(i\gamma_*\cancel{\mathcal{B}}_{\Upsilon}+P_L\mathcal{K}_{\Upsilon}+P_R\mathcal{H}_{\Upsilon})\mathcal{Z}_{\Upsilon)} -24\frac{\mathcal{C}_{W}}{\Upsilon^5}\bar{\mathcal{Z}}_{\Upsilon}(i\gamma_*\cancel{\mathcal{B}}_{\Upsilon}+P_L\mathcal{K}_{\Upsilon}+P_R\mathcal{H}_{\Upsilon})\mathcal{Z}_{\Upsilon}  \bigg],\nonumber\\
&& +\frac{1}{8} \bigg[ -\frac{24}{\Upsilon^5} \frac{4!}{3!} \bar{\mathcal{Z}}_{(W} P_L \mathcal{Z}_{\Upsilon} \bar{\mathcal{Z}}_{\Upsilon} P_R \mathcal{Z}_{\Upsilon)} +120\frac{\mathcal{C}_{W}}{\Upsilon^6}
\bar{\mathcal{Z}}_{\Upsilon} P_L \mathcal{Z}_{\Upsilon} \bar{\mathcal{Z}}_{\Upsilon} P_R \mathcal{Z}_{\Upsilon} 
\bigg] .
\end{eqnarray}
Note that we have to insert $\mathcal{K}_{W}=\Lambda_{W}=\mathcal{D}_{W}=0$ and $\mathcal{H}_{\bar{W}}=\Lambda_{\bar{W}}=\mathcal{D}_{\bar{W}}=0$ as given in Eqs. \eqref{Multiplet_W1} and \eqref{Multiplet_W2}. In addition, the complex conjugate multiplet $\bar{w}^2$ can be obtained by taking a replacement $Q \rightarrow \bar{Q}$ in the above expressions.

In the Liberated SUGRA gauge ($\Lambda_K=0$), the non-vanishing components are given by
\begin{eqnarray}
\mathcal{H}_{w}|_{\Lambda_K=0} &=& \frac{\mathcal{H}_{W}}{\Upsilon^2},\quad \mathcal{K}_{\bar{w}}|_{\Lambda_K=0} = \frac{\mathcal{K}_{\bar{W}}}{\Upsilon^2}\\
\Lambda_{w}|_{\Lambda_K=0} &=& -\frac{1}{\Upsilon^3} P_R \mathcal{H}_{W} \mathcal{Z}_{\Upsilon},\quad \Lambda_{\bar{w}}|_{\Lambda_K=0} = -\frac{1}{\Upsilon^3} P_L \mathcal{K}_{\bar{W}} \mathcal{Z}_{\Upsilon}\\
\mathcal{D}_{w}|_{\Lambda_K=0} &=& -\frac{1}{\Upsilon^3}\mathcal{K}_{\Upsilon}\mathcal{H}_{W} -\frac{3}{2\Upsilon^4} \bar{\mathcal{Z}}_{} P_R\mathcal{H}_{W} \mathcal{Z}_{\Upsilon},\quad \mathcal{D}_{\bar{w}}|_{\Lambda_K=0} = -\frac{1}{\Upsilon^3}\mathcal{K}_{\bar{W}}\mathcal{H}_{\Upsilon} -\frac{3}{2\Upsilon^4} \bar{\mathcal{Z}}_{} P_L\mathcal{K}_{\bar{W}} \mathcal{Z}_{\Upsilon}. \nonumber\\{} 
\end{eqnarray}

Furthermore, with the superconformal gauge choice $\mathcal{Z}_{\Upsilon}=0$, the non-vanishing components are
\begin{eqnarray}
\mathcal{H}_{w}|_{\Lambda_K=0,\mathcal{Z}_{\Upsilon}=0} &=& \frac{\mathcal{H}_{W}}{\Upsilon^2},\quad \mathcal{K}_{\bar{w}}|_{\Lambda_K=0,\mathcal{Z}_{\Upsilon}=0} = \frac{\mathcal{K}_{\bar{W}}}{\Upsilon^2}\\
\mathcal{D}_{w}|_{\Lambda_K=0,\mathcal{Z}_{\Upsilon}=0} &=& -\frac{1}{\Upsilon^3}\mathcal{K}_{\Upsilon}\mathcal{H}_{W},\quad \mathcal{D}_{\bar{w}}|_{\Lambda_K=0,\mathcal{Z}_{\Upsilon}=0} = -\frac{1}{\Upsilon^3}\mathcal{K}_{\bar{W}}\mathcal{H}_{\Upsilon}. \nonumber\\{} 
\end{eqnarray}

\subsubsection{\texorpdfstring{$T(\bar{w}^2),\bar{T}(w^2)$}{} chiral projection multiplets: (Weyl/Chiral) weights \texorpdfstring{$=(0,0)$}{}}

\begin{eqnarray}
T(\bar{w}^2) &=& \left( -\frac{1}{2}\mathcal{K}_{\bar{w}}, -\frac{1}{2} \sqrt{2} iP_L (\cancel{\mathcal{D}}\mathcal{Z}_{\bar{w}}+\Lambda_{\bar{w}}), \frac{1}{2}(\mathcal{D}_{\bar{w}}+\square^C \mathcal{C}_{\bar{w}} + i\mathcal{D}_a \mathcal{B}^a_{\bar{w}}) \right),\\
\bar{T}(w^2) &=& \left( -\frac{1}{2}\mathcal{K}_{\bar{w}}^{*}, \frac{1}{2} \sqrt{2} iP_R (\cancel{\mathcal{D}}\mathcal{Z}_{\bar{w}}^C+\Lambda_{\bar{w}}^C), \frac{1}{2}(\mathcal{D}_{\bar{w}}^{*}+\square^C \mathcal{C}_{\bar{w}}^{*} - i\mathcal{D}_a (\mathcal{B}^a_{\bar{w}})^{*}) \right) .
\end{eqnarray}
The superconformal setting for this and its complex conjugate are then given by
\begin{eqnarray}
T \equiv T(\bar{w}^2) &=& \{\mathcal{C}_T,\mathcal{Z}_T,\mathcal{H}_T,\mathcal{K}_T,\mathcal{B}_{\mu}^T,\Lambda_T,\mathcal{D}_T\} ,
\nonumber\\
\bar{T} \equiv \bar{T}(w^2) &=& \{\mathcal{C}_{\bar{T}},\mathcal{Z}_{\bar{T}},\mathcal{H}_{\bar{T}},\mathcal{K}_{\bar{T}},\mathcal{B}_{\mu}^{\bar{T}},\Lambda_{\bar{T}},\mathcal{D}_{\bar{T}}\},
\end{eqnarray}
where
\begin{eqnarray}
\mathcal{C}_T &=&  -\frac{1}{2} \mathcal{K}_{\bar{w}} = -\frac{1}{2\Upsilon^2}\mathcal{K}_{\bar{W}} +\frac{\mathcal{C}_{\bar{W}}}{\Upsilon^3}\mathcal{K}_{\Upsilon} +\frac{1}{4}\bigg[ -4\frac{1}{\Upsilon^3} (\bar{\mathcal{Z}}_{\bar{W}}P_R\mathcal{Z}_{\Upsilon}+\bar{\mathcal{Z}}_{\Upsilon}P_R\mathcal{Z}_{\bar{W}})+6\frac{\mathcal{C}_{\bar{W}}}{\Upsilon^4}\bar{\mathcal{Z}}_{\Upsilon}P_R\mathcal{Z}_{\Upsilon}\bigg],\nonumber\\{} \\
\mathcal{Z}_T &=& -P_L(\cancel{\mathcal{D}}\mathcal{Z}_{\bar{w}}+\Lambda_{\bar{w}}),\\
\mathcal{H}_T &=& -(\mathcal{D}_{\bar{w}}+\square^C \mathcal{C}_{\bar{w}} + i\mathcal{D}_a \mathcal{B}^a_{\bar{w}}),\\
\mathcal{K}_T &=& 0,\\
\mathcal{B}^T_{\mu} &=& -\frac{1}{2} i \mathcal{D}_{\mu} \mathcal{K}_{\bar{w}},\\
\Lambda_T &=& 0 ,\\
\mathcal{D}_T &=& 0.
\end{eqnarray}
and
\begin{eqnarray}
\mathcal{C}_{\bar{T}} &=&  -\frac{1}{2} \mathcal{K}_{\bar{w}}^{*},\\
\mathcal{Z}_{\bar{T}} &=& -P_R(\cancel{\mathcal{D}}\mathcal{Z}_{\bar{w}}^C+\Lambda_{\bar{w}}^C),\\
\mathcal{H}_{\bar{T}} &=& 0,\\
\mathcal{K}_{\bar{T}} &=& -(\mathcal{D}_{\bar{w}}^{*}+\square^C \mathcal{C}_{\bar{w}}^{*} - i\mathcal{D}_a (\mathcal{B}^a_{\bar{w}})^{*}),\\
\mathcal{B}^{\bar{T}}_{\mu} &=& \frac{1}{2} i \mathcal{D}_{\mu} \mathcal{K}_{\bar{w}}^{*},\\
\Lambda_{\bar{T}} &=& 0 ,\\
\mathcal{D}_{\bar{T}} &=& 0.
\end{eqnarray}

In the Liberated SUGRA gauge, the non-vanishing components are given by
\begin{eqnarray}
\mathcal{C}_T|_{\Lambda_K=0} &=&  -\frac{1}{2} \mathcal{K}_{\bar{w}}, \\
\mathcal{Z}_T|_{\Lambda_K=0} &=& -P_L\Lambda_{\bar{w}},\\
\mathcal{H}_T|_{\Lambda_K=0} &=& -\mathcal{D}_{\bar{w}},\\
\mathcal{B}^T_{\mu}|_{\Lambda_K=0} &=& -\frac{1}{2} i \mathcal{D}_{\mu} \mathcal{K}_{\bar{w}}.
\end{eqnarray}
With the superconformal gauge choice $\mathcal{Z}_{\Upsilon} =0$, we have 
\begin{eqnarray}
\mathcal{C}_T|_{\Lambda_K=0,\mathcal{Z}_{\Upsilon} =0} &=&  -\frac{1}{2} \mathcal{K}_{\bar{w}},\qquad \mathcal{C}_{\bar{T}}|_{\Lambda_K=0,\mathcal{Z}_{\Upsilon} =0} =  -\frac{1}{2} \mathcal{H}_{w} = -\frac{1}{2} \mathcal{K}_{\bar{w}}^{*} \\
\mathcal{H}_T|_{\Lambda_K=0,\mathcal{Z}_{\Upsilon} =0} &=& -\mathcal{D}_{\bar{w}},\qquad \mathcal{K}_{\bar{T}}|_{\Lambda_K=0,\mathcal{Z}_{\Upsilon} =0} = -\mathcal{D}_{w}=-\mathcal{D}_{\bar{w}}^{*}\\
\mathcal{B}^T_{\mu}|_{\Lambda_K=0,\mathcal{Z}_{\Upsilon} =0} &=& -\frac{1}{2} i \mathcal{D}_{\mu} \mathcal{K}_{\bar{w}},\qquad \mathcal{B}^{\bar{T}}_{\mu}|_{\Lambda_K=0,\mathcal{Z}_{\Upsilon} =0} = \frac{1}{2} i \mathcal{D}_{\mu} \mathcal{H}_{w}= \frac{1}{2} i \mathcal{D}_{\mu} \mathcal{K}_{\bar{w}}^{*}.
\end{eqnarray}

\subsubsection{Final form of the composite multiplet of liberated \texorpdfstring{$\mathcal{N}=1$}{} supergravity}

To construct the final composite multiplet which will give the new term in the action of liberated supergravity, we first collect the 
following chiral multiplets:
\begin{eqnarray}
\Big( X^i \equiv  \{S_0,z^I,T(\bar{w}^2),\mathcal{W}^2(K)\},~\chi^i \equiv  \{\chi^0,\chi^I,\chi^T,\chi^W\},~F^i \equiv \{F^{0},F^{I},F^{T},F^{W}\} \Big),\nonumber\\{}
\end{eqnarray}
where
\begin{eqnarray}
S_0 &=& \left(s_0, P_L\chi^0, F^0 \right),\qquad  
\bar{S}_0 = \left(s_0^{*}, P_R\chi^0, F^{0*} \right),\\
z^I &=& \left( z^I, P_L\chi^I, F^I \right),\qquad  
 \bar{z}^{\bar{I}} = \left( \bar{z}^{\bar{I}}, P_R\chi^{\bar{I}}, \bar{F}^{\bar{I}} \right),
\end{eqnarray}

\begin{eqnarray}
\mathcal{W}^2(K) &=& \left( \Bar{\Lambda}_K P_L \Lambda_K,~ \sqrt{2} P_L (-\gamma\cdot \hat{F}_K + 2i\mathcal{D}_K)\Lambda_K,~ 2\Bar{\Lambda}_KP_L \cancel{\mathcal{D}}\Lambda_K + \hat{F}_K^-\cdot \hat{F}_K^- - \mathcal{D}^2_K\right) \nonumber\\
&\equiv& \left( X^W, P_L\chi^W, F^W \right),\\
\bar{\mathcal{W}}^2(K) &=& \left( \Bar{\Lambda}_K P_R \Lambda_K,~ \sqrt{2} P_R (-\gamma\cdot \hat{F}_K - 2i\mathcal{D}_K^{*})\Lambda_K^C,~ 2\Bar{\Lambda}_KP_R \cancel{\mathcal{D}}\Lambda_K + \hat{F}_K^+\cdot \hat{F}_K^+ - (\mathcal{D}^{*}_K)^2\right) \nonumber\\
&\equiv& \left( \bar{X}^{\bar{W}}, P_R\chi^{\bar{W}}, \bar{F}^{\bar{W}} \right),
\end{eqnarray}

\begin{eqnarray}
T(\bar{w}^2) &=& \left( -\frac{1}{2}\mathcal{K}_{\bar{w}}, -\frac{1}{2} \sqrt{2} iP_L (\cancel{\mathcal{D}}\mathcal{Z}_{\bar{w}}+\Lambda_{\bar{w}}), \frac{1}{2}(\mathcal{D}_{\bar{w}}+\square^C \mathcal{C}_{\bar{w}} + i\mathcal{D}_a \mathcal{B}^a_{\bar{w}}) \right) \equiv \left( X^T, P_L\chi^T, F^T \right),\nonumber\\{}\\
\bar{T}(w^2) &=& \left( -\frac{1}{2}\mathcal{K}_{\bar{w}}^{*}, \frac{1}{2} \sqrt{2} iP_R (\cancel{\mathcal{D}}\mathcal{Z}_{\bar{w}}^C+\Lambda_{\bar{w}}^C), \frac{1}{2}(\mathcal{D}_{\bar{w}}^{*}+\square^C \mathcal{C}_{\bar{w}}^{*} - i\mathcal{D}_a (\mathcal{B}^a_{\bar{w}})^{*}) \right)
\equiv \left( \bar{X}^{\bar{T}}, P_R\chi^{\bar{T}}, \bar{F}^{\bar{T}} \right). \nonumber\\{}
\end{eqnarray}

Then, the lowest component of the final composite real multiplet is
\begin{eqnarray}
N &\equiv& \Upsilon^2 \frac{w^2\bar{w}^2}{T(\bar{w}^2)\bar{T}(w^2)}\mathcal{U} = \Upsilon^{-2}\dfrac{\mathcal{W}^2(K)\bar{\mathcal{W}}^2(K)}{T(w^2)\bar{T}(\bar{w}^2)} \mathcal{U} \nonumber\\
&=&  (X^0\bar{X}^0e^{-K(z^I,\bar{z}^{\bar{I}})/3})^{-2} X^W\bar{X}^{\bar{W}} (X^T)^{-1} (\bar{X}^{\bar{T}})^{-1}\mathcal{U}(z^I,\bar{z}^{\bar{I}}).
\end{eqnarray}
Then, since $N$ is now expressed as a function of superconformal chiral multiplets, we can represent the full off-shell expression of the superconformal Lagrangian of the liberated $\mathcal{N}=1$ supergravity via Eq. (17.19) of Freedman and Van Proeyen~\cite{fvp} as follows:
\begin{eqnarray}
  \mathcal{L}_{NEW} &\equiv& [\mathbb{N}]_De^{-1} \nonumber\\
  &=&
 \frac{1}{2}\mathcal{D}_N - \frac{1}{4}\bar{\psi}\cdot \gamma i\gamma_* \Lambda_N -\frac{1}{6} \mathcal{C}_N R(\omega) + \frac{1}{12} \left(\mathcal{C}_N\bar{\psi}_{\mu} \gamma^{\mu\rho\sigma} - i \bar{\mathcal{Z}}_N  \gamma^{\rho\sigma} \gamma_*\right) R_{\rho\sigma}'(Q)\nonumber\\
&&+\frac{1}{8} \varepsilon^{abcd} \bar{\psi}_{a}\gamma_b \psi_c (\mathcal{B}_{d}^N -\frac{1}{2}\bar{\psi}_d \mathcal{Z}_N) 
,\\
 &=& N_{i\bar{j}} \bigg( -\mathcal{D}_{\mu}X^i\mathcal{D}^{\mu}\bar{X}^{\bar{j}} - \frac{1}{2} \bar{\chi}^i \cancel{\mathcal{D}} \chi^{\bar{j}} - \frac{1}{2} \bar{\chi}^{\bar{j}}\cancel{\mathcal{D}}\chi^i + F^i\bar{F}^{\bar{j}}\bigg)
\nonumber\\
&& +\frac{1}{2}\bigg[ N_{ij\bar{k}} \Big( -\bar{\chi}^i\chi^j \bar{F}^{\bar{k}} + \bar{\chi}^i (\cancel{\mathcal{D}}X^j)\chi^{\bar{k}} \Big) +h.c. \bigg]
+ \frac{1}{4}N_{ij\bar{k}\bar{l}} \bar{\chi}^i\chi^j \bar{\chi}^{\bar{k}}\chi^{\bar{l}}
\nonumber\\
&& +\bigg[\frac{1}{2\sqrt{2}}\bar{\psi}\cdot \gamma 
\bigg( N_{i\bar{j}} F^i \chi^{\bar{j}} - N_{i\bar{j}} \cancel{\mathcal{D}}\bar{X}^{\bar{j}}\chi^i -\frac{1}{2}N_{ij\bar{k}} \chi^{\bar{k}}\bar{\chi}^i\chi^j  \bigg)  \nonumber\\
&& +\frac{1}{8} i\varepsilon^{\mu\nu\rho\sigma} \bar{\psi}_{\mu} \gamma_{\nu} \psi_{\rho} \bigg( N_i \mathcal{D}_{\sigma}X^i + \frac{1}{2} N_{i\bar{j}} \bar{\chi}^i \gamma_{\sigma} \chi^{\bar{j}} + \frac{1}{\sqrt{2}} N_i \bar{\psi}_{\sigma} \chi^i  \bigg)+h.c. \bigg] 
\nonumber\\
&&+\frac{1}{6} N \left( -R(\omega) +\frac{1}{2} \bar{\psi}_{\mu} \gamma^{\mu \nu\rho} R'_{\nu\rho}(Q) \right)
-\frac{1}{6\sqrt{2}} \left( N_i\bar{\chi}^i + N_{\bar{i}}\bar{\chi}^{\bar{i}} \right)\gamma^{\mu\nu} R'_{\mu\nu}(Q),\nonumber\\{}\label{NEW_Lagrangian}
\end{eqnarray}
where $i,j = 0,I,W,T$ and $\bar{j},\bar{k},\bar{l} = \Bar{0},\Bar{I},\Bar{W},\Bar{T}$ and
\begin{eqnarray}
&& R_{\mu\nu ab}(\omega) \equiv  \partial_{\mu} \omega_{\nu ab} - \partial_{\nu} \omega_{\mu ab} + \omega_{\mu ac}\omega_{\nu ~ b}^{~c} - \omega_{\nu ac}\omega_{\mu ~ b}^{~c} ,\\
&& R_{\mu\nu}'(Q) \equiv 2 \left(\partial_{[\mu} + \frac{1}{2} b_{[\mu} -\frac{3}{2} A_{[\mu} \gamma_* + \frac{1}{4} \omega_{[\mu}^{ab}(e,b,\psi)\gamma_{ab} \right)\psi_{\nu]},\\
&& \omega_{\mu}^{ab}(e,b,\psi) = \omega_{\mu}^{ab}(e,b) + \frac{1}{2} \psi_{\mu} \gamma^{[a} \psi^{b]} + \frac{1}{4} \bar{\psi}^a \gamma_{\mu} \psi^b.    
\end{eqnarray}

\subsection{Bosonic Lagrangians}

The bosonic contribution to the scalar potential can be found from the term $\mathcal{D}_N$. Using the above results and Eq. \eqref{NEW_Lagrangian}, we find the equivalent bosonic contribution to the scalar potential as follows: 
\begin{eqnarray}
\mathcal{L}_B \supset N_{i\bar{j}}F^i\bar{F}^{\bar{j}} \sim N_{W\bar{W}}F^W\bar{F}^{\bar{W}} = \Upsilon^{-2}\frac{1}{\mathcal{C}_{T}\mathcal{C}_{\bar{T}}}\mathcal{U} F^W\bar{F}^{\bar{W}}.
\end{eqnarray}
Since $\mathcal{C}_{T} =- \frac{1}{2}\mathcal{K}_{\bar{w}} \sim -\frac{1}{2} \frac{\mathcal{K}_{\bar{W}}}{\Upsilon^2} =-\frac{1}{2} \frac{(-2\bar{F}^{\bar{W}})}{\Upsilon^2} = \frac{\bar{F}^{\bar{W}}}{\Upsilon^2}$ and $\mathcal{C}_{\bar{T}} = \frac{F^W}{\Upsilon^2}$, we have
\begin{eqnarray}
\mathcal{L}_B \supset \Upsilon^{-2}\frac{\Upsilon^2\Upsilon^2}{\bar{F}^{\bar{W}}F^W}\mathcal{U} F^W\bar{F}^{\bar{W}} = \Upsilon^2 \mathcal{U}.
\end{eqnarray}

In the superconformal gauge $\Upsilon=1$, we get
\begin{eqnarray}
V_{NEW} =  \mathcal{U}.
\end{eqnarray}

\subsection{Fermionic Lagrangians}

In this section, we investigate fermionic terms in the liberated $\mathcal{N}=1$ supergravity Lagrangian. We focus in particular 
on the matter-coupling and on the most divergent fermionic terms, in order to explore interesting interactions and check the limits
of validity of the liberated $\mathcal{N}=1$ supergravity as an effective theory. 

First of all, let us recall that the final composite multiplet $N$ in terms of the superconformal chiral multiplets $\{S_0, z^I, W \equiv \mathcal{W}^2(K), T \equiv T(\bar{w}^2)\}$ are:
\begin{eqnarray}
N = \left(s_0s_0^{*}e^{-K(z^I,\bar{z}^{\bar{I}})/3}\right)^{-2}\dfrac{W\bar{W}}{T\bar{T}}\mathcal{U}(z^I,\bar{z}^{\bar{I}}).
\end{eqnarray}
We we also denote their lowest components as $W \equiv \bar{\Lambda}_KP_L\Lambda_K$, $\bar{W} \equiv \bar{\Lambda}_KP_R\Lambda_K$, $T\equiv \mathcal{C}_{T}$, and $\bar{T} \equiv \mathcal{C}_{\bar{T}}$. Note that the final composite multiplet consists of the four superconformal chiral multiplets only. Remember that $\mathcal{C}_{T} \sim \dfrac{\bar{F}^{\bar{W}}}{\Upsilon^2}$ and $F^W \propto \mathcal{D}_K^2|_{\textrm{boson}}\equiv \tilde{\mathcal{F}}^2$. 

Generically, the matter couplings are found from the following contributions:
\begin{eqnarray}
\mathcal{L}_{F}^{\textrm{matter}}|_{\psi=0} &=& N_{i\bar{j}} \bigg( -\mathcal{D}_{\mu}X^i\mathcal{D}^{\mu}\bar{X}^{\bar{j}} - \frac{1}{2} \bar{\chi}^i \cancel{\mathcal{D}} \chi^{\bar{j}} - \frac{1}{2} \bar{\chi}^{\bar{j}}\cancel{\mathcal{D}}\chi^i + F^i\bar{F}^{\bar{j}}\bigg) - \frac{N}{6}R(\omega)|_{\psi=0}
\nonumber\\
&& +\frac{1}{2}\bigg[ N_{ij\bar{k}} \Big( -\bar{\chi}^i\chi^j \bar{F}^{\bar{k}} + \bar{\chi}^i (\cancel{\mathcal{D}}X^j)\chi^{\bar{k}} \Big) +h.c. \bigg]
+ \frac{1}{4}N_{ij\bar{k}\bar{l}} \bar{\chi}^i\chi^j \bar{\chi}^{\bar{k}}\chi^{\bar{l}}\bigg|_{\psi=0},\nonumber\\
&=& \mathcal{L}_{F1}+\mathcal{L}_{F2}+\bar{\mathcal{L}}_{F2}+\mathcal{L}_{F3} + (\mathcal{L}_{F4}+\mathcal{L}_{F5}+h.c.)+\mathcal{L}_{F6} + \mathcal{L}_{F7}
\end{eqnarray}
where
\begin{eqnarray}
 \mathcal{D}_{\mu} X^i|_{\psi=0} &=& (\partial_{\mu} -w_ib_{\mu} -w_iA_{\mu})X^i,\nonumber\\
 \mathcal{D}_{\mu} P_L\chi^i|_{\psi=0} &=& \left(\partial_{\mu} +\frac{1}{4}\omega_{\mu}^{ab}\gamma_{ab}-(w_i+1/2)b_{\mu} + (w_i-3/2)iA_{\mu}\right) P_L\chi^i -\sqrt{2}w_iX^i   P_L\phi_{\mu}. \nonumber
\end{eqnarray}

Note that the matter couplings of fermions can be classified into seven types:
\begin{eqnarray}
\mathcal{L}_{F1} &\equiv&  -N_{i\bar{j}}\mathcal{D}_{\mu}X^i\mathcal{D}^{\mu}\bar{X}^{\bar{j}}\Big|_{\psi=0},\\
\mathcal{L}_{F2} &\equiv&  -\frac{1}{2} N_{i\bar{j}} \bar{\chi}^i \cancel{\mathcal{D}} \chi^{\bar{j}}\Big|_{\psi=0}  ,\\
\mathcal{L}_{F3} &\equiv&   -N_{i\bar{j}} F^i\bar{F}^{\bar{j}} \Big|_{\psi=0},\\
\mathcal{L}_{F4} &\equiv& -\frac{1}{2} N_{ij\bar{k}} \bar{\chi}^i\chi^j \bar{F}^{\bar{k}}\Big|_{\psi=0},\\
\mathcal{L}_{F5} &\equiv& \frac{1}{2}  N_{ij\bar{k}} \bar{\chi}^i (\cancel{\mathcal{D}}X^j)\chi^{\bar{k}}\Big|_{\psi=0} ,\\
\mathcal{L}_{F6} &\equiv&  \frac{1}{4}N_{ij\bar{k}\bar{l}} \bar{\chi}^i\chi^j \bar{\chi}^{\bar{k}}\chi^{\bar{l}} \Big|_{\psi=0} ,\\
\mathcal{L}_{F7} &\equiv& - \frac{N}{6}R(\omega)|_{\psi=0}.
\end{eqnarray}

The derivatives of the $N$ are given in general by
\begin{eqnarray}
N^{(r=q+p+m+k)}_{q,p,m,k} &=& (\partial_{0}^{q}\partial_{W}^{p}\partial_{T}^m \partial_{I}^{k} N ) \nonumber\\
&=& \bigg[(\partial_0^q\partial_I^{(k-n)}\Upsilon^{-2})_{(N)}(\Upsilon^{2})_{(\mathcal{C}_T)}(\Upsilon^{2})_{(\mathcal{C}_{\bar{T}})}
\nonumber\\
&&
\times (\Upsilon^{2m_1})_{(\partial_T^{m_1}\mathcal{C}_T)}(\Upsilon^{2m_2})_{(\partial_{\bar{T}}^{m_2}\mathcal{C}_{\bar{T}})} 
(W)^{1-p_1}(\bar{W})^{1-p_2}\mathcal{U}^{(n)} \bigg]\nonumber\\
&&/\bigg[ (\tilde{\mathcal{F}}^{2})_{(\mathcal{C}_{T})}(\tilde{\mathcal{F}}^{2})_{(\mathcal{C}_{\bar{T}})}(\tilde{\mathcal{F}}^{2m_1})_{(\partial_T^{m_1}\mathcal{C}_T)}(\tilde{\mathcal{F}}^{2m_2})_{(\partial_{\bar{T}}^{m_2}\mathcal{C}_{\bar{T}})} \bigg]. \nonumber\\
&=& (\partial_0^q\partial_I^{(k-n)}\Upsilon^{-2})\Upsilon^{4+2m_1+2m_2}\frac{\mathcal{U}^{(n)}}{\tilde{\mathcal{F}}^{2+2+2m_1+2m_2}}W^{1-p_1}\bar{W}^{1-p_2} \nonumber\\
&=& (\partial_0^q\partial_I^{(k-n)}\Upsilon^{-2})\Upsilon^{4+2m}\frac{\mathcal{U}^{(n)}}{\tilde{\mathcal{F}}^{4+2m}}W^{1-p_1}\bar{W}^{1-p_2},
\end{eqnarray}
where $\tilde{\mathcal{F}} \equiv  2K_{I\bar{\jmath}} \left( -\partial_{\mu} z^I \partial^{\mu} \bar{z}^{\bar{\jmath}} +F^I\bar{F}^{\bar{\jmath}}\right)$; 
$\mathcal{U}^{(n)}$ ($0 \leq n \leq 4$) is the $n$-th derivative of the 
function $\mathcal{U}(z^I,\bar{z}^{\bar{\imath}})$ with respect to $z^I,\bar{z}^I$, 
which are the lowest component of the matter chiral  multiplets; $q=q_1+q_2$ where $q_1$ 
($q_2$) is the order of the derivative with respect to the compensator scalar $s_0$ ($s_0^{*}$); $p=p_1+p_2$ where $p_1$ ($p_2$) is the order of the derivative with respect to the field strength multiplet scalar $W$ ($\bar{W}$); $m=m_1+m_2$ where $m_1$ ($m_2$) is the 
order of the derivative with respect to the chiral projection multiplet scalar $T(\bar{w}^2)$ ($\bar{T}(w^2)$); $k$ is the order of the derivative with respect to the matter multiplet scalar $z^I$; 
$n$ is the order of the derivative acting on the new term $\mathcal{U}$ with respect to the matter multiplet; $q$ is the total order of 
derivative with respect to the compensator scalars $s_0$ and $s_0^{*}$. An explicit form of the derivatives of $N$ is given in
Appendix A.

The mass dimension\footnote{the mass dimensions of the multiplets' lowest components are $[s_0]=1,[z^I]=0,[W]=3,[T]=0$, which gives $[\tilde{\mathcal{F}}]=2$.} of the derivatives of the $N$ is $[N^{(r=q+p+m+k)}_{q,p,m,k}]=-3p-4m-2$. This implies that the mass dimension of the operator coupled to $N^{(r=q+p+m+k)}_{q,p,m,k}$ must be equal to $3p+4m+2$.

Now, let us focus on the case such that $q=0$ and $k=n$ which gives the most singular fermionic terms in the limit
that the D-term vanishes. The most singular terms are those that contain the highest power of the auxiliary field $D$ in the 
denominator and therefore are the nonrenormalizable operators associated with the smallest UV cutoff mass scale. That is, we consider that the matter scalar derivatives act only on the new term $\mathcal{U}$ and there are no the derivatives with respect to the compensator scalar. Then, we have 
\begin{eqnarray}
N^{(r=p+m+n)}_{p,m,n} = \Upsilon^{2+2m}\frac{\mathcal{U}^{(n)}}{\tilde{\mathcal{F}}^{4+2m}}W^{1-p_1}\bar{W}^{1-p_2}.
\end{eqnarray}
In particular, since $r = p+m+n$ ($0 \leq r \leq 4$), it reduces to
\begin{eqnarray}
N_{i \ldots l}^{(r)} = N^{(r)}_{p,m,n} = \Upsilon^{2(1+r-n-p)}\frac{\mathcal{U}^{(n)}}{\tilde{\mathcal{F}}^{2(2+r-n-p)}}W^{1-p_1}\bar{W}^{1-p_2}.
\end{eqnarray}
Remember that we called $r$ the total order of the derivatives acting on the $N$. Here $n$ is the number of 
derivatives acting on the 
matter scalars in the new term $\mathcal{U}$; they produce $\mathcal{U}^{(n)}$. Finally, $p=p_1+p_2$ is the sum of the number of derivatives w.r.t the multiplets $(W,\bar{W})$ acting on $\mathcal{U}$.

\subsubsection{Structure of the fermionic components}

Next, let us explore the detailed structure of the chiral fermions of the superconformal multiplets. First of all, the compensator and matter fermions are given by
\begin{eqnarray}
 \chi^0 &=& P_L\chi^0,\qquad \chi^I = P_L\chi^I.
 \end{eqnarray}
Note that $\chi^0$ and $\chi^I$ are fundamental fermions, and later in the $S$-gauge, the compensator chiral fermions will be replaced by the matter ones according to: $P_L\chi^0 = \frac{1}{3}s_0 K_I P_L\chi^I$ and $P_R\chi^0 = \frac{1}{3}s_0^{*} K_{\bar{I}} P_R\chi^{\bar{I}}$.
 
On the contrary, the other fermions $\chi^W$ and $\chi^T$ are composite. Hence, we need to find their specific structure.
 
First, the $\mathcal{W}(K)^2$-multiplet fermion, say $\chi^W$, is found to be 
\begin{eqnarray}
 \chi^W &=& P_L \chi^W = \sqrt{2}  P_L(-\gamma\cdot \hat{F}_K + 2i\mathcal{D}_K)\Lambda_K
 \nonumber\\
 &=&  -2\sqrt{2} P_L\gamma^{\mu\nu} \partial_{[\mu}\mathcal{B}_{\nu]}^K\Lambda_K-\sqrt{2}P_L\gamma^{\mu\nu} \Bar{\psi}_{[\mu} \gamma_{\nu]}\Lambda_K\Lambda_K + 2\sqrt{2}iP_L\mathcal{D}_K\Lambda_K
 \nonumber\\
 &=& -2\sqrt{2} P_L\gamma^{\mu\nu} \partial_{[\mu}(iK_I\mathcal{D}_{\nu]} z^I -iK_{\bar{I}}\mathcal{D}_{\nu]} \bar{z}^{\bar{I}} + i K_{I\bar{J}}\bar{\chi}^I \gamma_{\nu]} \chi^{\bar{J}})\Lambda_K+ 2\sqrt{2}iP_L\mathcal{D}_K\Lambda_K
 \nonumber\\ 
 &=& 2 i \gamma^{\mu\nu} \partial_{[\mu}(K_I \Bar{\psi}_{\nu]}\chi^I -K_{\bar{I}}\Bar{\chi}^{\bar{I}}\psi_{\nu]} - \sqrt{2} K_{I\bar{J}}\bar{\chi}^I \gamma_{\nu]} \chi^{\bar{J}})P_L\Lambda_K
 \nonumber\\&& 
 + 2\sqrt{2}i\bigg[2K_{I\bar{J}} 
\bigg( -\mathcal{D}_{\mu} z^I \mathcal{D}^{\mu} \bar{z}^{\bar{J}} -\frac{1}{2} \bar{\chi}^I P_L \cancel{\mathcal{D}}\chi^{\bar{J}} -\frac{1}{2} \bar{\chi}^{\bar{J}} P_R \cancel{\mathcal{D}} \chi^I + F^I\bar{F}^{\bar{J}}\bigg) \nonumber\\
&&+ K_{IJ\bar{K}} \bigg( -\bar{\chi}^I \chi^J \bar{F}^{\bar{K}} + \bar{\chi}^I (\cancel{\mathcal{D}}z^J)\chi^{\bar{K}}         \bigg) + K_{\bar{I}\bar{J}K} \bigg( -\bar{\chi}^{\bar{I}} \chi^{\bar{J}} F^{K} + \bar{\chi}^{\bar{I}} (\cancel{\mathcal{D}}\bar{z}^{\bar{J}})\chi^{K}\bigg) \nonumber\\
&& + \frac{1}{2} K_{IJ\bar{K}\bar{L}} (\bar{\chi}^IP_L\chi^J )(\bar{\chi}^{\bar{K}} P_R \chi^{\bar{L}})\bigg]P_L\Lambda_K
 \nonumber\\ 
 &=& 2\sqrt{2}i\tilde{\mathcal{F}} (P_L\Lambda_K)_{1f} + \cdots + 7~\textrm{fermions}\nonumber\\
 &=& 2\sqrt{2}i \tilde{\mathcal{F}} (-\sqrt{2}iK_{\bar{I}J}[(\cancel{\mathcal{D}}z^{J})_{0f}\chi^{\bar{I}}-\bar{F}^{\bar{I}}\chi^J]) + \cdots + 7~\textrm{fermions} \nonumber\\
 &=& 4 \tilde{\mathcal{F}} K_{\bar{I}J}[(\cancel{\partial}z^{J})\chi^{\bar{I}}-\bar{F}^{\bar{I}}\chi^J] + \cdots + 7~\textrm{fermions},
\end{eqnarray}
where $P_L \Lambda_K = -\sqrt{2}iK_{\bar{I}J}[(\cancel{\mathcal{D}}z^{J})\chi^{\bar{I}}-\bar{F}^{\bar{I}}\chi^J] -\frac{i}{\sqrt{2}}K_{\bar{I}\bar{J}K}\chi^{K}\bar{\chi}^{\bar{I}}\chi^J$. Note that the composite fermion $\chi^W$ contains powers of the
matter fermions ranging from one to seven, and it is nonvanishing on-shell. 

Second, the chiral projection multiplet $T(\bar{w}^2)$ fermions are found to be as follows:
\begin{eqnarray}
 \chi^T &=& P_L \chi^T = -\frac{i}{\sqrt{2}} P_L(\cancel{\mathcal{D}}\mathcal{Z}_{\bar{w}}+\Lambda_{\bar{w}}) = -\frac{i}{\sqrt{2}}\bigg[ \frac{(\cancel{\mathcal{D}}\mathcal{Z}_{\bar{W}})}{\Upsilon^2} - \frac{(\cancel{\mathcal{D}}\Upsilon)\mathcal{Z}_{\bar{W}}}{\Upsilon^3} -2\frac{\mathcal{C}_{\bar{W}}}{\Upsilon^3}P_L\Lambda_{\Upsilon} -\frac{i\cancel{\mathcal{B}}_{\Upsilon}\mathcal{Z}_{\bar{W}}}{\Upsilon^3} \bigg]_{\mathcal{Z}_{\Upsilon}=0}\nonumber\\
  &=& \frac{1}{\Upsilon^2}\bigg[~ \cancel{\mathcal{D}}\chi^{\bar{W}} - \frac{(\cancel{\mathcal{D}}\Upsilon)\chi^{\bar{W}}}{\Upsilon} -2\frac{\Bar{\Lambda}_KP_R\Lambda_K}{\Upsilon}P_L\Lambda_{\Upsilon} -\frac{i\cancel{\mathcal{B}}_{\Upsilon}\chi^{\bar{W}}}{\Upsilon}~ \bigg]_{\mathcal{Z}_{\Upsilon}=0}
  \nonumber\\
  &=& \frac{1}{\Upsilon^2}\bigg[~ (\cancel{\mathcal{D}}\chi^{\bar{W}})_{1f} - \frac{[(\cancel{\mathcal{D}}\Upsilon)_{0f}+i(\cancel{\mathcal{B}}_{\Upsilon})_{0f}]}{\Upsilon}(\chi^{\bar{W}})_{1f}~ \bigg] + \cdots + 9~\textrm{fermions} ,\\
  &=& \frac{1}{\Upsilon^2}\bigg[~  4 (\cancel{\partial}\tilde{\mathcal{F}}) K_{\bar{I}J}[(\cancel{\partial}z^{J})\chi^{\bar{I}}-\bar{F}^{\bar{I}}\chi^J] \nonumber\\
  &&- \Big(\frac{2}{s_0^{*}}\cancel{\partial}s_0^{*} -\frac{2}{3}K_{\bar{K}}\cancel{\partial}\bar{z}^{\bar{K}} - 2\gamma^{\mu}(b_{\mu}+iA_{\mu})\Big)4 \tilde{\mathcal{F}} K_{\bar{I}J}[(\cancel{\partial}z^{J})\chi^{\bar{I}}-\bar{F}^{\bar{I}}\chi^J] ~ \bigg]_{1f}+ \cdots + 9~\textrm{fermions} ,\nonumber\\
  &=& \frac{4}{\Upsilon^2}  (\cancel{\partial}\tilde{\mathcal{F}}) K_{\bar{I}J}[(\cancel{\partial}z^{J})\chi^{\bar{I}}-\bar{F}^{\bar{I}}\chi^J] + (\chi^T)_{1f}|_{\tilde{\mathcal{F}}^1} + \cdots + 9~\textrm{fermions},
\end{eqnarray}
where $\tilde{\mathcal{F}} \equiv  2K_{I\bar{J}} \left( -\partial_{\mu} z^I \partial^{\mu} \bar{z}^{\bar{J}} +F^I\bar{F}^{\bar{J}}\right)$; ``$|_{\tilde{\mathcal{F}}^1}$'' denotes the terms proportional to $\tilde{\mathcal{F}}^1$, and
\begin{eqnarray}
\cancel{\partial} \tilde{\mathcal{F}}&=& 2(\cancel{\partial}K_{I\bar{J}}) \left( -\partial_{\mu} z^I \partial^{\mu} \bar{z}^{\bar{J}} +F^I\bar{F}^{\bar{J}}\right)
\nonumber\\
&&+2K_{I\bar{J}} \left( -(\cancel{\partial}\partial_{\mu} z^I) \partial^{\mu} \bar{z}^{\bar{J}}-\partial_{\mu} z^I (\cancel{\partial}\partial^{\mu} \bar{z}^{\bar{J}}) +(\cancel{\partial}F^I)\bar{F}^{\bar{J}}+F^I(\cancel{\partial}\bar{F}^{\bar{J}})\right).
\end{eqnarray}
Note that $\chi^T$ is also nonvanishing on-shell.

In the above calculation we have used the formula:
\begin{eqnarray}
\mathcal{D}_{\mu} P_L\chi^W &=&  \left(\partial_{\mu} +\frac{1}{4}\omega_{\mu}^{ab}\gamma_{ab}-\frac{7}{2}b_{\mu} + \frac{3}{2}iA_{\mu}\right)P_L\chi^W -\frac{1}{\sqrt{2}}P_L(\cancel{\mathcal{D}}W + F^W)\psi_{\mu} \nonumber\\
&&-3\sqrt{2}W  P_L\phi_{\mu}.\\
\end{eqnarray}
to find
\begin{eqnarray}
(\cancel{\mathcal{D}}\chi^W)_{1f} &=& (\cancel{\mathcal{D}}P_L\chi^W)_{1f} \nonumber\\
&=& \gamma^{\mu}\left(\partial_{\mu} +\frac{1}{4}\omega_{\mu}^{ab}\gamma_{ab}-\frac{7}{2}b_{\mu} + \frac{3}{2}iA_{\mu}\right)(P_L\chi^W)_{1f} + \frac{1}{\sqrt{2}}P_L (F^W)_{0f} \gamma^{\mu} \psi_{\mu} \nonumber\\
&=& \cancel{\partial}(P_L\chi^W)_{1f}+\gamma^{\mu}\left(\frac{1}{4}\omega_{\mu}^{ab}\gamma_{ab}-\frac{7}{2}b_{\mu} + \frac{3}{2}iA_{\mu}\right)(P_L\chi^W)_{1f} + \frac{1}{\sqrt{2}}P_L \tilde{\mathcal{F}}^2 \gamma^{\mu} \psi_{\mu}\nonumber\\
&=& 4 (\cancel{\partial}\tilde{\mathcal{F}}) K_{\bar{I}J}[(\cancel{\partial}z^{J})\chi^{\bar{I}}-\bar{F}^{\bar{I}}\chi^J] + 4 \tilde{\mathcal{F}} (\cancel{\partial}K_{\bar{I}J})[(\cancel{\partial}z^{J})\chi^{\bar{I}}-\bar{F}^{\bar{I}}\chi^J]\nonumber\\
&&+4 \tilde{\mathcal{F}} K_{\bar{I}J}[(\square z^{J})\chi^{\bar{I}}+\cancel{\partial}z^{J}\cancel{\partial}\chi^{\bar{I}}-(\cancel{\partial}\bar{F}^{\bar{I}})\chi^J-\bar{F}^{\bar{I}}(\cancel{\partial}\chi^J)] \nonumber\\
&&+\gamma^{\mu}\left(\frac{1}{4}\omega_{\mu}^{ab}\gamma_{ab}-\frac{7}{2}b_{\mu} + \frac{3}{2}iA_{\mu}\right)(4 \tilde{\mathcal{F}} K_{\bar{I}J}[(\cancel{\partial}z^{J})\chi^{\bar{I}}-\bar{F}^{\bar{I}}\chi^J] ) + \frac{1}{\sqrt{2}}P_L \tilde{\mathcal{F}}^2 \gamma^{\mu} \psi_{\mu} \nonumber\\
&=& 4 (\cancel{\partial}\tilde{\mathcal{F}}) K_{\bar{I}J}[(\cancel{\partial}z^{J})\chi^{\bar{I}}-\bar{F}^{\bar{I}}\chi^J] + (\cancel{\mathcal{D}}\chi^W)_{1f}|_{\tilde{\mathcal{F}}^1} + (\cancel{\mathcal{D}}\chi^W)_{1f}|_{\tilde{\mathcal{F}}^2}.
\end{eqnarray}
where $F^W = 2\Bar{\Lambda}_KP_L \cancel{\mathcal{D}}\Lambda_K + \hat{F}_K^-\cdot \hat{F}_K^- - \mathcal{D}^2_K$, $(\hat{F}_K^-)_{0f}=0$, and $(F^W)_{0f} = -(\mathcal{D}_K^2)_{0f} = -\tilde{\mathcal{F}}^2$, and
\begin{eqnarray}
(\cancel{\partial} \chi^W)_{1f} &=& (\cancel{\partial} P_L\chi^W)_{1f} = 4 (\cancel{\partial}\tilde{\mathcal{F}}) K_{\bar{I}J}[(\cancel{\partial}z^{J})\chi^{\bar{I}}-\bar{F}^{\bar{I}}\chi^J] + 4 \tilde{\mathcal{F}} (\cancel{\partial}K_{\bar{I}J})[(\cancel{\partial}z^{J})\chi^{\bar{I}}-\bar{F}^{\bar{I}}\chi^J]\nonumber\\
&&+4 \tilde{\mathcal{F}} K_{\bar{I}J}[(\cancel{\partial}\cancel{\partial}z^{J})\chi^{\bar{I}}+\cancel{\partial}z^{J}\cancel{\partial}\chi^{\bar{I}}-(\cancel{\partial}\bar{F}^{\bar{I}})\chi^J-\bar{F}^{\bar{I}}(\cancel{\partial}\chi^J)]\nonumber\\
&=& 4 (\cancel{\partial}\tilde{\mathcal{F}}) K_{\bar{I}J}[(\cancel{\partial}z^{J})\chi^{\bar{I}}-\bar{F}^{\bar{I}}\chi^J] + 4 \tilde{\mathcal{F}} (\cancel{\partial}K_{\bar{I}J})[(\cancel{\partial}z^{J})\chi^{\bar{I}}-\bar{F}^{\bar{I}}\chi^J]\nonumber\\
&&+4 \tilde{\mathcal{F}} K_{\bar{I}J}[(\square z^{J})\chi^{\bar{I}}+\cancel{\partial}z^{J}\cancel{\partial}\chi^{\bar{I}}-(\cancel{\partial}\bar{F}^{\bar{I}})\chi^J-\bar{F}^{\bar{I}}(\cancel{\partial}\chi^J)]
\nonumber\\
&\approx& 4 (\cancel{\partial}\tilde{\mathcal{F}}) K_{\bar{I}J}[(\cancel{\partial}z^{J})\chi^{\bar{I}}-\bar{F}^{\bar{I}}\chi^J] + 4 \tilde{\mathcal{F}} (\cancel{\partial}K_{\bar{I}J})[(\cancel{\partial}z^{J})\chi^{\bar{I}}-\bar{F}^{\bar{I}}\chi^J]\nonumber\\
&&-4 \tilde{\mathcal{F}} K_{\bar{I}J}(\cancel{\partial}\bar{F}^{\bar{I}})\chi^J .
\end{eqnarray}
Here $\approx$ means equality up to terms proportional to the equations of motion of free massless matter fields. Such 
terms produce only terms that contain additional factors of the matter field masses in the numerator and therefore give rise to either renormalizable operators or nonrenormalizable operators weighted by a mass scale higher than that associated to those
terms that do not vanish on shell. The details of  the above calculation is given in Appendix A.1.

Finally, we present here the chiral fermions of the superconformal multiplets up to multiple fermion terms.
\begin{eqnarray}
\chi^i =
\begin{cases}
 \chi^0,\\
 \chi^I,\\
 \chi^W = 4 \tilde{\mathcal{F}} K_{\bar{I}J}[(\cancel{\partial}z^{J})\chi^{\bar{I}}-\bar{F}^{\bar{I}}\chi^J] + \cdots 7~\textrm{fermions},\\
 \chi^T = \frac{1}{\Upsilon^2}\bigg[~  4 (\cancel{\partial}\tilde{\mathcal{F}}) K_{\bar{I}J}[(\cancel{\partial}z^{J})\chi^{\bar{I}}-\bar{F}^{\bar{I}}\chi^J]- \Big(\frac{2}{s_0^{*}}\cancel{\partial}s_0^{*} -\frac{2}{3}K_{\bar{K}}\cancel{\partial}\bar{z}^{\bar{K}} - 2\gamma^{\mu}(b_{\mu}+iA_{\mu})\Big)
 \\ \qquad\qquad\quad  \times 4 \tilde{\mathcal{F}} K_{\bar{I}J}[(\cancel{\partial}z^{J})\chi^{\bar{I}}-\bar{F}^{\bar{I}}\chi^J] ~ \bigg]_{1f}+ \cdots + 9~\textrm{fermions} ,
\end{cases}
 \nonumber
\end{eqnarray}
where 
\begin{eqnarray}
\tilde{\mathcal{F}} &=&  2K_{I\bar{J}} \left( -\partial_{\mu} z^I \partial^{\mu} \bar{z}^{\bar{J}} +F^I\bar{F}^{\bar{J}}\right)
\nonumber\\
\cancel{\partial} \tilde{\mathcal{F}}&=& 2(\cancel{\partial}K_{I\bar{J}}) \left( -\partial_{\mu} z^I \partial^{\mu} \bar{z}^{\bar{J}} +F^I\bar{F}^{\bar{J}}\right)
\nonumber\\
&&+2K_{I\bar{J}} \left( -(\cancel{\partial}\partial_{\mu} z^I) \partial^{\mu} \bar{z}^{\bar{J}}-\partial_{\mu} z^I (\cancel{\partial}\partial^{\mu} \bar{z}^{\bar{J}}) +(\cancel{\partial}F^I)\bar{F}^{\bar{J}}+F^I(\cancel{\partial}\bar{F}^{\bar{J}})\right).\nonumber
\end{eqnarray}
Notice that none of the chiral fermions $\chi^i$ vanish on-shell, and that only $\chi^T$ dependens on $\tilde{\mathcal{F}}$ and includes the factor of $\Upsilon^{-2}$. All of these properties affect the mass dimension of the expansion coefficients of the nonrenormalizable Lagrangians.

We finally expand $W$ and $\Bar{W}$ as follows:
\begin{eqnarray}
 W &=&  -2K_{\bar{\imath}J}[\bar{\chi}^{J}(\overline{\cancel{\mathcal{D}}\bar{z}^{\bar{\imath}}})-\bar{F}^{\bar{\imath}}\bar{\chi}^{J}]K_{{\bar{\imath}}'J'}[(\cancel{\mathcal{D}}\bar{z}^{{\bar{\imath}}'})\chi^{J'}-F^{J'}\chi^{{\bar{\imath}}'}] 
 \nonumber\\
 && - K_{{\bar{\imath}}J}[\bar{\chi}^{J}(\overline{\cancel{\mathcal{D}}\bar{z}^{\bar{\imath}}})-\bar{F}^{\bar{\imath}}\bar{\chi}^{J}] K_{\bar{\imath}'\bar{\jmath}'K'}[\chi^{K'}\bar{\chi}^{\bar{\imath}'}\chi^{\bar{\jmath}'}] \nonumber\\
&& - K_{\bar{\imath}\bar{\jmath}K}[\bar{\chi}^{\bar{\jmath}}\chi^{\bar{\imath}}\bar{\chi}^{K}]K_{{\bar{\imath}}'J'}[(\cancel{\mathcal{D}}\bar{z}^{{\bar{\imath}}'})\chi^{J'}-F^{J'}\chi^{{\bar{\imath}}'}]  \nonumber\\
&&- \frac{1}{2}K_{\bar{\imath}\bar{\jmath}K}[\bar{\chi}^{\bar{\jmath}}\chi^{\bar{\imath}}\bar{\chi}^{K}]K_{\bar{\imath}'\bar{\jmath}'K'}[\chi^{K'}\bar{\chi}^{\bar{\imath}'}\chi^{\bar{\jmath}'}],\\
\bar{W} &=& (W)^{*}.
\end{eqnarray}
Notice that $W$ and $\Bar{W}$ can be represented by products of two, four, and six fundamental fermions. 

\subsection{Additional gauge fixing for physical theory in the liberated supergravity}

First of all, we consider the conventional superconformal gauge which is chosen by
\begin{eqnarray}
&&\mathcal{C}_{\Upsilon} = \Upsilon = 1 \Longleftrightarrow s_0\bar{s}_0e^{-K/3}=1,\\ &&\mathcal{Z}_{\Upsilon} = 0 \implies P_L\chi^0 - \frac{1}{3}s_0 K_I P_L\chi^I =0 \quad \& \quad P_R\chi^0 - \frac{1}{3}s_0^{*} K_{\bar{I}} P_R\chi^{\bar{I}} =0, \label{conv1}\\
&& s_0 = \bar{s}_0 \implies s_0 = \bar{s}_0 = e^{K/6}, \label{conv2} \\
&& b_{\mu} = 0. \label{conv3}
\end{eqnarray}
Note that the first condition is the $D$-gauge fixing which gets us to the Einstein frame; the second one is the improved $S$-gauge fixing; the third one is the $A$-gauge fixing; the last one is the $K$-gauge fixing.

To compare our results with the formulation of liberated supergravity in~\cite{fkr}, we choose a gauge given by
\beq
\Lambda_K =0 \Longleftrightarrow \chi^{W}=\chi^{T}=0.
\eeq{gauge-fixing-lib}

In both the conventional superconformal gauge~(\ref{conv1},\ref{conv2},\ref{conv3})  and in~\eqref{gauge-fixing-lib}, the relevant multiplets are
\begin{eqnarray}
S_0 &=& \left(e^{K/6}, \frac{1}{3}e^{K/6} K_I P_L\chi^I, F^0 \right),\qquad  
\bar{S}_0 = \left(e^{K/6}, \frac{1}{3}e^{K/6} K_{\bar{I}} P_R\chi^{\bar{I}}, F^{0*} \right),\\
z^I &=& \left( z^I, P_L\chi^I, F^I \right),\qquad  
 \bar{z}^{\bar{I}} = \left( \bar{z}^{\bar{I}}, P_R\chi^{\bar{I}}, \bar{F}^{\bar{I}} \right),\\
\mathcal{W}^2(K) &=&  \left( W, P_L\chi^W, F^W \right)=\left( 0, 0, \hat{F}_K^-\cdot \hat{F}_K^- - \mathcal{D}^2_K\right),\\
\bar{\mathcal{W}}^2(K) &=&  \left( \bar{W}, P_R\chi^{\bar{W}}, \bar{F}^{\bar{W}} \right)= \left( 0,0, \hat{F}_K^+ \cdot \hat{F}_K^+ - (\mathcal{D}^{*}_K)^2\right),\\
T(\bar{w}^2) &=& \left( T, P_L\chi^T, F^T \right) = \left( -\frac{1}{2}\mathcal{K}_{\bar{w}}, 0, \frac{1}{2}\mathcal{D}_{\bar{w}} \right),\\
\bar{T}(w^2) &=& \left( \bar{X}^{\bar{T}}, P_R\chi^{\bar{T}}, \bar{F}^{\bar{T}} \right) = \left( -\frac{1}{2}\mathcal{K}_{\bar{w}}^{*}, 0, \frac{1}{2}\mathcal{D}_{\bar{w}}^{*} \right). 
\end{eqnarray}
Let us further observe that all the terms coupled to $N$ and its derivatives with respect to $0,I,T$ vanish because $N$ contains the product of $W\bar{W}$, which is zero in the liberated gauge, i.e. $W=\bar{W}=0$. We see that all the fermionic terms coupled to the derivatives of $N$ with respect to $W$ or $T$ vanish as well, because these terms always couple to the fermions $\chi^W$ and $\chi^T$ which vanish in the gauge. The only non-vanishing contribution is given by only the bosonic term, $N_{W\bar{W}}F^W\bar{F}^{\bar{W}}$ which gives us the new term $\mathcal{U}$ to the scalar potential. Therefore, in both gauges, the liberated supergravity Lagrangian is specified by
\begin{eqnarray}
\mathcal{L}_{Lib} = \mathcal{L}_{SUGRA} + \mathcal{L}_{NEW},
\end{eqnarray}
where $\mathcal{L}_{SUGRA}$ is the standard supergravity Lagrangian which contains the auxiliary fields $F^0,F^I$ and $\mathcal{L}_{NEW} = \mathcal{U}(z^I,\bar{z}^{\bar{I}})$. Then, with this Lagrangian, after solving the equations of motion for the auxiliary fields, we can obtain the usual supergravity action in addition to the new term.

\section{Component action of new FI term in superconformal tensor calculus}

In this section, we review the component action of a new, K\"{a}hler invariant Fayet-Iliopoulos term proposed by Antoniadis, Chatrabhuti, Isono, and Knoops \cite{acik}, using again the superconformal tensor calculus. The full Lagrangian with the new FI terms is given by 
\begin{eqnarray}
\mathcal{L} &=& -3[S_0\bar{S}_0e^{-K(z,\bar{z})}]_D + ([S_0^3W(z)]_F -\frac{1}{4} [\bar{\lambda}P_L\lambda]_F+h.c.) \nonumber\\
&&- \xi \left[(S_0\bar{S}_0e^{-K(z,\bar{z})})^{-3}\frac{(\bar{\lambda}P_L\lambda)(\bar{\lambda}P_R\lambda)}{T(\bar{w}'^2)\bar{T}(w'^2)}(V)_D\right]_D,
\end{eqnarray}
where the last term coupled to the parameter $\xi$ corresponds to the new FI terms.

\subsection{Component action of the new FI term}

We consider matter chiral multiplets $Z^i$, the chiral compensator $S_0$, a real multiplet $V$, and another real multiplet 
$(V)_D$, whose lowest component is the auxiliary D term of the real multiplet $V$. Their superconformal multiplets are given as follows:
\begin{eqnarray}
&& V = \{0,0,0,0,A_{\mu},\lambda,D\} ~\textrm{in the Wess-Zumino gauge,~i.e.}~v=\zeta=\mathcal{H}=0, \\
&& Z^i = (z^i,-i\sqrt{2}P_L\chi^i,-2F^i,0,+i\mathcal{D}_{\mu}z^i,0,0) = \{ z^i, P_L\chi^i,F^i\},\\
&& \bar{Z}^{\bar{i}} = (\bar{z}^{\bar{i}},+i\sqrt{2}P_R\chi^{\bar{i}},0,-2\bar{F}^{\bar{i}},-i\mathcal{D}_{\mu}\bar{z}^{\bar{i}},0,0) = \{ \bar{z}^{\bar{i}}, P_R\chi^{\bar{i}},\bar{F}^{\bar{i}}\},\\
&& S_0 = (s_0,-i\sqrt{2}P_L\chi^0,-2F_0,0,+i\mathcal{D}_{\mu}s_0,0,0) = \{ s_0, P_L\chi^0,F_0\},\\
&& \bar{S}_0 = (\bar{s}_0,+i\sqrt{2}P_R\chi^0,0,-2\bar{F}_0,-i\mathcal{D}_{\mu}\bar{s}_0,0,0) = \{\bar{s}_0, P_R\chi^0,\bar{F}_0\},\\
&& \bar{\lambda}P_L\lambda = (\bar{\lambda}P_L\lambda,-i\sqrt{2}P_L\Lambda,2D_-^2,0,+i\mathcal{D}_{\mu}(\bar{\lambda}P_L\lambda),0,0) = \{\bar{\lambda}P_L\lambda, P_L\Lambda,-D_-^2\},\\
&& \bar{\lambda}P_R\lambda = (\bar{\lambda}P_R\lambda,+i\sqrt{2}P_R\Lambda,0,2D_+^2,-i\mathcal{D}_{\mu}(\bar{\lambda}P_R\lambda),0,0) = \{\bar{\lambda}P_R\lambda, P_R\Lambda,-D_+^2\},\\
&& (V)_D = (D,\cancel{\mathcal{D}}\lambda,0,0,\mathcal{D}^{b}\hat{F}_{ab},-\cancel{\mathcal{D}}\cancel{\mathcal{D}}\lambda,-\square^CD),
\end{eqnarray}
where
\begin{eqnarray}
&& P_L\Lambda \equiv \sqrt{2}P_L(-\frac{1}{2}\gamma\cdot \hat{F} + iD)\lambda,\qquad P_R\Lambda \equiv \sqrt{2}P_R(-\frac{1}{2}\gamma\cdot \hat{F} - iD)\lambda,\\
&& D_-^2 \equiv D^2 - \hat{F}^-\cdot\hat{F}^- - 2  \bar{\lambda}P_L\cancel{\mathcal{D}}\lambda,\qquad D_+^2 \equiv D^2 - \hat{F}^+\cdot\hat{F}^+ - 2  \bar{\lambda}P_R\cancel{\mathcal{D}}\lambda,\\
&& \mathcal{D}_{\mu}\lambda \equiv \bigg(\partial_{\mu}-\frac{3}{2}b_{\mu}+\frac{1}{4}w_{\mu}^{ab}\gamma_{ab}-\frac{3}{2}i\gamma_*\mathcal{A}_{\mu}\bigg)\lambda - \bigg(\frac{1}{4}\gamma^{ab}\hat{F}_{ab}+\frac{1}{2}i\gamma_* D\bigg)\psi_{\mu}
\\
 && \hat{F}_{ab} \equiv F_{ab} + e_a^{~\mu}e_b^{~\nu} \bar{\psi}_{[\mu}\gamma_{\nu]}\lambda,\qquad F_{ab} \equiv e_a^{~\mu}e_b^{~\nu} (2\partial_{[\mu}A_{\nu]}),\\
 && \hat{F}^{\pm}_{\mu\nu} \equiv \frac{1}{2}(\hat{F}_{\mu\nu}\pm \tilde{\hat{F}}_{\mu\nu}), \qquad \tilde{\hat{F}}_{\mu\nu} \equiv -\frac{1}{2} i\epsilon_{\mu\nu\rho\sigma}\hat{F}^{\rho\sigma} .
\end{eqnarray}

\subsubsection{\texorpdfstring{$w'^2,\bar{w}'^2$}{} Composite Complex Multiplets: (Weyl/Chiral) weights \texorpdfstring{$= (-1,3)$}{} and \texorpdfstring{$(-1,-3)$}{}}

We show here the components of the first superconformal {\it composite} complex multiplets $w'^2$ and $\Bar{w}'^2$ with Weyl/chiral weights $(-1,3)$ and $(-1,-3)$ respectively. These composite multiplets are defined to be
\begin{eqnarray}
&& w'^2 \equiv \frac{\bar{\lambda}P_L\lambda}{(S_0\bar{S}_0e^{-K/3})^2} = \{\mathcal{C}_w,\mathcal{Z}_w,\mathcal{H}_w,\mathcal{K}_w,\mathcal{B}^w_{\mu},\Lambda_w,\mathcal{D}_w\} \\
&& \bar{w}'^2 \equiv \frac{\bar{\lambda}P_R\lambda}{(S_0\bar{S}_0e^{-K/3})^2}
= \{\mathcal{C}_{\bar{w}},\mathcal{Z}_{\bar{w}},\mathcal{H}_{\bar{w}},\mathcal{K}_{\bar{w}},\mathcal{B}^{\bar{w}}_{\mu},\Lambda_{\bar{w}},\mathcal{D}_{\bar{w}}\}.
\end{eqnarray}
where
\begin{eqnarray}
\mathcal{C}_w &=& h \equiv \frac{\bar{\lambda}P_L\lambda}{(s_0\bar{s}_0e^{-K(z,\bar{z})/3})^2},\\
\mathcal{Z}_w &=& i\sqrt{2}(-h_a\Omega^a + h_{\bar{a}}\Omega^{\bar{a}}),\\
\mathcal{H}_w &=& -2h_aF^a + h_{ab}\bar{\Omega}^a\Omega^b,\\ 
\mathcal{K}_w &=& -2h_{\bar{a}}F^{\bar{a}} + h_{\bar{a}\bar{b}}\bar{\Omega}^{\bar{a}}\Omega^{\bar{b}},\\ 
\mathcal{B}^w_{\mu} &=& ih_a\mathcal{D}_{\mu}X^a-ih_{\bar{a}}\mathcal{D}_{\mu}\bar{X}^{\bar{a}}+ih_{a\bar{b}}\bar{\Omega}^{a}\gamma_{\mu}\Omega^{\bar{b}},\\ 
P_L\Lambda_w &=& -\sqrt{2}ih_{\bar{a}b}[(\cancel{\mathcal{D}}X^b)\Omega^{\bar{a}}-F^{\bar{a}}\Omega^b]-\frac{i}{\sqrt{2}}h_{\bar{a}\bar{b}c}\Omega^c\bar{\Omega}^{\bar{a}}\Omega^{\bar{b}},\\
P_R\Lambda_w &=& \sqrt{2}ih_{a\bar{b}}[(\cancel{\mathcal{D}}\bar{X}^{\bar{b}})\Omega^{a}-F^{a}\Omega^{\bar{b}}]+\frac{i}{\sqrt{2}}h_{ab\bar{c}}\Omega^{\bar{c}}\bar{\Omega}^{a}\Omega^{b},\\
\mathcal{D}_w &=& 2h_{a\bar{b}}\Big(-\mathcal{D}_{\mu}X^a\mathcal{D}^{\mu}\bar{X}^{\bar{b}}-\frac{1}{2}\bar{\Omega}^aP_L\cancel{\mathcal{D}}\Omega^{\bar{b}}-\frac{1}{2}\bar{\Omega}^{\bar{b}}P_R\cancel{\mathcal{D}}\Omega^a+F^aF^{\bar{b}}\Big) \nonumber\\
&&+h_{ab\bar{c}}(-\bar{\Omega}^a\Omega^bF^{\bar{c}}+\bar{\Omega}^a(\cancel{\mathcal{D}}X^b)\Omega^{\bar{c}})+ h_{\bar{a}\bar{b}c}(-\bar{\Omega}^{\bar{a}}\Omega^{\bar{b}}F^{c}+\bar{\Omega}^{\bar{a}}(\cancel{\mathcal{D}}\bar{X}^{\bar{b}})\Omega^{c}) \nonumber\\
&&+ \frac{1}{2}h_{ab\bar{c}\bar{d}}(\bar{\Omega}^aP_L\Omega^b)(\bar{\Omega}^{\bar{c}}P_R\Omega^{\bar{d}}).
\end{eqnarray}
Notice that when finding the multiplet $\bar{w}'^2$, we can just replace $h$ by its complex conjugate $h^{*}$.

\subsubsection{\texorpdfstring{$T(\bar{w}'^2),\bar{T}(w'^2)$}{} chiral projection multiplets: (Weyl/Chiral) weights \texorpdfstring{$=(0,0)$}{}}

The second superconformal multiplets that we need are the {\it composite} chiral projection multiplets $T(\Bar{w}'^2)$ and $\Bar{T}(w'^2)$ with Weyl/chiral weights $(0,0)$. From their component supermultiplets defined by 
\begin{eqnarray}
T(\bar{w}'^2) &=& \left( -\frac{1}{2}\mathcal{K}_{\bar{w}}, -\frac{1}{2} \sqrt{2} iP_L (\cancel{\mathcal{D}}\mathcal{Z}_{\bar{w}}+\Lambda_{\bar{w}}), \frac{1}{2}(\mathcal{D}_{\bar{w}}+\square^C \mathcal{C}_{\bar{w}} + i\mathcal{D}_a \mathcal{B}^a_{\bar{w}}) \right),\\
\bar{T}(w'^2) &=& \left( -\frac{1}{2}\mathcal{K}_{\bar{w}}^{*}, \frac{1}{2} \sqrt{2} iP_R (\cancel{\mathcal{D}}\mathcal{Z}_{\bar{w}}^C+\Lambda_{\bar{w}}^C), \frac{1}{2}(\mathcal{D}_{\bar{w}}^{*}+\square^C \mathcal{C}_{\bar{w}}^{*} - i\mathcal{D}_a (\mathcal{B}^a_{\bar{w}})^{*}) \right)
\end{eqnarray}
we find the corresponding superconformal multiplets and their complex conjugates as follows:
\begin{eqnarray}
T \equiv T(\bar{w}'^2) &=& \{\mathcal{C}_T,\mathcal{Z}_T,\mathcal{H}_T,\mathcal{K}_T,\mathcal{B}_{\mu}^T,\Lambda_T,\mathcal{D}_T\} \nonumber\\
\bar{T} \equiv \bar{T}(w'^2) &=& \{\mathcal{C}_{\bar{T}},\mathcal{Z}_{\bar{T}},\mathcal{H}_{\bar{T}},\mathcal{K}_{\bar{T}},\mathcal{B}_{\mu}^{\bar{T}},\Lambda_{\bar{T}},\mathcal{D}_{\bar{T}}\},
\end{eqnarray}
whose superconformal components are given by
\begin{eqnarray}
\mathcal{C}_T &=&  -\frac{1}{2} \mathcal{K}_{\bar{w}} = h^{*}_{\bar{a}}F^{\bar{a}} -\frac{1}{2} h^{*}_{\bar{a}\bar{b}}\bar{\Omega}^{\bar{a}}\Omega^{\bar{b}} \equiv C_T\\
\mathcal{Z}_T &=& -\sqrt{2}iP_L\bigg[\cancel{\mathcal{D}}(-h^{*}_a\Omega^a + h^{*}_{\bar{a}}\Omega^{\bar{a}})-h^{*}_{\bar{a}b}[(\cancel{\mathcal{D}}X^b)\Omega^{\bar{a}}-F^{\bar{a}}\Omega^b]-\frac{1}{2}h^{*}_{\bar{a}\bar{b}c}\Omega^c\bar{\Omega}^{\bar{a}}\Omega^{\bar{b}}\nonumber\\
&&\qquad\qquad+h^{*}_{a\bar{b}}[(\cancel{\mathcal{D}}\bar{X}^{\bar{b}})\Omega^{a}-F^{a}\Omega^{\bar{b}}]+\frac{1}{2}h^{*}_{ab\bar{c}}\Omega^{\bar{c}}\bar{\Omega}^{a}\Omega^{b}\bigg] \equiv -\sqrt{2}iP_L\Omega_T ,\\
\mathcal{H}_T &=& -2\bigg[h^{*}_{a\bar{b}}\Big(-\mathcal{D}_{\mu}X^a\mathcal{D}^{\mu}\bar{X}^{\bar{b}}-\frac{1}{2}\bar{\Omega}^aP_L\cancel{\mathcal{D}}\Omega^{\bar{b}}-\frac{1}{2}\bar{\Omega}^{\bar{b}}P_R\cancel{\mathcal{D}}\Omega^a+F^aF^{\bar{b}}\Big) \nonumber\\
&&+\frac{1}{2}h^{*}_{ab\bar{c}}(-\bar{\Omega}^a\Omega^bF^{\bar{c}}+\bar{\Omega}^a(\cancel{\mathcal{D}}X^b)\Omega^{\bar{c}})+\frac{1}{2} h^{*}_{\bar{a}\bar{b}c}(-\bar{\Omega}^{\bar{a}}\Omega^{\bar{b}}F^{c}+\bar{\Omega}^{\bar{a}}(\cancel{\mathcal{D}}\bar{X}^{\bar{b}})\Omega^{c}) \nonumber\\
&&+ \frac{1}{4}h^{*}_{ab\bar{c}\bar{d}}(\bar{\Omega}^aP_L\Omega^b)(\bar{\Omega}^{\bar{c}}P_R\Omega^{\bar{d}})+\frac{1}{2}\square^C h^{*} + \frac{1}{2}i\mathcal{D}^{\mu} (ih^{*}_a\mathcal{D}_{\mu}X^a-ih^{*}_{\bar{a}}\mathcal{D}_{\mu}\bar{X}^{\bar{a}}+ih^{*}_{a\bar{b}}\bar{\Omega}^{a}\gamma_{\mu}\Omega^{\bar{b}})\bigg] \nonumber\\
&\equiv& -2F_T,\\
\mathcal{K}_T &=& 0,\\
\mathcal{B}^T_{\mu} &=& -i\mathcal{D}_{\mu}\mathcal{C}_T,\\
\Lambda_T &=& 0 ,\\
\mathcal{D}_T &=& 0,
\end{eqnarray}
where we used $a,b,c,d = 0,i(\equiv z^i),W (\equiv \bar{\lambda}P_L\lambda)$. This gives the superfield components of the chiral projection multiplet $T$:
\begin{eqnarray}
T(\bar{w}'^2) = ( C_T, P_L\Omega_T, F_T )
\end{eqnarray}
where 
\begin{eqnarray}
C_T &=&  h^{*}_{\bar{a}}F^{\bar{a}} -\frac{1}{2} h^{*}_{\bar{a}\bar{b}}\bar{\Omega}^{\bar{a}}\Omega^{\bar{b}},\\
P_L\Omega_T &=& \cancel{\mathcal{D}}(-h^{*}_a\Omega^a + h^{*}_{\bar{a}}\Omega^{\bar{a}})-h^{*}_{\bar{a}b}[(\cancel{\mathcal{D}}X^b)\Omega^{\bar{a}}-F^{\bar{a}}\Omega^b]-\frac{1}{2}h^{*}_{\bar{a}\bar{b}c}\Omega^c\bar{\Omega}^{\bar{a}}\Omega^{\bar{b}}\nonumber\\
&&+h^{*}_{a\bar{b}}[(\cancel{\mathcal{D}}\bar{X}^{\bar{b}})\Omega^{a}-F^{a}\Omega^{\bar{b}}]+\frac{1}{2}h^{*}_{ab\bar{c}}\Omega^{\bar{c}}\bar{\Omega}^{a}\Omega^{b},\\
F_T &=&  h^{*}_{a\bar{b}}\Big(-\mathcal{D}_{\mu}X^a\mathcal{D}^{\mu}\bar{X}^{\bar{b}}-\frac{1}{2}\bar{\Omega}^aP_L\cancel{\mathcal{D}}\Omega^{\bar{b}}-\frac{1}{2}\bar{\Omega}^{\bar{b}}P_R\cancel{\mathcal{D}}\Omega^a+F^aF^{\bar{b}}\Big) \nonumber\\
&&+\frac{1}{2}h^{*}_{ab\bar{c}}(-\bar{\Omega}^a\Omega^bF^{\bar{c}}+\bar{\Omega}^a(\cancel{\mathcal{D}}X^b)\Omega^{\bar{c}})+\frac{1}{2} h^{*}_{\bar{a}\bar{b}c}(-\bar{\Omega}^{\bar{a}}\Omega^{\bar{b}}F^{c}+\bar{\Omega}^{\bar{a}}(\cancel{\mathcal{D}}\bar{X}^{\bar{b}})\Omega^{c}) \nonumber\\
&&+ \frac{1}{4}h^{*}_{ab\bar{c}\bar{d}}(\bar{\Omega}^aP_L\Omega^b)(\bar{\Omega}^{\bar{c}}P_R\Omega^{\bar{d}})+\frac{1}{2}\square^C h^{*} - \frac{1}{2}\mathcal{D}^{\mu} (h^{*}_a\mathcal{D}_{\mu}X^a-h^{*}_{\bar{a}}\mathcal{D}_{\mu}\bar{X}^{\bar{a}}+h^{*}_{a\bar{b}}\bar{\Omega}^{a}\gamma_{\mu}\Omega^{\bar{b}}). \nonumber\\{}
\end{eqnarray}

Morever,
\begin{eqnarray}
\bar{T}(w'^2) = \{ C_T^{*}, P_R\Omega_T, F_T^{*}\}
\end{eqnarray}
where
\begin{eqnarray}
C_T^{*} &=&  h_{a}F^{a} -\frac{1}{2} h_{ab}\bar{\Omega}^{a}\Omega^{b},\\
P_R\Omega_T &=& \cancel{\mathcal{D}}(-h_a\Omega^a + h_{\bar{a}}\Omega^{\bar{a}})-h_{\bar{a}b}[(\cancel{\mathcal{D}}X^b)\Omega^{\bar{a}}-F^{\bar{a}}\Omega^b]-\frac{1}{2}h_{\bar{a}\bar{b}c}\Omega^c\bar{\Omega}^{\bar{a}}\Omega^{\bar{b}}\nonumber\\
&&+h_{a\bar{b}}[(\cancel{\mathcal{D}}\bar{X}^{\bar{b}})\Omega^{a}-F^{a}\Omega^{\bar{b}}]+\frac{1}{2}h_{ab\bar{c}}\Omega^{\bar{c}}\bar{\Omega}^{a}\Omega^{b},\\
F_T^{*} &=&  h_{a\bar{b}}\Big(-\mathcal{D}_{\mu}X^a\mathcal{D}^{\mu}\bar{X}^{\bar{b}}-\frac{1}{2}\bar{\Omega}^aP_L\cancel{\mathcal{D}}\Omega^{\bar{b}}-\frac{1}{2}\bar{\Omega}^{\bar{b}}P_R\cancel{\mathcal{D}}\Omega^a+F^aF^{\bar{b}}\Big) \nonumber\\
&&+\frac{1}{2}h_{ab\bar{c}}(-\bar{\Omega}^a\Omega^bF^{\bar{c}}+\bar{\Omega}^a(\cancel{\mathcal{D}}X^b)\Omega^{\bar{c}})+\frac{1}{2} h_{\bar{a}\bar{b}c}(-\bar{\Omega}^{\bar{a}}\Omega^{\bar{b}}F^{c}+\bar{\Omega}^{\bar{a}}(\cancel{\mathcal{D}}\bar{X}^{\bar{b}})\Omega^{c}) \nonumber\\
&&+ \frac{1}{4}h_{ab\bar{c}\bar{d}}(\bar{\Omega}^aP_L\Omega^b)(\bar{\Omega}^{\bar{c}}P_R\Omega^{\bar{d}})+\frac{1}{2}\square^C h - \frac{1}{2}\mathcal{D}^{\mu} (h_a\mathcal{D}_{\mu}X^a-h_{\bar{a}}\mathcal{D}_{\mu}\bar{X}^{\bar{a}}+h_{a\bar{b}}\bar{\Omega}^{\bar{b}}\gamma_{\mu}\Omega^{a}). \nonumber\\{}
\end{eqnarray}

\subsubsection{Composite real multiplet \texorpdfstring{$\mathcal{R}$}{}: (Weyl/Chiral) weights \texorpdfstring{$=(2,0)$}{}}

We present here a superconformal composite real multiplet $\mathcal{R}$ with Weyl/chiral weights $(2,0)$. Defining some chiral multiplets $\mathcal{X}^A \equiv \{X^A,P_L\Omega^A,F^A\}$ where $A=\{ S_0,Z^i,\bar{\lambda}P_L\lambda,T(\bar{w}'^2)$\} and their conjugates, we represent the composite one $\mathcal{R}$ as
\begin{eqnarray}
\mathcal{R} \equiv (S_0\bar{S}_0e^{-K/3})^{-3} \frac{(\bar{\lambda}P_L\lambda)(\bar{\lambda}P_R\lambda)}{T(\bar{w}'^2) \bar{T}(w'^2)} 
\end{eqnarray}
whose lowest component is 
\begin{eqnarray}
\mathcal{C}_{\mathcal{R}}  \equiv (s_0\bar{s}_0e^{-K/3})^{-3}\frac{(\bar{\lambda}P_L\lambda)(\bar{\lambda}P_R\lambda)}{C_TC_{\bar{T}}}\equiv f(X^A,\bar{X}^{\bar{A}}) \label{def-f}
\end{eqnarray}
where $C_T = -D_+^2 \Delta^{-2}$; $C_{\bar{T}} = -D_-^2\Delta^{-2}$, and $\Delta \equiv s_0\bar{s}_0e^{-K/3}$. Then, the superconformal multiplet of the new Fayet-Iliopoulos term can be written by using
\begin{eqnarray}
\mathcal{R}\cdot (V)_D = \{\tilde{\mathcal{C}},\tilde{\mathcal{Z}},\tilde{\mathcal{H}},\tilde{\mathcal{K}},\tilde{\mathcal{B}}_{\mu},\tilde{\Lambda},\tilde{\mathcal{D}}\},
\end{eqnarray}
whose superconformal multiplet components are as follows:
\begin{eqnarray}
\tilde{\mathcal{C}} &=& Df,\\
\tilde{\mathcal{Z}} &=& f\cancel{\mathcal{D}}\lambda+Di\sqrt{2}(-f_{A}\Omega^A+f_{\bar{A}}\Omega^{\bar{A}}),\\
\tilde{\mathcal{H}} &=& D(-2f_AF^A + f_{AB}\bar{\Omega}^A\Omega^B)-i\sqrt{2}(-f_{A}\bar{\Omega}^A+f_{\bar{A}}\bar{\Omega}^{\bar{A}})P_L\cancel{\mathcal{D}}\lambda,\\
\tilde{\mathcal{K}} &=& D(-2f_{\bar{A}}F^{\bar{A}} + f_{\bar{A}\bar{B}}\bar{\Omega}^{\bar{A}}\Omega^{\bar{B}})-i\sqrt{2}(-f_{A}\bar{\Omega}^A+f_{\bar{A}}\bar{\Omega}^{\bar{A}})P_R\cancel{\mathcal{D}}\lambda,\\
\tilde{\mathcal{B}} &=& (\mathcal{D}^{\nu}\hat{F}_{\mu\nu})f+D(if_A\mathcal{D}_{\mu}X^A-if_{\bar{A}}\mathcal{D}_{\mu}\bar{X}^{\bar{A}}+if_{A\bar{B}}\bar{\Omega}^A\gamma_{\mu}\Omega^{\bar{B}}),\\
\tilde{\Lambda} &=& -f\cancel{\mathcal{D}}\cancel{\mathcal{D}}\lambda + D(P_L\Lambda^f+P_R\Lambda^f)+\frac{1}{2}\Big(\gamma_*(-f_A\cancel{\mathcal{D}}X^A+f_{\bar{A}}\cancel{\mathcal{D}}\bar{X}^{\bar{A}}-f_{A\bar{B}}\bar{\Omega}^A\cancel{\gamma}\Omega^{\bar{B}})\nonumber\\
&&+P_L(-2f_{\bar{A}}F^{\bar{A}} + f_{\bar{A}\bar{B}}\bar{\Omega}^{\bar{A}}\Omega^{\bar{B}})+P_R(-2f_AF^A + f_{AB}\bar{\Omega}^A\Omega^B) -\cancel{\mathcal{D}}f\Big)\cancel{\mathcal{D}}\lambda\nonumber\\
&& +\frac{1}{2}\Big( i\gamma_*\gamma^{\mu}\mathcal{D}^{\nu}\hat{F}_{\mu\nu}   -\cancel{\mathcal{D}}D\Big)i\sqrt{2}(-f_{A}\Omega^A+f_{\bar{A}}\Omega^{\bar{A}}) ,\\
\tilde{\mathcal{D}} &=&-f\square^C D + D 
\bigg\{ 2f_{A\bar{B}}(-\mathcal{D}_{\mu}X^A\mathcal{D}^{\mu}\bar{X}^{\bar{B}}-\frac{1}{2}\bar{\Omega}^AP_L\cancel{\mathcal{D}}\Omega^{\bar{B}}-\frac{1}{2}\cancel{\Omega}^{\bar{B}}P_R\cancel{\mathcal{D}}\Omega^A+F^AF^{\bar{B}}) \nonumber\\
&&+f_{AB\bar{C}}(-\bar{\Omega}^A\Omega^B F^{\bar{C}} + \bar{\Omega}^A(\cancel{\mathcal{D}}X^B)\Omega^{\bar{C}}) 
+f_{\bar{A}\bar{B}C}(-\bar{\Omega}^{\bar{A}}\Omega^{\bar{B}} F^C + \bar{\Omega}^{\bar{A}}(\cancel{\mathcal{D}}\bar{X}^{\bar{B}})\Omega^C) \nonumber\\
&&+\frac{1}{2}f_{AB\bar{C}\bar{D}} (\bar{\Omega}^AP_L\Omega^B)(\bar{\Omega}^{\bar{C}}P_R\Omega^{\bar{D}}) \bigg\}\nonumber\\
&& -(\mathcal{D}_{\nu}\hat{F}^{\mu\nu})(if_A\mathcal{D}_{\mu}X^A-if_{\bar{A}}\mathcal{D}_{\mu}\bar{X}^{\bar{A}}+if_{A\bar{B}}\bar{\Omega}^A\gamma_{\mu}\Omega^{\bar{B}}) \nonumber\\
&& + \bigg( \sqrt{2}if_{\bar{A}B}[(\cancel{\mathcal{D}}X^B)\Omega^{\bar{A}}-F^{\bar{A}}\Omega^B]  +\frac{i}{\sqrt{2}}f_{\bar{A}\bar{B}C} \Omega^C\bar{\Omega}^{\bar{A}}\Omega^{\bar{B}} \bigg)\cancel{\mathcal{D}}\lambda\nonumber\\
&& - \bigg( \sqrt{2}if_{A\bar{B}}[(\cancel{\mathcal{D}}\bar{X}^{\bar{B}})\Omega^{A}-F^{A}\Omega^{\bar{B}}]  +\frac{i}{\sqrt{2}}f_{AB\bar{C}} \Omega^{\bar{C}}\bar{\Omega}^{A}\Omega^{B} \bigg)\cancel{\mathcal{D}}\lambda\nonumber\\
&&-(\mathcal{D}_{\mu}f)(\mathcal{D}^{\mu}D)-\frac{1}{2}\cancel{\mathcal{D}}[i\sqrt{2}(-f_{A}\Omega^A+f_{\bar{A}}\Omega^{\bar{A}})](\cancel{\mathcal{D}}\lambda)+\frac{1}{2}i\sqrt{2}(-f_{A}\Omega^A+f_{\bar{A}}\Omega^{\bar{A}})(\cancel{\mathcal{D}}\cancel{\mathcal{D}}\lambda),
\end{eqnarray}
where the indices $A,B,C,D$ run over $0,i,W,T$. The component action of the new FI term is then given by the D-term density formula 
\begin{eqnarray}
\mathcal{L}_{NEW} \equiv -[\xi \mathcal{R}\cdot (V)_D]_D &=& -\frac{\xi}{4}\int d^4x e \bigg[  \tilde{\mathcal{D}} -\frac{1}{2}\bar{\psi}\cdot \gamma i\gamma_* \tilde{\Lambda} -\frac{1}{3}\tilde{\mathcal{C}}R(\omega)\nonumber\\
&&+\frac{1}{6}\Big(\tilde{\mathcal{C}}\bar{\psi}_{\mu}\gamma^{\mu\rho\sigma}-i\bar{\tilde{\mathcal{Z}}}\gamma^{\rho\sigma}\gamma_*\Big)R'_{\rho\sigma}(Q)\nonumber\\
&&+\frac{1}{4}\varepsilon^{abcd}\bar{\psi}_{a}\gamma_b\psi_c\Big(\tilde{\mathcal{B}}_{d}-\frac{1}{2}\bar{\psi}_d\tilde{\mathcal{Z}}\Big)\bigg]+\textrm{h.c.}.
\end{eqnarray}

\subsubsection{Bosonic term of the new FI term}

The new FI term is obtained from the term $D^2f_{W\bar{W}}F^WF^{\bar{W}}$ inside the D-term $\tilde{\mathcal{D}}$. Thus, we get
\begin{eqnarray}
 \mathcal{L}_{NEW} &\supset& -\xi D f_{W\bar{W}}F^WF^{\bar{W}} = -\xi D \frac{ (s_0\bar{s}_0e^{-K/3})^{-3}}{C_TC_T^{*}}F^WF^{\bar{W}}
 \supset
-\xi D\frac{ (s_0\bar{s}_0e^{-K/3})^{-3}}{(h_WF^W)(h^{*}_{\bar{W}}F^{\bar{W}})}F^WF^{\bar{W}}
\nonumber\\&=& - \xi D\frac{ (s_0\bar{s}_0e^{-K/3})^{-3}}{(s_0\bar{s}_0e^{-K/3})^{-4}F^{W}F^{\bar{W}}}F^WF^{\bar{W}} = -\xi D(s_0\bar{s}_0e^{-K/3}).
\end{eqnarray}
Hence, in the superconformal gauge ($s_0\bar{s}_0e^{-K/3}=M_{pl}^2=1$), we obtain 
\begin{eqnarray}
 \mathcal{L}_{\textrm{new FI}}/e = -\xi D,
\end{eqnarray}
or
\begin{eqnarray}
 \mathcal{L}_{\textrm{new FI}}/e = -M_{pl}^2\xi D.
\end{eqnarray}

\section{Consistency check of supergravity as effective field theory through superconformal tensor calculus}

This section studies the fermionic nonrenormalizable terms in the Lagrangians containing either liberated $\mathcal{N}=1$ terms 
or
new FI terms. We will make full use of the superconformal tensor calculus to find the limits of validity of those effective
field theories.

\subsection{Constraining liberated \texorpdfstring{$\mathcal{N}=1$}{} supergravity}

We investigate first the nonrenormalizable fermionic terms in ``liberated'' supergravity, to find possible constraints on the liberated term $\mathcal{U}$. To begin with, we rewrite  the coefficients 
\begin{eqnarray}
N^{(r=q+p+m+k)}_{q,p,m,k}
&=& (\partial_0^q\partial_I^{(k-n)}\Upsilon^{-2})\Upsilon^{4+2m}\frac{\mathcal{U}^{(n)}}{\tilde{\mathcal{F}}^{4+2m}}W^{1-p_1}\bar{W}^{1-p_2},\nonumber
\end{eqnarray}
 as
\begin{eqnarray}
N^{(r=q+p+m+k)}_{q,p,m,k}
&=& ((-1)^{q_1+q_2}(q_1+1)!(q_2+1)!s_0^{-(2+q_1)}(s_0^{*})^{-(2+q_2)}(\partial_{I}^{(k-n)}e^{2K/3}))\nonumber\\
&&\times \Upsilon^{4+2m}\frac{\mathcal{U}^{(n)}}{\Tilde{\mathcal{F}}^{4+2m}}W^{1-p_1}\Bar{W}^{1-p_2}
\end{eqnarray}
where we used $\Upsilon = s_0s_0^{*}e^{-K/3}$. 

The most powerful constraints on $\mathcal{U}$ come from the most singular fermionic terms in the 
Lagrangian. Such singular terms are those containing the highest power of $\tilde{\mathcal{F}}$ in the denominator. These
powers are linearly proportional to $m$ because they are due to taking derivatives of $N$ with respect to the lowest component of the
 multiplet $T(\bar{w}^2)$. Therefore, we must investigate the fermionic interactions containing only derivatives with respect to the chiral projection
 and matter scalar indices, i.e. $T$ and $I$, to obtain the terms that are coupled to $\mathcal{U}^{(n)}$ and contain the 
 maximal inverse powers of $\tilde{\mathcal{F}}$. Such terms are those with $q=p=0$ and $k=n$.

In particular, we find by direct inspection of the fermionic interactions that the most singular terms are given by the couplings to the derivatives proportional to $N_{T\bar{T}}$, $N_{WT\bar{T}}$, $N_{W\bar{W}T\bar{T}}$  if our theory has a single chiral matter multiplet, while they are given by $N_{TT\bar{T}\bar{T}}$ for two or more chiral matter multiplets. The latter terms vanish identically for a single multiplet because of Fermi statistics. This has been investigated in detail in Ref. \cite{jp1}. Here we merely recall the results from Ref. \cite{jp1}: 
\begin{itemize}
    \item For the  single chiral matter multiplet case, 
    \begin{eqnarray}
     \mathcal{L}_{F2}|_{q=p=0,k=n} \supset  M_{pl}^{2(n-4)} \frac{\mathcal{U}^{(n)}}{\mathcal{F}^{2(4-n)}} \mathcal{O}_F'^{(2(6-n))}.\label{SingleChiralMatterCase_Vac}
    \end{eqnarray}
    \item For two or more chiral matter multiplets, 
    \begin{eqnarray}
     \mathcal{L}_{F6}|_{q=p=0,k=n} \supset c'M_{pl}^{2(n-6)}\frac{\mathcal{U}^{(n)}}{\mathcal{F}^{2(6-n)}} \mathcal{O}_F'^{(2(8-n))}.\label{GeneralChiralMatterCases_Vac}
    \end{eqnarray}
    where $ \mathcal{O}(10^{-2}) \leq c' \leq  \mathcal{O}(1)$.
\end{itemize} 
From this, by calling $\Lambda_{cut}$ the cutoff UV scale of liberated supergravity and using the inequality $\mathcal{C}^{(4-\delta)} \lesssim \frac{1}{\Lambda_{cut}^{\delta-4}}$ ($\delta = 2(6-n)$ for $N_{mat}=1$ and $\delta = 2(8-n)$ for $N_{mat}\geq 2$) we
obtain
\begin{eqnarray}
 \mathcal{U}^{(n)} \lesssim \begin{cases}
  \mathcal{F}^{2(4-n)} \left(\dfrac{M_{pl}}{\Lambda_{cut}}\right)^{2(4-n)}  \quad \textrm{where}\quad0\leq n \leq 2\quad\textrm{for}~N_{mat} =1,\\
\mathcal{F}^{2(6-n)}\left(\dfrac{M_{pl}}{\Lambda_{cut}}\right)^{2(6-n)}  \quad \textrm{where}\quad0\leq n \leq 4\quad\textrm{for}~N_{mat} \geq 2.
  \end{cases}\label{pre_inequality}
\end{eqnarray}

A conventional definition of the supersymmetry breaking scale $M_S$ is given in terms of F-term expectation value; therefore,
 we can define 
$M_S^4=M_{pl}^4\mathcal{F}$ and the constraints on $\mathcal{U}^{(n)}$ then become 
\begin{eqnarray}
 \mathcal{U}^{(n)} \lesssim \begin{cases}
  \left(\dfrac{M_S}{M_{pl}}\right)^{8(4-n)}\left(\dfrac{M_{pl}}{\Lambda_{cut}}\right)^{2(4-n)} \quad \textrm{where}\quad0\leq n \leq 2\quad\textrm{for}~N_{mat} =1,\\
\left(\dfrac{M_S}{M_{pl}}\right)^{8(6-n)}\left(\dfrac{M_{pl}}{\Lambda_{cut}}\right)^{2(6-n)} \quad\textrm{where}\quad0\leq n \leq 4\quad\textrm{for}~N_{mat} \geq 2.
  \end{cases}\label{generic_constraints}
\end{eqnarray}
The consequences of these constraints were studied in Ref.~\cite{jp1}.

\subsection{Constraining new Fayet-Iliopoulos terms}

In this section\footnote{This subsection also complements the detailed derivation of the constraint on the new FI parameter $\xi$ omitted in Ref. \cite{jp2}.}, we explore the possible constraints on the new  FI parameter $\xi$. First of all, let us find the fermionic terms coming from the new FI terms. From its Lagrangian, we observe that the most singular terms come from the derivatives of the function $f$, defined in Eq.~\eqref{def-f} and below, with respect to the multiplet $T$ since they produce the inverse powers of the D-term `$D$'. The function $f$ was defined by 
\begin{eqnarray}
 f = (s_0\bar{s}_0e^{-K/3})^{-3}\frac{(\bar{\lambda}P_L\lambda)(\bar{\lambda}P_R\lambda)}{C_TC_{\bar{T}}},
\end{eqnarray}
where $C_T = -D_+^2 \Delta^{-2}$; $C_{\bar{T}} = -D_-^2\Delta^{-2}$, and $\Delta \equiv s_0\bar{s}_0e^{-K/3}$.

Then, the most singular terms in the Lagrangians has the generic form
\begin{eqnarray}
 \mathcal{L}_F \supset -c(\xi D^p \cdot \partial_T^{m_1} \partial_{\bar{T}}^{m_2} f) \mathcal{O}^{(\delta)}_F,
\end{eqnarray}
where $c$ is a dimensionless constant of order $\mathcal{O}(1)$; $\xi$ is the dimensionless new FI constant, $\mathcal{O}^{(\delta)}_F$ is an effective field operator of dimension $\delta$ independent of $D$,  $m_1,m_2$ are the order of derivative with respect to the $T,\bar{T}$, and $p$ is a power of $D$ where $p=0,1$.   

The most singular terms appear in the Lagrangian as follows:
\begin{eqnarray}
 \mathcal{L}_F &\supset& -(m_1!m_2!c) \xi M_{pl}^{4m+2}D^{-2m-4+p} (\bar{\lambda}P_L\lambda)(\bar{\lambda}P_R\lambda) \mathcal{O}_F^{(\delta)} 
 \nonumber\\
 &=& -(m_1!m_2!c) \xi M_{pl}^{4m+2}D^{-2m-4+p} \mathcal{O}_F^{(\delta+6)},
\end{eqnarray}
where $\mathcal{O}_F^{(\delta+6)} \equiv (\bar{\lambda}P_L\lambda)(\bar{\lambda}P_R\lambda) \mathcal{O}_F^{(\delta)}$ and $\delta = 4-2p$ so that $[\mathcal{L}_f]=4$.
Therefore, we have 
\begin{eqnarray}
 \mathcal{L}_F \supset \xi M_{pl}^{4m+2}D^{-2m-4+p} \mathcal{O}_F^{(10-2p)} ,
\end{eqnarray}
where we omitted $-(m_1!m_2!c)$ since this is $ \mathcal{O}(1)$.

Next, we explore the constraints coming from the new FI term. First of all, we recall that the new FI term proposed by Antoniadis et al. can be added to the standard FI term, when the theory contains also a gauged U(1) R-symmetry  $U(1)_R$. Then, we have the following auxiliary D-term Lagrangian
\begin{eqnarray}
\mathcal{L}_{\textrm{aux D}}= \frac{1}{2}D^2 -i(G_ik^i-G_{\bar{i}}k^{\bar{i}})D - \xi D \equiv \frac{1}{2}D^2-(\xi'+\xi)D,
\end{eqnarray}
where $\xi$ is the new FI constant,  $\xi' \equiv i(G_ik^i-G_{\bar{i}}k^{\bar{i}})$. Here $k=k^i\partial_i$ is the Killing vector
of a standard R-symmetry $U_R(1)$  and $G$ is the supergravity G-function. 
If we restore explicit powers of the gauge coupling $g$, then we get
\begin{eqnarray}
\mathcal{L}_{\textrm{aux D}}= \frac{1}{2g^2}D^2 -\xi'D - \xi  D = \frac{1}{2g^2}D^2-(\xi'+\xi)D .
\end{eqnarray}

After solving the equation of motion for $D$, we have
\begin{eqnarray}
 D = g^2M_{pl}^2(\xi+\xi') .
\end{eqnarray}
Plugging this result into the Lagrangian again, we obtain
\begin{eqnarray}
  \mathcal{L}_F &\supset& \xi M_{pl}^{4m+2}(g^2M_{pl}^2(\xi+\xi'))^{-2m-4+p} \mathcal{O}_F'^{(10-2p)} \nonumber\\
  &=& \xi(g^2(\xi+\xi'))^{-2m-4+p} M_{pl}^{-6+2p} \mathcal{O}_F'^{(10-2p)}.
\end{eqnarray}
This implies $\xi(g^2(\xi+\xi'))^{-2m-4+p} M_{pl}^{-6+2p}\lesssim \frac{1}{\Lambda_{cut}^{6-2p}}$ so we arrive at
\begin{eqnarray}
 \xi(g^2(\xi+\xi'))^{-2m-4+p} \lesssim \left(\frac{M_{pl}}{\Lambda_{cut}}\right)^{6-2p}.
\end{eqnarray}
This is a constraint on the new FI term constant $\xi$, written in terms of the UV cutoff $\Lambda_{cut}$.

\section{Summary}

In Sec. 2 of this work we computed the component action of the liberated $\mathcal{N}=1$ supergravity using the superconformal tensor calculus. In Sec. 3, we revisited the component action of a K\"{a}hler-invariant new FI term (called ``ACIK-FI term'') in the superconformal formalism. In Sec. 4, we used the superconformal tensor calculus to constrain the size of new supergravity 
terms that are present in liberated supergravity and in the ACIK-FI term, but absent in standard supergravity. The computations performed in Secs. 2, 3, and 4 spell out the results of~\cite{jp1,jp2}. specifically, we relate the ``liberated'' scalar potential 
$\mathcal{U}$ to the UV cutoff and we derive the constraints on the ACIK-FI term used in~\cite{jp2}. 
What makes our constraints powerful is that differently from standard supergravity, both liberated supergravity and the ACIK-FI terms introduce nonrenormalizable interactions proportional to inverse powers of the supersymmetry breaking scale $M_S$. This makes it impossible to send $M_S$
to zero while keeping the UV cutoff of the theory finite. The most singular nonrenormalizable interactions in the limit $M_S\rightarrow 0$ are cumbersome, multi-fermion operators, but they can be found and studies using the superconformal tensor calculus in a systematic and economical way. Superconformal calculus techniques are general so can be applied to any supergravity theory. Our analysis is based on standard 
Effective Field Theory methods and it is thus complementary to swampland~\cite{swampland} constraints.

\appendix
\section{Some detailed calculations in liberated supergravity}
\subsection{Some details}
\begin{eqnarray}
\mathcal{D}_{\mu} z^I &=& \partial_{\mu}z^I -\frac{1}{\sqrt{2}}\bar{\psi}_{\mu} \chi^I,\\
\mathcal{C}_{\bar{W}}  &\equiv& \bar{W} = \Bar{\Lambda}_KP_R\Lambda_K =  2~\textrm{fermions} + 4~\textrm{fermions} + 6~\textrm{fermions} ,\\
P_L \Lambda_K &=& -\sqrt{2}iK_{\bar{I}J}[(\cancel{\mathcal{D}}z^{J})\chi^{\bar{I}}-\bar{F}^{\bar{I}}\chi^J] -\frac{i}{\sqrt{2}}K_{\bar{I}\bar{J}K}\chi^{K}\bar{\chi}^{\bar{I}}\chi^J,\\
(\cancel{\mathcal{D}}\chi^{\bar{W}})_{1f} &=& 2\sqrt{2}i(\cancel{\mathcal{D}}\tilde{\mathcal{F}})_{0f}(P_L\Lambda_K)_{1f} + 2\sqrt{2}i \tilde{\mathcal{F}}  (P_L\cancel{\mathcal{D}}\Lambda_K)_{1f},\\
 \cancel{\mathcal{D}}\Lambda_K &=& \gamma \cdot \mathcal{D} \Lambda_K,\\
 \mathcal{D}_{\mu} \Lambda_K &=& \Big(\partial_{\mu} -\frac{3}{2}b_{\mu} + \frac{1}{4}\omega_{\mu}^{ab} \gamma_{ab} -\frac{3}{2}i\gamma_* \mathcal{A}_{\mu} \Big) \Lambda_K - \Big(\frac{1}{4}\gamma \cdot  \hat{F}^K + \frac{1}{2}i\gamma_* \mathcal{D}_K \Big)\psi_{\mu},\\
 (P_L\cancel{\mathcal{D}}\Lambda_K)_{1f} &=& \gamma^{\mu}\mathcal{D}_{\mu}|_{\psi=0} (P_R\Lambda_K)_{1f} + \frac{i}{2} \tilde{\mathcal{F}} \gamma^{\mu}\psi_{\mu},\\
 (\gamma \cdot \hat{F}^K)_{0f} &=& 0,\qquad \mathcal{D}_K|_{0f} = \tilde{\mathcal{F}}\nonumber\\
 \end{eqnarray}
 \begin{eqnarray}
 (\hat{F}^K_{\mu\nu})_{0f} &=& (2\partial_{[\mu} \mathcal{B}_{\nu]}^K)_{0f} = 2i\partial_{[\mu}(K_I\partial_{\nu]} z^I -K_{\bar{I}}\partial_{\nu]} \bar{z}^{\bar{I}}) = 2i(\partial_{[\mu}K_I\partial_{\nu]} z^I -\partial_{[\mu}K_{\bar{I}}\partial_{\nu]} \bar{z}^{\bar{I}}) \nonumber\\
 &=& 2i(K_{IJ}\partial_{[\mu}z^{(J}\partial_{\nu]} z^{I)} -K_{\bar{I}\bar{J}}\partial_{[\mu}z^{(\bar{J}}\partial_{\nu]} \bar{z}^{\bar{I})}) = 0,\\
 (\cancel{\mathcal{D}}\Upsilon)_{0f} &=&  (\cancel{\partial} -2\gamma^{\mu}b_{\mu} -2i\gamma^{\mu}A_{\mu})\Upsilon \nonumber\\
 &=& (\cancel{\partial}s_0)s_0^{*}e^{-K/3} + s_0(\cancel{\partial}s_0^{*})e^{-K/3} + s_0s_0^{*}e^{-K/3}(-\frac{1}{3}\cancel{\partial}K) -2\gamma^{\mu}(b_{\mu}+iA_{\mu})\Upsilon \nonumber\\
 &=&  \Upsilon(\frac{1}{s_0}\cancel{\partial} s_0 + \frac{1}{s_0^{*}} \cancel{\partial} s_0^{*} - \frac{1}{3}K_{I}\cancel{\partial}z^I-\frac{1}{3}K_{\bar{I}}\cancel{\partial}\bar{z}^{\bar{I}}-2\gamma^{\mu}(b_{\mu}+iA_{\mu})) 
 ,\\
 i(\cancel{\mathcal{B}}_{\Upsilon})_{0f} &=&  \Upsilon(-\frac{1}{s_0}\cancel{\partial} s_0 + \frac{1}{s_0^{*}} \cancel{\partial} s_0^{*} + \frac{1}{3}K_{I}\cancel{\partial}z^I-\frac{1}{3}K_{\bar{I}}\cancel{\partial}\bar{z}^{\bar{I}}),\nonumber\\
\frac{(\cancel{\mathcal{D}}\Upsilon)_{0f}+i(\cancel{\mathcal{B}}_{\Upsilon})_{0f} }{\Upsilon}  &=& \frac{2}{s_0^{*}}\cancel{\partial}s_0^{*} -\frac{2}{3}K_{\bar{I}}\cancel{\partial}\bar{z}^{\bar{I}} - 2\gamma^{\mu}(b_{\mu}+iA_{\mu}).
\end{eqnarray}

\subsection{Derivatives of \texorpdfstring{$N$}{}}
In this appendix we give explicit formulas for the derivatives of $N$ with respect to the multiplets $i = 0,I,W,T$.
The first derivative of $N$ is
\begin{eqnarray}
N_{i} &=& \delta^{0}_{i}\partial_{0}+\delta^{I}_{i}\partial_{I}+\delta^{W}_{i}\partial_{W}+ \delta^{T}_{i}\partial_{T}, \quad N_{\bar{i}} =\delta^{\bar{0}}_{\bar{i}}\partial_{\bar{0}}+\delta^{\bar{I}}_{\bar{i}}\partial_{\bar{I}}+\delta^{\bar{W}}_{\bar{i}}\partial_{\bar{W}}+ \delta^{\bar{T}}_{\bar{i}}\partial_{\bar{T}},
\end{eqnarray}
the second derivative of $N$ is
\begin{eqnarray}
N_{i\bar{j}} &=& \delta^{T}_{i}\delta^{\Bar{0}}_{\bar{j}}\partial_{\bar{0}} \partial_{T}+\delta^{W}_{i}\delta^{\Bar{0}}_{\bar{j}}\partial_{\bar{0}} \partial_{W}
+\delta^{0}_{i}\delta^{\Bar{0}}_{\bar{j}}\partial_{0} \partial_{\bar{0}}+\delta^{I}_{i}\delta^{\Bar{0}}_{\bar{j}}\partial_{I} \partial_{\bar{0}}
+\delta^{T}_{i}\delta^{\Bar{J}}_{\bar{j}}\partial_{\bar{J}} \partial_{T}+\delta^{W}_{i}\delta^{\Bar{J}}_{\bar{j}}\partial_{\bar{J}} \partial_{W}
\nonumber\\
&&+\delta^{0}_{i}\delta^{\Bar{J}}_{\bar{j}}\partial_{0} \partial_{\bar{J}}+\delta^{I}_{i}\delta^{\Bar{J}}_{\bar{j}}\partial_{I} \partial_{\bar{J}}+\delta^{T}_{i}\delta^{\Bar{T}}_{\bar{j}}\partial_{\bar{T}} \partial_{T}+\delta^{W}_{i}\delta^{\Bar{T}}_{\bar{j}}\partial_{\bar{T}} \partial_{W}+\delta^{0}_{i}\delta^{\Bar{T}}_{\bar{j}}\partial_{0} \partial_{\bar{T}}+\delta^{I}_{i}\delta^{\Bar{T}}_{\bar{j}}\partial_{I} \partial_{\bar{T}}\nonumber\\
&&+\delta^{T}_{i}\delta^{\Bar{W}}_{\bar{j}}\partial_{\bar{W}} \partial_{T}+\delta^{W}_{i}\delta^{\bar{W}}_{\bar{j}}\partial_{\bar{W}} \partial_{W}+\delta^{0}_{i}\delta^{\Bar{W}}_{\bar{j}}\partial_{0} \partial_{\bar{W}}+\delta^{I}_{i}\delta^{\Bar{W}}_{\bar{j}}\partial_{I} \partial_{\bar{W}},
\end{eqnarray}
the third derivative of $N$ is
\begin{eqnarray}
N_{ij\bar{k}} &=& \delta^{T}_{i}\delta^{J}_{j}\delta^{\Bar{0}}_{\Bar{k}} \partial_{\bar{0}} \partial_{J} \partial_{T}+\delta^{W}_{i}\delta^{J}_{j}\delta^{\Bar{0}}_{\Bar{k}}\partial_{\bar{0}} \partial_{J} \partial_{W}+\delta^{0}_{i}\delta^{J}_{j}\delta^{\Bar{0}}_{\Bar{k}}\partial_{0} \partial_{\bar{0}} \partial_{J}+\delta^{I}_{i}\delta^{J}_{j}\delta^{\Bar{0}}_{\Bar{k}}\partial_{I} \partial_{\bar{0}} \partial_{J}\nonumber\\
&&+2\delta^{(W}_{i}\delta^{T)}_{j}\delta^{\Bar{0}}_{\Bar{k}} \partial_{\bar{0}} \partial_{T} \partial_{W}
+2 \delta^{(0}_{i}\delta^{T)}_{j}\delta^{\Bar{0}}_{\Bar{k}}\partial_{0} \partial_{\bar{0}} \partial_{T}+\delta^{T}_{i}\delta^{T}_{j}\delta^{\Bar{0}}_{\Bar{k}}\partial_{\bar{0}} \partial_{T}^2
+\delta^{I}_{i}\delta^{T}_{j}\delta^{\Bar{0}}_{\Bar{k}}\partial_{I} \partial_{\bar{0}} \partial_{T}
\nonumber\\
&&+2 \delta^{(0}_{i}\delta^{W)}_{j}\delta^{\Bar{0}}_{\Bar{k}}\partial_{0} \partial_{\bar{0}} \partial_{W}+\delta^{I}_{i}\delta^{W}_{j}\delta^{\Bar{0}}_{\Bar{k}}\partial_{I} \partial_{\bar{0}} \partial_{W}+\delta^{0}_{i}\delta^{0}_{j}\delta^{\Bar{0}}_{\Bar{k}}\partial_{0}^2 \partial_{\bar{0}}+\delta^{I}_{i}\delta^{0}_{j}\delta^{\Bar{0}}_{\Bar{k}}\partial_{I} \partial_{0} \partial_{\bar{0}}
\nonumber\\
&&+\delta^{T}_{i}\delta^{J}_{j}\delta^{\Bar{K}}_{\Bar{k}}\partial_{\bar{K}} \partial_{J} \partial_{T}+\delta^{W}_{i}\delta^{J}_{j}\delta^{\Bar{K}}_{\Bar{k}}\partial_{\bar{K}} \partial_{J} \partial_{W}+\delta^{0}_{i}\delta^{J}_{j}\delta^{\Bar{K}}_{\Bar{k}}\partial_{0} \partial_{\bar{K}} \partial_{J}+\delta^{I}_{i}\delta^{J}_{j}\delta^{\Bar{K}}_{\Bar{k}}\partial_{I} \partial_{\bar{K}} \partial_{J}
\nonumber\\
&&+2\delta^{(W}_{i}\delta^{T)}_{j}\delta^{\Bar{K}}_{\Bar{k}} \partial_{\bar{K}} \partial_{T} \partial_{W}
+2 \delta^{(0}_{i}\delta^{T)}_{j}\delta^{\Bar{K}}_{\Bar{k}}\partial_{0} \partial_{\bar{K}} \partial_{T}+\delta^{T}_{i}\delta^{T}_{j}\delta^{\Bar{K}}_{\Bar{k}}\partial_{\bar{K}} \partial_{T}^2+\delta^{I}_{i}\delta^{T}_{j}\delta^{\Bar{K}}_{\Bar{k}}\partial_{I} \partial_{\bar{K}} \partial_{T}
\nonumber\\
&&+2 \delta^{(0}_{i}\delta^{W)}_{j}\delta^{\Bar{K}}_{\Bar{k}}\partial_{0} \partial_{\bar{K}} \partial_{W}+\delta^{I}_{i}\delta^{W}_{j}\delta^{\Bar{K}}_{\Bar{k}}\partial_{I} \partial_{\bar{K}} \partial_{W}+\delta^{0}_{i}\delta^{0}_{j}\delta^{\Bar{K}}_{\Bar{k}}\partial_{0}^2 \partial_{\bar{K}}+\delta^{I}_{i}\delta^{0}_{j}\delta^{\Bar{K}}_{\Bar{k}}\partial_{I} \partial_{0} \partial_{\bar{K}}
\nonumber\\
&&+\delta^{T}_{i}\delta^{J}_{j}\delta^{\bar{T}}_{\Bar{k}}\partial_{\bar{T}} \partial_{J} \partial_{T}+\delta^{W}_{i}\delta^{J}_{j}\delta^{\bar{T}}_{\Bar{k}}\partial_{\bar{T}} \partial_{J} \partial_{W}+\delta^{0}_{i}\delta^{J}_{j}\delta^{\bar{T}}_{\Bar{k}}\partial_{0} \partial_{\bar{T}} \partial_{J}+\delta^{I}_{i}\delta^{J}_{j}\delta^{\Bar{T}}_{\Bar{k}}\partial_{I} \partial_{\bar{T}} \partial_{J}
\nonumber\\
&&+2 \delta^{(W}_{i}\delta^{T)}_{j}\delta^{\Bar{T}}_{\Bar{k}}\partial_{\bar{T}} \partial_{T} \partial_{W}
+2 \delta^{(0}_{i}\delta^{T)}_{j}\delta^{\bar{T}}_{\Bar{k}}\partial_{0} \partial_{\bar{T}} \partial_{T}+\delta^{T}_{i}\delta^{T}_{j}\delta^{\bar{T}}_{\Bar{k}}\partial_{\bar{T}} \partial_{T}^2+\delta^{I}_{i}\delta^{T}_{j}\delta^{\bar{T}}_{\Bar{k}}\partial_{I} \partial_{\bar{T}} \partial_{T}
\nonumber\\
&&+2 \delta^{(0}_{i}\delta^{W)}_{j}\delta^{\bar{T}}_{\Bar{k}}\partial_{0} \partial_{\bar{T}} \partial_{W}+\delta^{I}_{i}\delta^{W}_{j}\delta^{\bar{T}}_{\Bar{k}}\partial_{I} \partial_{\bar{T}} \partial_{W}+\delta^{0}_{i}\delta^{0}_{j}\delta^{\bar{T}}_{\Bar{k}}\partial_{0}^2 \partial_{\bar{T}}+\delta^{I}_{i}\delta^{0}_{j}\delta^{\bar{T}}_{\Bar{k}}\partial_{I} \partial_{0} \partial_{\bar{T}}
\nonumber\\
&&+\delta^{T}_{i}\delta^{J}_{j}\delta^{\bar{W}}_{\Bar{k}}\partial_{\bar{W}} \partial_{J} \partial_{T}+\delta^{W}_{i}\delta^{J}_{j}\delta^{\bar{W}}_{\Bar{k}}\partial_{\bar{W}} \partial_{J} \partial_{W}+\delta^{0}_{i}\delta^{J}_{j}\delta^{\bar{W}}_{\Bar{k}}\partial_{0} \partial_{\bar{W}} \partial_{J}+\delta^{I}_{i}\delta^{J}_{j}\delta^{\bar{W}}_{\Bar{k}}\partial_{I} \partial_{\bar{W}} \partial_{J}
\nonumber\\
&&+2 \delta^{(W}_{i}\delta^{T)}_{j}\delta^{\bar{W}}_{\Bar{k}}\partial_{\bar{W}} \partial_{T} \partial_{W}
+2 \delta^{(0}_{i}\delta^{T)}_{j}\delta^{\bar{W}}_{\Bar{k}}\partial_{0} \partial_{\bar{W}} \partial_{T}+\delta^{T}_{i}\delta^{T}_{j}\delta^{\bar{W}}_{\Bar{k}}\partial_{\bar{W}} \partial_{T}^2+\delta^{I}_{i}\delta^{T}_{j}\delta^{\bar{W}}_{\Bar{k}}\partial_{I} \partial_{\bar{W}} \partial_{T}
\nonumber\\
&&+2 \delta^{(0}_{i}\delta^{W)}_{j}\delta^{\bar{W}}_{\Bar{k}}\partial_{0} \partial_{\bar{W}} \partial_{W}+\delta^{I}_{i}\delta^{W}_{j}\delta^{\bar{W}}_{\Bar{k}}\partial_{I} \partial_{\bar{W}} \partial_{W}+\delta^{0}_{i}\delta^{0}_{j}\delta^{\bar{W}}_{\Bar{k}}\partial_{0}^2 \partial_{\bar{W}}+\delta^{I}_{i}\delta^{0}_{j}\delta^{\bar{W}}_{\Bar{k}}\partial_{I} \partial_{0} \partial_{\bar{W}},\qquad \qquad
\end{eqnarray}
and the fourth derivative of $N$ is
\begin{eqnarray}
N_{ij\bar{k}\bar{l}} = \partial_{i}\partial_{j}\partial_{\bar{k}}\partial_{\bar{l}}N = \circled{1} + \circled{2} + \circled{3} + \circled{4} + \circled{5} + \circled{6} 
\end{eqnarray}

\begin{eqnarray}
&& \circled{1} =\delta^{0}_{i}\delta^{0}_{j}\delta^{\bar{0}}_{\Bar{k}}\delta^{\bar{0}}_{\Bar{l}}
\partial_{\bar{0}}^2 \partial_{0}^2
+\delta^{0}_{i}\delta^{0}_{j}\delta^{\bar{T}}_{\Bar{k}}\delta^{\bar{T}}_{\Bar{l}}\partial_{\bar{T}}^2 \partial_{0}^2
+\delta^{0}_{i}\delta^{0}_{j}\delta^{\bar{K}}_{\Bar{k}}\delta^{0}_{\Bar{l}}\partial_{\bar{0}} \partial_{\bar{K}} \partial_{0}^2
+\delta^{0}_{i}\delta^{0}_{j}\delta^{\bar{0}}_{\Bar{k}}\delta^{\bar{L}}_{\Bar{l}}\partial_{\bar{0}} \partial_{\bar{L}} \partial_{0}^2
\nonumber\\
&&
+\delta^{0}_{i}\delta^{0}_{j}\delta^{\bar{K}}_{\Bar{k}}\delta^{\bar{L}}_{\Bar{l}}\partial_{\bar{K}} \partial_{\bar{L}} \partial_{0}^2
+2\delta^{0}_{i}\delta^{0}_{j}\delta^{(\bar{0}}_{\Bar{k}}\delta^{\bar{T})}_{\Bar{l}}\partial_{\bar{0}} \partial_{\bar{T}} \partial_{0}^2
+\delta^{0}_{i}\delta^{0}_{j}\delta^{\bar{K}}_{\Bar{k}}\delta^{\bar{T}}_{\Bar{l}}\partial_{\bar{K}} \partial_{\bar{T}} \partial_{0}^2
+\delta^{0}_{i}\delta^{0}_{j}\delta^{\bar{T}}_{\Bar{k}}\delta^{\bar{L}}_{\Bar{l}}\partial_{\bar{L}} \partial_{\bar{T}} \partial_{0}^2
\nonumber\\
&&
+2\delta^{0}_{i}\delta^{0}_{j}\delta^{(\bar{0}}_{\Bar{k}}\delta^{\bar{W})}_{\Bar{l}} \partial_{\bar{0}} \partial_{\bar{W}} \partial_{0}^2
+\delta^{0}_{i}\delta^{0}_{j}\delta^{\bar{K}}_{\Bar{k}}\delta^{\bar{W}}_{\Bar{l}}\partial_{\bar{K}} \partial_{\bar{W}} \partial_{0}^2
+\delta^{0}_{i}\delta^{0}_{j}\delta^{\bar{W}}_{\Bar{k}}\delta^{\bar{L}}_{\Bar{l}}\partial_{\bar{L}} \partial_{\bar{W}} \partial_{0}^2
+2\delta^{0}_{i}\delta^{0}_{j}\delta^{(\bar{W}}_{\Bar{k}}\delta^{\bar{T})}_{\Bar{l}} \partial_{\bar{T}} \partial_{\bar{W}} \partial_{0}^2
\nonumber\\
&&
+\delta^{I}_{i}\delta^{0}_{j}\delta^{\bar{0}}_{\Bar{k}}\delta^{\bar{0}}_{\Bar{l}}\partial_{\bar{0}}^2 \partial_{I} \partial_{0}
+\delta^{I}_{i}\delta^{0}_{j}\delta^{\bar{T}}_{\Bar{k}}\delta^{\bar{T}}_{\Bar{l}}\partial_{\bar{T}}^2 \partial_{I} \partial_{0}
+\delta^{I}_{i}\delta^{0}_{j}\delta^{\bar{K}}_{\Bar{k}}\delta^{\bar{0}}_{\Bar{l}}\partial_{\bar{0}} \partial_{\bar{K}} \partial_{I} \partial_{0}
+\delta^{I}_{i}\delta^{0}_{j}\delta^{\bar{0}}_{\Bar{k}}\delta^{\bar{L}}_{\Bar{l}}\partial_{\bar{0}} \partial_{\bar{L}} \partial_{I} \partial_{0}
\nonumber\\
&&
+\delta^{I}_{i}\delta^{0}_{j}\delta^{\bar{K}}_{\Bar{k}}\delta^{\bar{L}}_{\Bar{l}}\partial_{\bar{K}} \partial_{\bar{L}} \partial_{I} \partial_{0}
+2\delta^{I}_{i}\delta^{0}_{j}\delta^{(\bar{0}}_{\Bar{k}}\delta^{\bar{T})}_{\Bar{l}} \partial_{\bar{0}} \partial_{\bar{T}} \partial_{I} \partial_{0}
+\delta^{I}_{i}\delta^{0}_{j}\delta^{\bar{K}}_{\Bar{k}}\delta^{\bar{T}}_{\Bar{l}}\partial_{\bar{K}} \partial_{\bar{T}} \partial_{I} \partial_{0}
+\delta^{I}_{i}\delta^{0}_{j}\delta^{\bar{T}}_{\Bar{k}}\delta^{\bar{L}}_{\Bar{l}}\partial_{\bar{L}} \partial_{\bar{T}} \partial_{I} \partial_{0}
\nonumber\\
&&
+2 \delta^{I}_{i}\delta^{0}_{j}\delta^{(\bar{0}}_{\Bar{k}}\delta^{\bar{W})}_{\Bar{l}}\partial_{\bar{0}} \partial_{\bar{W}} \partial_{I} \partial_{0}
+\delta^{I}_{i}\delta^{0}_{j}\delta^{\bar{K}}_{\Bar{k}}\delta^{\bar{W}}_{\Bar{l}}\partial_{\bar{K}} \partial_{\bar{W}} \partial_{I} \partial_{0}
+\delta^{I}_{i}\delta^{0}_{j}\delta^{\bar{W}}_{\Bar{k}}\delta^{\bar{L}}_{\Bar{l}}\partial_{\bar{L}} \partial_{\bar{W}} \partial_{I} \partial_{0}
\nonumber\\
&&
+2 \delta^{I}_{i}\delta^{0}_{j}\delta^{(\bar{W}}_{\Bar{k}}\delta^{\bar{T})}_{\Bar{l}}\partial_{\bar{T}} \partial_{\bar{W}} \partial_{I} \partial_{0}
+\delta^{0}_{i}\delta^{J}_{j}\delta^{\bar{0}}_{\Bar{k}}\delta^{\bar{0}}_{\Bar{l}}\partial_{\bar{0}}^2 \partial_{J} \partial_{0}
+\delta^{0}_{i}\delta^{J}_{j}\delta^{\bar{T}}_{\Bar{k}}\delta^{\bar{T}}_{\Bar{l}}\partial_{\bar{T}}^2 \partial_{J} \partial_{0}
\nonumber\\
&&
+\delta^{0}_{i}\delta^{J}_{j}\delta^{\bar{K}}_{\Bar{k}}\delta^{\bar{0}}_{\Bar{l}}\partial_{\bar{0}} \partial_{\bar{K}} \partial_{J} \partial_{0}
+\delta^{0}_{i}\delta^{J}_{j}\delta^{\bar{0}}_{\Bar{k}}\delta^{\bar{L}}_{\Bar{l}}\partial_{\bar{0}} \partial_{\bar{L}} \partial_{J} \partial_{0}
+\delta^{0}_{i}\delta^{J}_{j}\delta^{\bar{K}}_{\Bar{k}}\delta^{\bar{L}}_{\Bar{l}}\partial_{\bar{K}} \partial_{\bar{L}} \partial_{J} \partial_{0}
\end{eqnarray}
\begin{eqnarray}
&& \circled{2} =
2\delta^{0}_{i}\delta^{J}_{j}\delta^{(\bar{0}}_{\Bar{k}}\delta^{\bar{T})}_{\Bar{l}} \partial_{\bar{0}} \partial_{\bar{T}} \partial_{J} \partial_{0}
+\delta^{0}_{i}\delta^{J}_{j}\delta^{\bar{K}}_{\Bar{k}}\delta^{\bar{T}}_{\Bar{l}}\partial_{\bar{K}} \partial_{\bar{T}} \partial_{J} \partial_{0}
+\delta^{0}_{i}\delta^{J}_{j}\delta^{\bar{T}}_{\Bar{k}}\delta^{\bar{L}}_{\Bar{l}}\partial_{\bar{L}} \partial_{\bar{T}} \partial_{J} \partial_{0}
\nonumber\\
&&
+2\delta^{0}_{i}\delta^{J}_{j}\delta^{(\bar{0}}_{\Bar{k}}\delta^{\bar{W})}_{\Bar{l}} \partial_{\bar{0}} \partial_{\bar{W}} \partial_{J} \partial_{0}
+\delta^{0}_{i}\delta^{J}_{j}\delta^{\bar{K}}_{\Bar{k}}\delta^{\bar{W}}_{\Bar{l}}\partial_{\bar{K}} \partial_{\bar{W}} \partial_{J} \partial_{0}
+\delta^{0}_{i}\delta^{J}_{j}\delta^{\bar{W}}_{\Bar{k}}\delta^{\bar{L}}_{\Bar{l}}\partial_{\bar{L}} \partial_{\bar{W}} \partial_{J} \partial_{0}
\nonumber\\
&&
+2 \delta^{0}_{i}\delta^{J}_{j}\delta^{(\bar{W}}_{\Bar{k}}\delta^{\bar{T})}_{\Bar{l}}\partial_{\bar{T}} \partial_{\bar{W}} \partial_{J} \partial_{0}
+2 \delta^{(0}_{i}\delta^{T)}_{j}\delta^{\bar{0}}_{\Bar{k}}\delta^{\bar{0}}_{\Bar{l}}\partial_{\bar{0}}^2 \partial_{T} \partial_{0}
+2\delta^{(0}_{i}\delta^{T)}_{j}\delta^{\bar{T}}_{\Bar{k}}\delta^{\bar{T}}_{\Bar{l}} \partial_{\bar{T}}^2 \partial_{T} \partial_{0}
\nonumber\\
&&
+2 \delta^{(0}_{i}\delta^{T)}_{j}\delta^{\bar{K}}_{\Bar{k}}\delta^{\bar{0}}_{\Bar{l}}\partial_{\bar{0}} \partial_{\bar{K}} \partial_{T} \partial_{0}
+2 \delta^{(0}_{i}\delta^{T)}_{j}\delta^{\bar{0}}_{\Bar{k}}\delta^{\bar{L}}_{\Bar{l}}\partial_{\bar{0}} \partial_{\bar{L}} \partial_{T} \partial_{0}
+2 \delta^{(0}_{i}\delta^{T)}_{j}\delta^{\bar{K}}_{\Bar{k}}\delta^{\bar{L}}_{\Bar{l}}\partial_{\bar{K}} \partial_{\bar{L}} \partial_{T} \partial_{0}
\nonumber\\
&&
+4 \delta^{(0}_{i}\delta^{T)}_{j}\delta^{(\bar{0}}_{\Bar{k}}\delta^{\bar{T})}_{\Bar{l}}\partial_{\bar{0}} \partial_{\bar{T}} \partial_{T} \partial_{0}
+2 \delta^{(0}_{i}\delta^{T)}_{j}\delta^{\bar{K}}_{\Bar{k}}\delta^{\bar{T}}_{\Bar{l}}\partial_{\bar{K}} \partial_{\bar{T}} \partial_{T} \partial_{0}
+2\delta^{(0}_{i}\delta^{T)}_{j}\delta^{\bar{T}}_{\Bar{k}}\delta^{\bar{L}}_{\Bar{l}} \partial_{\bar{L}} \partial_{\bar{T}} \partial_{T} \partial_{0}
\nonumber\\
&&
+4\delta^{(0}_{i}\delta^{T)}_{j}\delta^{(\bar{0}}_{\Bar{k}}\delta^{\bar{W})}_{\Bar{l}} \partial_{\bar{0}} \partial_{\bar{W}} \partial_{T} \partial_{0}
+2\delta^{(0}_{i}\delta^{T)}_{j}\delta^{\bar{K}}_{\Bar{k}}\delta^{\bar{W}}_{\Bar{l}} \partial_{\bar{K}} \partial_{\bar{W}} \partial_{T} \partial_{0}
+2\delta^{(0}_{i}\delta^{T)}_{j}\delta^{\bar{W}}_{\Bar{k}}\delta^{\bar{L}}_{\Bar{l}} \partial_{\bar{L}} \partial_{\bar{W}} \partial_{T} \partial_{0}
\nonumber\\
&&
+4\delta^{(0}_{i}\delta^{T)}_{j}\delta^{(\bar{W}}_{\Bar{k}}\delta^{\bar{T})}_{\Bar{l}} \partial_{\bar{T}} \partial_{\bar{W}} \partial_{T} \partial_{0}
+2\delta^{(0}_{i}\delta^{W)}_{j}\delta^{\bar{0}}_{\Bar{k}}\delta^{\bar{0}}_{\Bar{l}} \partial_{\bar{0}}^2 \partial_{W} \partial_{0}
+2\delta^{(0}_{i}\delta^{W)}_{j}\delta^{\bar{T}}_{\Bar{k}}\delta^{\bar{T}}_{\Bar{l}} \partial_{\bar{T}}^2 \partial_{W} \partial_{0}
\nonumber\\
&&
+2\delta^{(0}_{i}\delta^{W)}_{j}\delta^{\bar{K}}_{\Bar{k}}\delta^{\bar{0}}_{\Bar{l}} \partial_{\bar{0}} \partial_{\bar{K}} \partial_{W} \partial_{0}
+2 \delta^{(0}_{i}\delta^{W)}_{j}\delta^{\bar{0}}_{\Bar{k}}\delta^{\bar{L}}_{\Bar{l}}\partial_{\bar{0}} \partial_{\bar{L}} \partial_{W} \partial_{0}
+2\delta^{(0}_{i}\delta^{W)}_{j}\delta^{\bar{K}}_{\Bar{k}}\delta^{\bar{L}}_{\Bar{l}} \partial_{\bar{K}} \partial_{\bar{L}} \partial_{W} \partial_{0}
\end{eqnarray}
\begin{eqnarray}
&& \circled{3} =
4\delta^{(0}_{i}\delta^{W)}_{j}\delta^{(\bar{0}}_{\Bar{k}}\delta^{\bar{T})}_{\Bar{l}} \partial_{\bar{0}} \partial_{\bar{T}} \partial_{W} \partial_{0}
+2\delta^{(0}_{i}\delta^{W)}_{j}\delta^{\bar{K}}_{\Bar{k}}\delta^{\bar{T}}_{\Bar{l}} \partial_{\bar{K}} \partial_{\bar{T}} \partial_{W} \partial_{0}
+2\delta^{(0}_{i}\delta^{W)}_{j}\delta^{\bar{T}}_{\Bar{k}}\delta^{\bar{L}}_{\Bar{l}} \partial_{\bar{L}} \partial_{\bar{T}} \partial_{W} \partial_{0}
\nonumber\\
&&
+4\delta^{(0}_{i}\delta^{W)}_{j}\delta^{(\bar{0}}_{\Bar{k}}\delta^{\bar{W})}_{\Bar{l}} \partial_{\bar{0}} \partial_{\bar{W}} \partial_{W} \partial_{0}
+2\delta^{(0}_{i}\delta^{W)}_{j}\delta^{\bar{K}}_{\Bar{k}}\delta^{\bar{W}}_{\Bar{l}} \partial_{\bar{K}} \partial_{\bar{W}} \partial_{W} \partial_{0}
+2\delta^{(0}_{i}\delta^{W)}_{j}\delta^{\bar{W}}_{\Bar{k}}\delta^{\bar{L}}_{\Bar{l}} \partial_{\bar{L}} \partial_{\bar{W}} \partial_{W} \partial_{0}
\nonumber\\
&&
+4 \delta^{(0}_{i}\delta^{W)}_{j}\delta^{(\bar{W}}_{\Bar{k}}\delta^{\bar{T})}_{\Bar{l}}\partial_{\bar{T}} \partial_{\bar{W}} \partial_{W} \partial_{0}
+\delta^{T}_{i}\delta^{T}_{j}\delta^{\bar{0}}_{\Bar{k}}\delta^{\bar{0}}_{\Bar{l}}\partial_{\bar{0}}^2 \partial_{T}^2
+\delta^{T}_{i}\delta^{T}_{j}\delta^{\bar{T}}_{\Bar{k}}\delta^{\bar{T}}_{\Bar{l}}\partial_{\bar{T}}^2 \partial_{T}^2
\nonumber\\
&&
+\delta^{T}_{i}\delta^{T}_{j}\delta^{\bar{K}}_{\Bar{k}}\delta^{\bar{0}}_{\Bar{l}}\partial_{\bar{0}} \partial_{\bar{K}} \partial_{T}^2
+\delta^{T}_{i}\delta^{T}_{j}\delta^{\bar{0}}_{\Bar{k}}\delta^{\bar{L}}_{\Bar{l}}\partial_{\bar{0}} \partial_{\bar{L}} \partial_{T}^2
+\delta^{T}_{i}\delta^{T}_{j}\delta^{\bar{K}}_{\Bar{k}}\delta^{\bar{L}}_{\Bar{l}}\partial_{\bar{K}} \partial_{\bar{L}} \partial_{T}^2
\nonumber\\
&&
+2\delta^{T}_{i}\delta^{T}_{j}\delta^{(\bar{0}}_{\Bar{k}}\delta^{\bar{T})}_{\Bar{l}} \partial_{\bar{0}} \partial_{\bar{T}} \partial_{T}^2
+\delta^{T}_{i}\delta^{T}_{j}\delta^{\bar{K}}_{\Bar{k}}\delta^{\bar{T}}_{\Bar{l}}\partial_{\bar{K}} \partial_{\bar{T}} \partial_{T}^2
+\delta^{T}_{i}\delta^{T}_{j}\delta^{\bar{T}}_{\Bar{k}}\delta^{\bar{L}}_{\Bar{l}}\partial_{\bar{L}} \partial_{\bar{T}} \partial_{T}^2
\nonumber\\
&&
+2 \delta^{T}_{i}\delta^{T}_{j}\delta^{(\bar{0}}_{\Bar{k}}\delta^{\bar{W})}_{\Bar{l}}\partial_{\bar{0}} \partial_{\bar{W}} \partial_{T}^2
+\delta^{T}_{i}\delta^{T}_{j}\delta^{\bar{K}}_{\Bar{k}}\delta^{\bar{W}}_{\Bar{l}}\partial_{\bar{K}} \partial_{\bar{W}} \partial_{T}^2
+\delta^{T}_{i}\delta^{T}_{j}\delta^{\bar{W}}_{\Bar{k}}\delta^{\bar{L}}_{\Bar{l}}\partial_{\bar{L}} \partial_{\bar{W}} \partial_{T}^2
\nonumber\\
&&
+2 \delta^{T}_{i}\delta^{T}_{j}\delta^{(\bar{W}}_{\Bar{k}}\delta^{\bar{T})}_{\Bar{l}}\partial_{\bar{T}} \partial_{\bar{W}} \partial_{T}^2
+\delta^{I}_{i}\delta^{J}_{j}\delta^{\bar{0}}_{\Bar{k}}\delta^{\bar{0}}_{\Bar{l}}\partial_{\bar{0}}^2 \partial_{I} \partial_{J}
+\delta^{I}_{i}\delta^{J}_{j}\delta^{\bar{T}}_{\Bar{k}}\delta^{\bar{T}}_{\Bar{l}}\partial_{\bar{T}}^2 \partial_{I} \partial_{J}
\nonumber\\
&&
+\delta^{I}_{i}\delta^{J}_{j}\delta^{\bar{K}}_{\Bar{k}}\delta^{\bar{0}}_{\Bar{l}}\partial_{\bar{0}} \partial_{\bar{K}} \partial_{I} \partial_{J}
+\delta^{I}_{i}\delta^{J}_{j}\delta^{\bar{0}}_{\Bar{k}}\delta^{\bar{L}}_{\Bar{l}}\partial_{\bar{0}} \partial_{\bar{L}} \partial_{I} \partial_{J}
+\delta^{I}_{i}\delta^{J}_{j}\delta^{\bar{K}}_{\Bar{k}}\delta^{\bar{L}}_{\Bar{l}}\partial_{\bar{K}} \partial_{\bar{L}} \partial_{I} \partial_{J}
\end{eqnarray}
\begin{eqnarray}
&& \circled{4} = 
2\delta^{I}_{i}\delta^{J}_{j}\delta^{(\bar{0}}_{\Bar{k}}\delta^{\bar{T})}_{\Bar{l}} \partial_{\bar{0}} \partial_{\bar{T}} \partial_{I} \partial_{J}
+\delta^{I}_{i}\delta^{J}_{j}\delta^{\bar{K}}_{\Bar{k}}\delta^{\bar{T}}_{\Bar{l}}\partial_{\bar{K}} \partial_{\bar{T}} \partial_{I} \partial_{J}
+\delta^{I}_{i}\delta^{J}_{j}\delta^{\bar{T}}_{\Bar{k}}\delta^{\bar{L}}_{\Bar{l}}\partial_{\bar{L}} \partial_{\bar{T}} \partial_{I} \partial_{J}
\nonumber\\
&&
+2 \delta^{I}_{i}\delta^{J}_{j}\delta^{(\bar{0}}_{\Bar{k}}\delta^{\bar{W})}_{\Bar{l}}\partial_{\bar{0}} \partial_{\bar{W}} \partial_{I} \partial_{J}
+\delta^{I}_{i}\delta^{J}_{j}\delta^{\bar{K}}_{\Bar{k}}\delta^{\bar{W}}_{\Bar{l}}\partial_{\bar{K}} \partial_{\bar{W}} \partial_{I} \partial_{J}
+\delta^{I}_{i}\delta^{J}_{j}\delta^{\bar{W}}_{\Bar{k}}\delta^{\bar{L}}_{\Bar{l}}\partial_{\bar{L}} \partial_{\bar{W}} \partial_{I} \partial_{J}
\nonumber\\
&&
+2\delta^{I}_{i}\delta^{J}_{j}\delta^{(\bar{W}}_{\Bar{k}}\delta^{\bar{T})}_{\Bar{l}} \partial_{\bar{T}} \partial_{\bar{W}} \partial_{I} \partial_{J}
+\delta^{I}_{i}\delta^{T}_{j}\delta^{\bar{0}}_{\Bar{k}}\delta^{\bar{0}}_{\Bar{l}}\partial_{\bar{0}}^2 \partial_{I} \partial_{T}
+\delta^{I}_{i}\delta^{T}_{j}\delta^{\bar{T}}_{\Bar{k}}\delta^{\bar{T}}_{\Bar{l}}\partial_{\bar{T}}^2 \partial_{I} \partial_{T}
\nonumber\\
&&
+\delta^{I}_{i}\delta^{T}_{j}\delta^{\bar{K}}_{\Bar{k}}\delta^{\bar{0}}_{\Bar{l}}\partial_{\bar{0}} \partial_{\bar{K}} \partial_{I} \partial_{T}
+\delta^{I}_{i}\delta^{T}_{j}\delta^{\bar{0}}_{\Bar{k}}\delta^{\bar{L}}_{\Bar{l}}\partial_{\bar{0}} \partial_{\bar{L}} \partial_{I} \partial_{T}
+\delta^{I}_{i}\delta^{T}_{j}\delta^{\bar{K}}_{\Bar{k}}\delta^{\bar{L}}_{\Bar{l}}\partial_{\bar{K}} \partial_{\bar{L}} \partial_{I} \partial_{T}
\nonumber\\
&&
+2\delta^{I}_{i}\delta^{T}_{j}\delta^{(\bar{0}}_{\Bar{k}}\delta^{\bar{T})}_{\Bar{l}} \partial_{\bar{0}} \partial_{\bar{T}} \partial_{I} \partial_{T}
+\delta^{I}_{i}\delta^{T}_{j}\delta^{\bar{K}}_{\Bar{k}}\delta^{\bar{T}}_{\Bar{l}}\partial_{\bar{K}} \partial_{\bar{T}} \partial_{I} \partial_{T}
+\delta^{I}_{i}\delta^{T}_{j}\delta^{\bar{T}}_{\Bar{k}}\delta^{\bar{L}}_{\Bar{l}}\partial_{\bar{L}} \partial_{\bar{T}} \partial_{I} \partial_{T}
\nonumber\\
&&
+2 \delta^{I}_{i}\delta^{T}_{j}\delta^{(\bar{0}}_{\Bar{k}}\delta^{\bar{W})}_{\Bar{l}}\partial_{\bar{0}} \partial_{\bar{W}} \partial_{I} \partial_{T}
+\delta^{I}_{i}\delta^{T}_{j}\delta^{\bar{K}}_{\Bar{k}}\delta^{\bar{W}}_{\Bar{l}}\partial_{\bar{K}} \partial_{\bar{W}} \partial_{I} \partial_{T}
+\delta^{I}_{i}\delta^{T}_{j}\delta^{\bar{W}}_{\Bar{k}}\delta^{\bar{L}}_{\Bar{l}}\partial_{\bar{L}} \partial_{\bar{W}} \partial_{I} \partial_{T}
\nonumber\\
&&
+2 \delta^{I}_{i}\delta^{T}_{j}\delta^{(\bar{W}}_{\Bar{k}}\delta^{\bar{T})}_{\Bar{l}}\partial_{\bar{T}} \partial_{\bar{W}} \partial_{I} \partial_{T}
+\delta^{T}_{i}\delta^{J}_{j}\delta^{\bar{0}}_{\Bar{k}}\delta^{\bar{0}}_{\Bar{l}}\partial_{\bar{0}}^2 \partial_{J} \partial_{T}
+\delta^{T}_{i}\delta^{J}_{j}\delta^{\bar{T}}_{\Bar{k}}\delta^{\bar{T}}_{\Bar{l}}\partial_{\bar{T}}^2 \partial_{J} \partial_{T}
\nonumber\\
&&
+\delta^{T}_{i}\delta^{J}_{j}\delta^{\bar{K}}_{\Bar{k}}\delta^{\bar{0}}_{\Bar{l}}\partial_{\bar{0}} \partial_{\bar{K}} \partial_{J} \partial_{T}
+\delta^{T}_{i}\delta^{J}_{j}\delta^{\bar{0}}_{\Bar{k}}\delta^{\bar{L}}_{\Bar{l}}\partial_{\bar{0}} \partial_{\bar{L}} \partial_{J} \partial_{T}
+\delta^{T}_{i}\delta^{J}_{j}\delta^{\bar{K}}_{\Bar{k}}\delta^{\bar{L}}_{\Bar{l}}\partial_{\bar{K}} \partial_{\bar{L}} \partial_{J} \partial_{T}
\end{eqnarray}
\begin{eqnarray}
&& \circled{5} =
2 \delta^{T}_{i}\delta^{J}_{j}\delta^{(\bar{0}}_{\Bar{k}}\delta^{\bar{T})}_{\Bar{l}}\partial_{\bar{0}} \partial_{\bar{T}} \partial_{J} \partial_{T}
+\delta^{T}_{i}\delta^{J}_{j}\delta^{\bar{K}}_{\Bar{k}}\delta^{\bar{T}}_{\Bar{l}}\partial_{\bar{K}} \partial_{\bar{T}} \partial_{J} \partial_{T}
+\delta^{T}_{i}\delta^{J}_{j}\delta^{\bar{T}}_{\Bar{k}}\delta^{\bar{L}}_{\Bar{l}}\partial_{\bar{L}} \partial_{\bar{T}} \partial_{J} \partial_{T}
\nonumber\\
&&
+2 \delta^{T}_{i}\delta^{J}_{j}\delta^{(\bar{0}}_{\Bar{k}}\delta^{\bar{W})}_{\Bar{l}}\partial_{\bar{0}} \partial_{\bar{W}} \partial_{J} \partial_{T}
+\delta^{T}_{i}\delta^{J}_{j}\delta^{\bar{K}}_{\Bar{k}}\delta^{\bar{W}}_{\Bar{l}}\partial_{\bar{K}} \partial_{\bar{W}} \partial_{J} \partial_{T}
+\delta^{T}_{i}\delta^{J}_{j}\delta^{\bar{W}}_{\Bar{k}}\delta^{\bar{L}}_{\Bar{l}}\partial_{\bar{L}} \partial_{\bar{W}} \partial_{J} \partial_{T}
\nonumber\\
&&
+2\delta^{T}_{i}\delta^{J}_{j}\delta^{(\bar{W}}_{\Bar{k}}\delta^{\bar{T})}_{\Bar{l}} \partial_{\bar{T}} \partial_{\bar{W}} \partial_{J} \partial_{T}
+\delta^{I}_{i}\delta^{W}_{j}\delta^{\bar{0}}_{\Bar{k}}\delta^{\bar{0}}_{\Bar{l}}\partial_{\bar{0}}^2 \partial_{I} \partial_{W}
+\delta^{I}_{i}\delta^{W}_{j}\delta^{\bar{T}}_{\Bar{k}}\delta^{\bar{T}}_{\Bar{l}}\partial_{\bar{T}}^2 \partial_{I} \partial_{W}
\nonumber\\
&&
+\delta^{I}_{i}\delta^{W}_{j}\delta^{\bar{K}}_{\Bar{k}}\delta^{\bar{0}}_{\Bar{l}}\partial_{\bar{0}} \partial_{\bar{K}} \partial_{I} \partial_{W}
+\delta^{I}_{i}\delta^{W}_{j}\delta^{\bar{0}}_{\Bar{k}}\delta^{\bar{L}}_{\Bar{l}}\partial_{\bar{0}} \partial_{\bar{L}} \partial_{I} \partial_{W}
+\delta^{I}_{i}\delta^{W}_{j}\delta^{\bar{K}}_{\Bar{k}}\delta^{\bar{L}}_{\Bar{l}}\partial_{\bar{K}} \partial_{\bar{L}} \partial_{I} \partial_{W}
\nonumber\\
&&
+2 \delta^{I}_{i}\delta^{W}_{j}\delta^{(\bar{0}}_{\Bar{k}}\delta^{\bar{T})}_{\Bar{l}}\partial_{\bar{0}} \partial_{\bar{T}} \partial_{I} \partial_{W}
+\delta^{I}_{i}\delta^{W}_{j}\delta^{\bar{K}}_{\Bar{k}}\delta^{\bar{T}}_{\Bar{l}}\partial_{\bar{K}} \partial_{\bar{T}} \partial_{I} \partial_{W}
+\delta^{I}_{i}\delta^{W}_{j}\delta^{\bar{T}}_{\Bar{k}}\delta^{\bar{L}}_{\Bar{l}}\partial_{\bar{L}} \partial_{\bar{T}} \partial_{I} \partial_{W}
\nonumber\\
&&
+2\delta^{I}_{i}\delta^{W}_{j}\delta^{(\bar{0}}_{\Bar{k}}\delta^{\bar{W})}_{\Bar{l}} \partial_{\bar{0}} \partial_{\bar{W}} \partial_{I} \partial_{W}
+\delta^{I}_{i}\delta^{W}_{j}\delta^{\bar{K}}_{\Bar{k}}\delta^{\bar{W}}_{\Bar{l}}\partial_{\bar{K}} \partial_{\bar{W}} \partial_{I} \partial_{W}
+\delta^{I}_{i}\delta^{W}_{j}\delta^{\bar{W}}_{\Bar{k}}\delta^{\bar{L}}_{\Bar{l}}\partial_{\bar{L}} \partial_{\bar{W}} \partial_{I} \partial_{W}
\nonumber\\
&&
+2 \delta^{I}_{i}\delta^{W}_{j}\delta^{(\bar{W}}_{\Bar{k}}\delta^{\bar{T})}_{\Bar{l}}\partial_{\bar{T}} \partial_{\bar{W}} \partial_{I} \partial_{W}
+\delta^{W}_{i}\delta^{J}_{j}\delta^{\bar{0}}_{\Bar{k}}\delta^{\bar{0}}_{\Bar{l}}\partial_{\bar{0}}^2 \partial_{J} \partial_{W}
+\delta^{W}_{i}\delta^{J}_{j}\delta^{\bar{T}}_{\Bar{k}}\delta^{\bar{T}}_{\Bar{l}}\partial_{\bar{T}}^2 \partial_{J} \partial_{W}
\nonumber\\
&&
+\delta^{W}_{i}\delta^{J}_{j}\delta^{\bar{K}}_{\Bar{k}}\delta^{\bar{0}}_{\Bar{l}}\partial_{\bar{0}}\partial_{\bar{K}} \partial_{J} \partial_{W}
+\delta^{W}_{i}\delta^{J}_{j}\delta^{\bar{0}}_{\Bar{k}}\delta^{\bar{L}}_{\Bar{l}}\partial_{\bar{0}} \partial_{\bar{L}} \partial_{J} \partial_{W}
+\delta^{W}_{i}\delta^{J}_{j}\delta^{\bar{K}}_{\Bar{k}}\delta^{\bar{L}}_{\Bar{l}}\partial_{\bar{K}} \partial_{\bar{L}} \partial_{J} \partial_{W}
\end{eqnarray}
\begin{eqnarray}
&& \circled{6} =
2\delta^{W}_{i}\delta^{J}_{j}\delta^{(\bar{0}}_{\Bar{k}}\delta^{\bar{T})}_{\Bar{l}} \partial_{\bar{0}} \partial_{\bar{T}} \partial_{J} \partial_{W}
+\delta^{W}_{i}\delta^{J}_{j}\delta^{\bar{K}}_{\Bar{k}}\delta^{\bar{T}}_{\Bar{l}}\partial_{\bar{K}} \partial_{\bar{T}} \partial_{J} \partial_{W}
+\delta^{W}_{i}\delta^{J}_{j}\delta^{\bar{T}}_{\Bar{k}}\delta^{\bar{L}}_{\Bar{l}}\partial_{\bar{L}} \partial_{\bar{T}} \partial_{J} \partial_{W}
\nonumber\\
&&
+2 \delta^{W}_{i}\delta^{J}_{j}\delta^{(\bar{0}}_{\Bar{k}}\delta^{\bar{W})}_{\Bar{l}}\partial_{\bar{0}} \partial_{\bar{W}} \partial_{J} \partial_{W}
+\delta^{W}_{i}\delta^{J}_{j}\delta^{\bar{K}}_{\Bar{k}}\delta^{\bar{W}}_{\Bar{l}}\partial_{\bar{K}} \partial_{\bar{W}} \partial_{J} \partial_{W}
+\delta^{W}_{i}\delta^{J}_{j}\delta^{\bar{W}}_{\Bar{k}}\delta^{\bar{L}}_{\Bar{l}}\partial_{\bar{L}} \partial_{\bar{W}} \partial_{J} \partial_{W}
\nonumber\\
&&
+2 \delta^{W}_{i}\delta^{J}_{j}\delta^{(\bar{W}}_{\Bar{k}}\delta^{\bar{T})}_{\Bar{l}}\partial_{\bar{T}} \partial_{\bar{W}} \partial_{J} \partial_{W}
+2 \delta^{(W}_{i}\delta^{T)}_{j}\delta^{\bar{0}}_{\Bar{k}}\delta^{\bar{0}}_{\Bar{l}}\partial_{\bar{0}}^2 \partial_{T} \partial_{W}
+2 \delta^{(W}_{i}\delta^{T)}_{j}\delta^{\bar{T}}_{\Bar{k}}\delta^{\bar{T}}_{\Bar{l}}\partial_{\bar{T}}^2 \partial_{T} \partial_{W}
\nonumber\\
&&
+2\delta^{(W}_{i}\delta^{T)}_{j}\delta^{\bar{K}}_{\Bar{k}}\delta^{\bar{0}}_{\Bar{l}} \partial_{\bar{0}} \partial_{\bar{K}} \partial_{T} \partial_{W}
+2 \delta^{(W}_{i}\delta^{T)}_{j}\delta^{\bar{0}}_{\Bar{k}}\delta^{\bar{L}}_{\Bar{l}}\partial_{\bar{0}} \partial_{\bar{L}} \partial_{T} \partial_{W}
+2\delta^{(W}_{i}\delta^{T)}_{j}\delta^{\bar{K}}_{\Bar{k}}\delta^{\bar{L}}_{\Bar{l}} \partial_{\bar{K}} \partial_{\bar{L}} \partial_{T} \partial_{W}
\nonumber\\
&&
+4\delta^{(W}_{i}\delta^{T)}_{j}\delta^{(\bar{0}}_{\Bar{k}}\delta^{\bar{T})}_{\Bar{l}} \partial_{\bar{0}} \partial_{\bar{T}} \partial_{T} \partial_{W}
+2\delta^{(W}_{i}\delta^{T)}_{j}\delta^{\bar{K}}_{\Bar{k}}\delta^{\bar{T}}_{\Bar{l}} \partial_{\bar{K}} \partial_{\bar{T}} \partial_{T} \partial_{W}
+2 \delta^{(W}_{i}\delta^{T)}_{j}\delta^{\bar{T}}_{\Bar{k}}\delta^{\bar{L}}_{\Bar{l}}\partial_{\bar{L}} \partial_{\bar{T}} \partial_{T} \partial_{W}
\nonumber\\
&&
+4\delta^{(W}_{i}\delta^{T)}_{j}\delta^{(\bar{0}}_{\Bar{k}}\delta^{\bar{W})}_{\Bar{l}} \partial_{\bar{0}} \partial_{\bar{W}} \partial_{T} \partial_{W}
+2\delta^{(W}_{i}\delta^{T)}_{j}\delta^{\bar{K}}_{\Bar{k}}\delta^{\bar{W}}_{\Bar{l}} \partial_{\bar{K}} \partial_{\bar{W}} \partial_{T} \partial_{W}
+2 \delta^{(W}_{i}\delta^{T)}_{j}\delta^{\bar{W}}_{\Bar{k}}\delta^{\bar{L}}_{\Bar{l}}\partial_{\bar{L}} \partial_{\bar{W}} \partial_{T} \partial_{W}
\nonumber\\
&&
+4 \delta^{(W}_{i}\delta^{T)}_{j}\delta^{(\bar{W}}_{\Bar{k}}\delta^{\bar{T})}_{\Bar{l}}\partial_{\bar{T}} \partial_{\bar{W}} \partial_{T} \partial_{W}
\end{eqnarray}




\subsection*{Acknowledgments} 
M.P.\ is supported in part by NSF grant PHY-1915219.



\begin{thebibliography}{99}
\bibitem{lhc}
M.~Aaboud \textit{et al.} [ATLAS],
``Search for supersymmetry in final states with two same-sign or three leptons and jets using 36 fb$^{-1}$ of $\sqrt{s}=13$ TeV $pp$ collision data with the ATLAS detector,''
JHEP \textbf{09} (2017), 084
[erratum: JHEP \textbf{08} (2019), 121]
[arXiv:1706.03731 [hep-ex]];
A.~M.~Sirunyan \textit{et al.} [CMS],
``Search for supersymmetry in proton-proton collisions at 13 TeV in final states with jets and missing transverse momentum,''
JHEP \textbf{10} (2019), 244
[arXiv:1908.04722 [hep-ex]].

\bibitem{fvp}
D.~Z.~Freedman and A.~Van Proeyen,
``Supergravity,''   Cambridge University Press (2012).



\bibitem{acik} I. Antoniadis, A. Chatrabhuti, H. Isono, and R. Knoops, ``The cosmological constant in supergravity,'' Eur. Phys. J. C (2018) {\bf 78}:718 [arXiv:1805.00852 [hep-th]]

\bibitem{ar} I. Antoniadis and F. Rondeau,  ``New K\"ahler invariant Fayet-Iliopoulos terms in supergravity and cosmological applications,'' Eur. Phys. J. C (2020) 80:346 [arXiv:1912.08117 [hep-th]]

\bibitem{oldACIK} I. Antoniadis, A. Chatrabhuti, H. Isono, and R. Knoops, ``Fayet-Iliopoulos terms in supergravity and D-term inflation,'' Eur. Phys. J. C (2018) 78:366 [arXiv:1803.03817 [hep-th]]
\bibitem{cftp}N. Cribiori, F. Farakos, M. Tournoy, A. Van Proeyen, ``Fayet-Iliopoulos terms in supergravity without gauged R-symmetry,'' J. High Energy Phys. 04 (2018) 032, arXiv:1712 .08601[hep -th].
\bibitem{akk} 
Y.~Aldabergenov, S.~V.~Ketov and R.~Knoops,
``General couplings of a vector multiplet in $N=1$ supergravity with new FI terms,'' Phys. Lett. B \textbf{785}, 284-287 (2018)
doi:10.1016/j.physletb.2018.07.072
[arXiv:1806.04290 [hep-th]].

\bibitem{Kuzenko} S. M. Kuzenko, ``Taking a vector supermultiplet apart: Alternative Fayet-Iliopoulos- type terms,'' arXiv:1801.04794 [hep-th].

 \bibitem{fkr} 
 F.~Farakos, A.~Kehagias and A.~Riotto,
``Liberated $ \mathcal{N} $ = 1 supergravity,''
JHEP \textbf{06}, 011 (2018)
[arXiv:1805.01877 [hep-th]].

\bibitem{jp1} H. Jang and M. Porrati, ``Constraining Liberated Supergravity,'' Phys. Rev. {\bf D 103}, 025008 (2021) [arXiv:2010.06789 [hep-th]].

\bibitem{jp2} 
H.~Jang and M.~Porrati,
``Inflation, gravity mediated supersymmetry breaking, and de Sitter vacua in supergravity with a K\"ahler-invariant Fayet-Iliopoulos term,''
Phys. Rev. D \textbf{103}, 105006 (2021) [arXiv:2102.11358 [hep-th]].

 \bibitem{cfgvnv} E.~Cremmer, B.~Julia, J.~Scherk, S.~Ferrara, L.~Girardello and P.~van Nieuwenhuizen,
``Spontaneous Symmetry Breaking and Higgs Effect in Supergravity Without Cosmological Constant,''
Nucl. Phys. B \textbf{147}, 105 (1979); E.~Cremmer, S.~Ferrara, L.~Girardello and A.~Van Proeyen,
``Yang-Mills Theories with Local Supersymmetry: Lagrangian, Transformation Laws and SuperHiggs Effect,''
Nucl. Phys. B \textbf{212}, 413 (1983).

\bibitem{Linear} S. Ferrara, R. Kallosh, A. V. Proyen, and T. Wrase, ``Linear versus non-linear supersymmetry, in general,'' JHEP \textbf{04}, 065 (2016) [arXiv:1603.02653 [hep-th]].




\bibitem{kyy} T. Kugo, R. Yokokura, and K. Yoshioka, ``Component versus superspace approaches to $D=4$, $\mathcal{N}=1$ conformal supergravity,'' Prog. Theor. Exp. Phys. \textbf{2016}, 073B07 (2016) [arXiv:1602.04441 [hep-th]].

\bibitem{SUGRAprimer}
M.~Rausch de Traubenberg and M.~Valenzuela, ``A Supergravity Primer: From Geometrical Principles to the Final Lagrangian,'' World Scientific (2020).

\bibitem{swampland} C. Vafa, “The String landscape and the swampland,” hep-th/0509212 [hep-th]; T. D. Brennan, F. Carta and C. Vafa,“The String Landscape, the Swampland, and the Missing Corner,” PoS TASI2017, 015 (2017) [arXiv:1711.00864 [hep-th]]; E. Palti,“The Swampland: Introduction and Review,” Fortsch. Phys. 67, no.6, 1900037 (2019) [arXiv:1903.06239 [hep-th]].

\end{thebibliography}
\end{document}